\documentclass{article}
\usepackage{graphicx} 
\usepackage{float}
\usepackage{xcolor}
\usepackage{natbib}
\usepackage{soul}
\usepackage{hyperref}
\usepackage{siunitx}
\usepackage{subcaption}
\usepackage{mathtools}

\title{AI Assisted Experiment Control and Calibration}
\author{
    T. Britton\textsuperscript{1},
    M. Goodrich\textsuperscript{1},
    N. Jarvis\textsuperscript{2},\\
    T. Jeske\textsuperscript{1}\thanks{roark@jlab.org - corresponding author},
    N. Kalra\textsuperscript{1}
    D. Lawrence\textsuperscript{1}\thanks{davidl@jlab.org - Principal Investigator},
    D. McSpadden\textsuperscript{1},
}

\date{December 2023}

\begin{document}

\maketitle

\vspace{-6.5cm}
\hspace*{\fill}v1.3
\vspace{6.5cm}

\noindent
\textsuperscript{1}Thomas Jefferson National Accelerator Facility\\
\textsuperscript{2}Carnegie Mellon University

\tableofcontents
\section{Introduction}
The Thomas Jefferson National Accelerator Facility (JLab) is home to the Continuous Electron Beam Accelerator Facility (CEBAF), four experimental halls, the Superconducting Radio Frequency Institute, and numerous other facilities. CEBAF delivers an approximately 12 GeV polarized electron beam to the experimental halls, each of which contains unique, complex detector systems. The primary focus of this work involves detector systems that comprise the GlueX  spectrometer \cite{ADHIKARI2021164807} located in Hall-D. Additionally, we include exploratory data analysis for detector systems that comprise the CLAS12 spectrometer \cite{BURKERT2020163419} located in Hall-B. 

Traditional calibrations take place in an offline setting and in most cases, after data has been taken. 
The timeline to complete calibrations can range from months to years, which presents a significant delay in publishing. 
Differences in calibration procedures and the frequency with which calibrations are done presented a significant roadblock for this project regarding Hall-B, as they do not calibrate every run, a distinction from the Hall-D practice. 
Since we would ideally train a model to predict calibration constants from previous run periods, the data used for training models for the CLAS12 drift chambers was unavailable. 
In contrast, Hall-D stores calibration constants from every run in a calibration constants database, which provides a wealth of target values with which to train a model. 

Given the novelty of the proposed project, it is important to have a team of scientists with diverse expertise. The project team is comprised of members from the Experimental Physics Software and Computing Infrastructure group (EPSCI): Torri Jeske, Thomas Britton, Michael Goodrich, and David Lawrence, Diana McSpadden from the newly formed Data Science Department, and Naomi Jarvis from Carnegie Mellon University. Diana McSpadden (data scientist) and Torri Jeske (physicist) were hired for this project in May and April of 2021, respectively. An additional data scientist, Nikhil Kalra, was hired in FY21Q4. 

\section{Project Objectives}
This project was submitted to the Department of Energy in response to the announcement ``Data, Artificial Intelligence, and Machine Learning at DOE Scientific User Facilities". The main objectives for the project are briefly summarized below: 
\begin{enumerate}
    \item Control the HV of the GlueX CDC and simultaneously generate calibration constants autonomously in near real-time. Resulting in:
    \begin{enumerate}
    \item Stabilization of the gain to within 5\% over a two-week period with no measurable degradation of the timing resolution. 
    \item Reduction in time required from experts for data monitoring and calibration.
    \item Reduction in time before physics analysis can begin
    \end{enumerate}
    \item Develop a similar control system for CLAS12 drift chambers
\end{enumerate}
The initial objectives regarding the GlueX Central Drift Chamber have been met. As stated previously, a similar control system for the CLAS12 drift chambers was not developed as the training data is not readily available. For this reason, milestones were updated throughout the duration of the project to determine other suitable detector systems for ML based calibration and control.

\subsection{Control of the GlueX Central Drift Chamber}
We proposed to control the high voltage (HV) of the drift chamber anode wires, dynamically, using an AI system that incorporates information from various sources in the experimental Hall during data taking. 
The model will predict a set of calibration values and adjust the HV such that the response of the detector is stable in the presence of varying environmental and experimental conditions. Such a system would decrease the time that experts are required to spend on calibration. 
Specifically, the drift chamber gain would be stabilized within 5\% over a two week time period with no measurable degradation of the timing resolution. 
A successful application with the GlueX CDC would be used to develop similar systems for other detectors. 

\subsection{Reduce time commitment required for data monitoring and calibration}
AI would reduce the time commitment required from the experts, who would then be free to turn their attention to data analysis and our physics program. 
An AI would also be able to take over some responsibilities from the shift crew in charge of the detectors and the data acquisition.
As the AI would be monitoring the environmental conditions, it would be aware of unusual weather events, such as the rapidly falling pressure that occurs when a storm front arrives, and it would be able to alert the crew of the need to stop a run early or take the appropriate corrective steps on its own and continue running, all without requiring expert intervention.

\subsection{Reduce delay between data taking and publications}

For this project, the objective is to reduce the number of iterations required to calibrate the
CDC or reduce the number of runs that must be calibrated.
Stabilizing the detector operation reduces or even eliminates the need to calibrate the data from the detector. 
This not only improves the quality of the data but reduces the time spent calibrating the data before the physics analysis can begin. 
This reduces the time between running the experiment and publishing the results. 

\subsection{Facilitate automation of other detector systems}
Other detector systems, namely the Forward Drift Chamber (FDC) and the Forward Calorimeter (FCAL) in Hall-D and the Electromagnetic Calorimeter (ECAL) in Hall-B were studied as part of this project.

\section{Traditional Detector Calibration and Operation}
The aim of this section is to describe the traditional calibrations in use for the operation of the central and forward drift chambers and the FCAL located in Hall-D. 
It is important to understand the existing calibration processes as the calibration constants obtained from them are the target values for the models that are developed. 

\subsection{GlueX Central Drift Chamber}

\begin{figure} 
    \centering
    \includegraphics[width=0.5\textwidth]{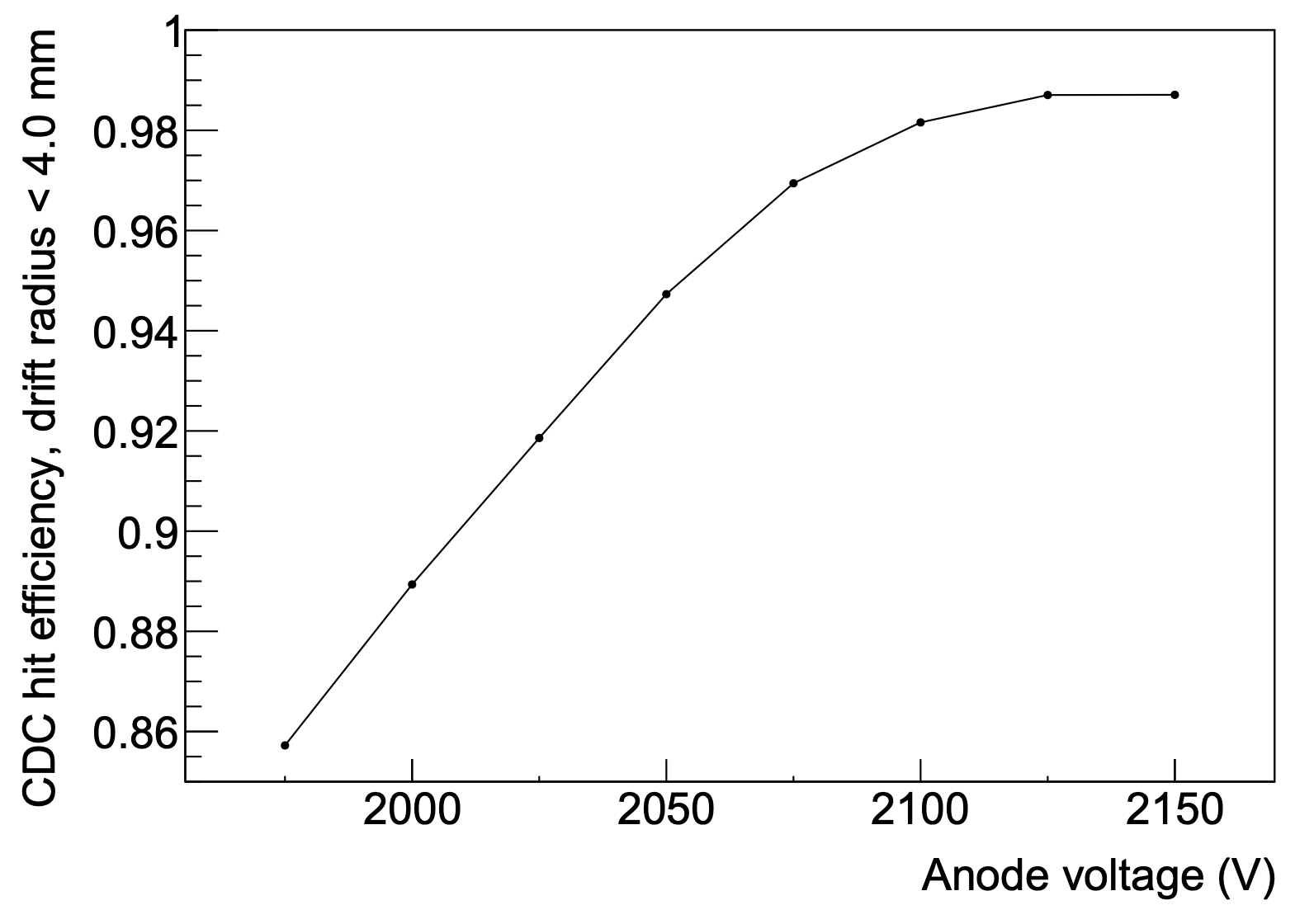}
    \caption{Hits per track efficiency as a function of anode voltage. The anode voltage of 2125 V was chosen to give the maximum hit efficiency for small drift radii.}
    \label{fig:cdcOperatingHV}
\end{figure}

The GlueX Central Drift Chamber is a cylindrical, straw tube wire chamber located inside the solenoid of the GlueX detector. It contains a 50:50 mixture of $Ar:CO_2$ gas at atmospheric pressure. The operating anode voltage is held at 2125 V. This voltage was determined using a HV scan and calculating the hits-per-track efficiency as a function of high voltage. 
The operating voltage was selected based on the location of the start of the plateau. Fig. \ref{fig:cdcOperatingHV} shows the relationship between the hits-per-track efficiency and high voltage. 
The CDC's purpose is to detect and track charged particles with momenta as low as 0.25 GeV/c; additionally, it provides particle identification via dE/dx.

\begin{figure} 
    \centering
    \includegraphics[width=0.9\textwidth]{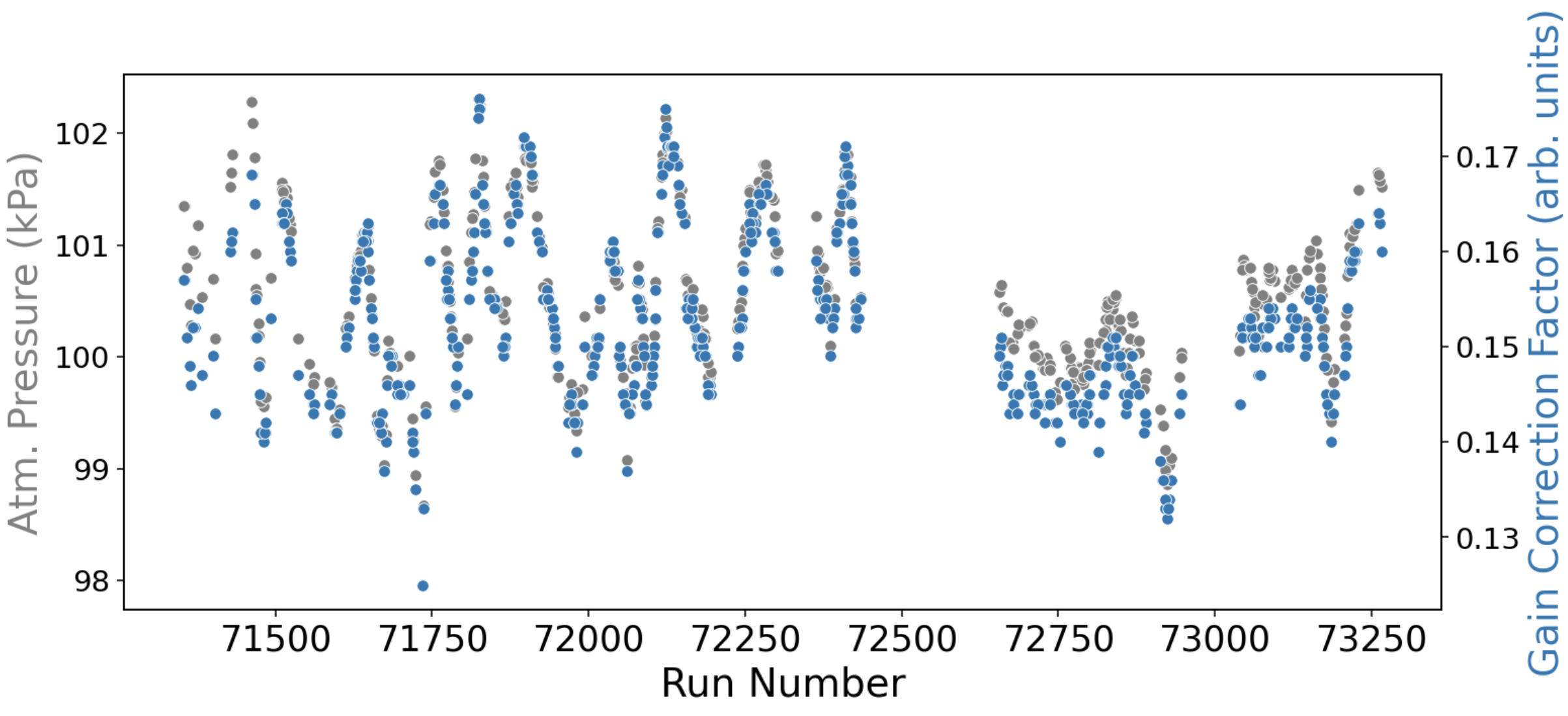}
    \caption{Mean atmospheric pressure (gray) and gain correction factor (blue) as a function of run number during the GlueX 2020 Run Period.} 
    \label{fig:pressure_gcf}
\end{figure}

The CDC has two main calibrations performed on a run-by-run basis: the chamber gain and drift-time to drift-distance. 
The chamber gain is sensitive to the atmospheric pressure, gas temperature, and current drawn from the high voltage boards. 
As the atmospheric pressure varies throughout data taking, this fluctuation is accounted for via calibration \textit{after} data taking. An example of this behavior for the GlueX 2020 run period is shown in Fig. \ref{fig:pressure_gcf}. 

The drift-time to drift-distance calibration requires many more iterations of the calibration software. The six calibration constants correspond to the fit parameters of the function: 
\begin{equation}
    d(t) = f_{\delta}\left(\frac{d_0(t)}{f_0}P + 1 - P\right)
\end{equation}
where 
\begin{equation}
    f_{\delta} = a\sqrt{t} + bt + ct^3,
\end{equation} 
where 
\begin{equation}
    f_0 = a_1\sqrt{t} + b_1t + ct^3,
\end{equation}
and where
\begin{subequations}
\begin{align}
    a &= a_1 + a_2|\delta| \\
    b &= b_1 + b_2|\delta| \\
    c &= c_1 + c_2|\delta|
\end{align}
\end{subequations}
where $\delta$ is the magnitude of the deflection for each straw.  
P is a piece-wise function of the form
\[ P=
    \begin{dcases}
        0 & t > T \\
        \frac{T-t}{T} & t \leq T\\
    \end{dcases}
\] where T=250 ns. This value is chosen because drift times less than this are not significantly affected by the distortion of the electric field due to the straw sag. 
A more detailed description of the calculation regarding the straw deformation can be found in \cite{JARVIS2020163727}. 

\begin{figure}
    \centering
    \includegraphics[width=0.9\textwidth]{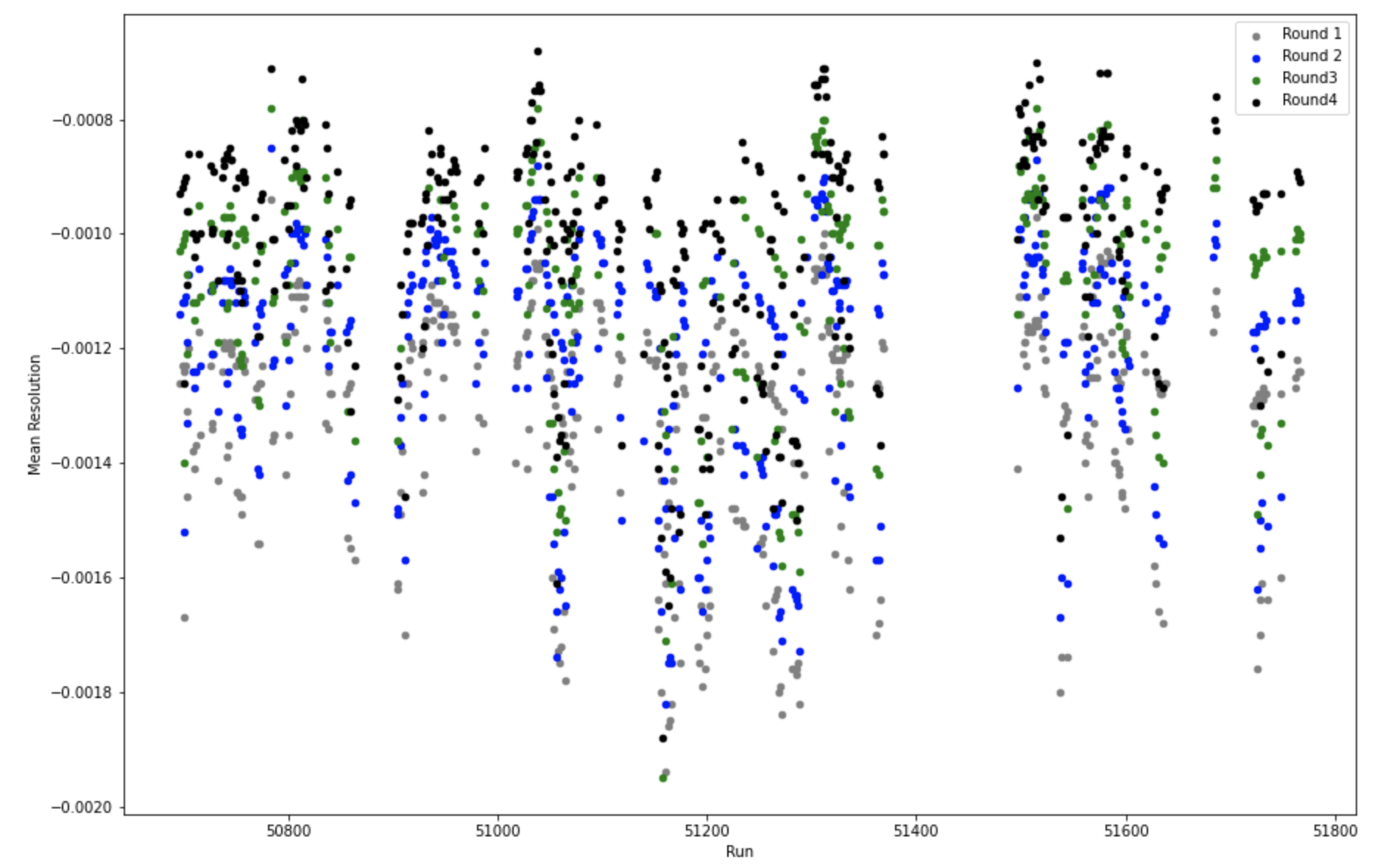}
    \caption{Resolution of fitted track residuals for 4 iterations as a function of run number.  }
    \label{fig:ttod_iterations}
\end{figure}

The default calibration values are updated with each iteration of the calibrations. 
Typically, time-to-distance calibrations are considered finished when the difference between the fit contours from the current iteration and the previous iteration is minimized. 
There are about 7-10 iterations of the time-to-distance calibration, representing a substantial amount of computing resources and expert time. 
In Fig. \ref{fig:ttod_iterations}, a sample of the drift-time to drift distance calibration constants is shown for four iterations. 
Once the drift time to drift distance calibrations are complete, the chamber gain is calibrated a final time.

Similarly to the gain, the time-to-distance fit parameters discussed are correlated with the gas density. Since the atmospheric pressure and gas temperature are measured continuously throughout data taking, it is possible to determine starting values at the beginning (or end) of a data taking session using the correlations between the fit parameters and gas density. 
We did not use this approach at first, as we did not know how adjusting the HV at the beginning of each run would affect the time-to-distance calibration. 

The chamber gain is calibrated by adjusting the fitted dE/dx peak position at $p=1.5$ GeV/c to match that obtained from simulation. The peak position obtained from simulation is 2.02 keV/cm. 




\section{Development of a CDC control system } \label{sec:dev_CDC_control}
This section describes the initial data extraction, feature engineering, and model development for the CDC control system. 
\subsection{Data analysis \& preparation} \label{sec:data_analysis}
Data were extracted from two sources: the Experimental Physics Industrial Controls System (EPICS) \cite{epics1989}, and the Hall-D Calibration Constants Database \cite{jeffersonlabccdb}.
Data specific to the CDC were extracted from the EPICS archive for use as input variables. 
A program maintained by Jefferson Lab called \texttt{myStats} is used to extract data from the EPICS archive. This command line tool outputs the minimum, mean, maximum, and standard deviation values for requested process variables. These process variables correspond to the input features used for training our model. Initially, the electron beam current, atmospheric pressure, gas temperature measured from four of the five (D1-D5, excluding D2) thermocouples located inside the CDC, current drawn by the CDC's 72 High Voltage Boards (HVB) in nine rings (A-I rings), and additional engineered features were evaluated for use in a model to predict the GCF for an individual run. 
To correlate the input features with the calibration constants, the input features were extracted from EPICS during the time period in which the experimental data was taken.

The CCDB contains the calibration values for all sub-detector systems and is the source of our target value, the CDC Gain Correction Factor (GCF), for each experimental run. 
Features from EPICS are readily available throughout data-taking, which is important for online control. 
Some features, e.g. particle momentum, while correlated with gain, were excluded as their calculation requires computationally expensive track reconstruction, making them unsuitable for real-time control. 

Balancing the distributions of atmospheric pressure between training and test data sets was motivated by the scarcity of extreme values in atmospheric pressure. 
A low pressure threshold of $\leq$ 99.27 kPa and a high pressure threshold of $\geq$ 101.57 kPa provided similar pressure distributions and adequate diversity of gas temperature and HV board current values for low, medium, and high pressure data within the pressure-balanced training and test data sets. 

\subsection{Model development}
Using data from experiments performed in 2018 and 2020, a deep feed-forward neural network was trained with 122 variables. 
Linear regression (LR), deep feed-forward artificial neural network (ANN), Random Forest (RF) \cite{breiman2001random}, XGBoost \cite{XGBoost}, and Gaussian process regression (GPR) \cite{NIPS1995_7cce53cf} algorithms were compared to predict GCF using different combinations of input features. 
Comparisons were conducted using both the 2018 and 2020 data and, alternatively, solely the 2020 data set. 

There is a noticeable offset in the relationships between the input features and target values between the 2018 and 2020 run periods. The alcohol solution used in the CDC gas system bath was changed from 1-propyl alcohol to isopropanol \cite{JARVIS2020163727} after the 2018 run period. The 2018 data was set aside until a procedure was developed to make the GCF values compatible.
Results from the ML algorithm comparisons, neural architecture search (NAS), and hyperparameter search are presented in Table \ref{tab:model_results}. 
Training and testing with only the 2020 dataset resulted in lower Mean Absolute Percent Error (MAPE) and significantly lower max percent error, attributed to the different alcohols used in the CDC during the 2018 and 2020 run periods.

The Gaussian Process Regression (GPR) model with 11 features exhibited the lowest MAPE. 
Notably, two of the input features in this model are sourced from the Pair Spectrometer (PS). 
If PS is unused or unavailable, we do not want the CDC control system to be impacted; thus, we deliberately made this choice to eliminate this model from further investigation.
Additionally, features from other sub-detectors were found to be correlated with GCF, but were excluded from further consideration to ensure the CDC control system is self-contained, ensuring no data from other detectors is used. 
This is especially important if some sub-detectors are not used in future experiments, due to experimental design, maintenance, or other other unforeseen reasons.

Shapley \cite{shapely} values were calculated to measure the average marginal contribution of each feature across possible combinations of features. Shapley analysis identified three variables to predict GCF. 
Reassuringly, these were consistent with known drift chamber behavior: the mean atmospheric pressure, the maximum gas temperature, and an engineered feature:  the sum of the maximum gas temperatures and mean current drawn by the innermost high voltage boards on the CDC (a CDC-specific proxy for the rate of charged particles passing through the detector). Again, the statistical calculations are from the JLab \texttt{myStats} data extraction software. They represent the mean and maximum calculation of EPICS variables for the date and time ranges of the data files used for an experimental run's traditional calibration.
The GPR with three features resulted in the lowest max percent error, and utilized three input features: 
\begin{enumerate}
    \item the mean atmospheric pressure,
    \item the max temperature from the D1 thermocouple,
    \item and an engineered feature of the sum of the max temperature from the D1 thermocouple and the max HV board current from the CDC’s A HV board.
\end{enumerate}

These three input features agree with the CDC expert's knowledge of the operational behavior of the CDC's gain in response to experimental and environmental conditions.
After model comparison and analysis, consultation with the CDC expert, and discussions with Hall-D operations staff regarding implementation considerations, three input features were finalized:
\begin{enumerate}
    \item MEAN\_PRESSURE: the mean atmospheric pressure (kPa)
    \item MAX\_TEMP: maximum gas temperature (K)
    \item MEAN\_HVBI: mean current drawn by the innermost high voltage (HV) boards on the CDC ($\mu$A)
\end{enumerate}

\begin{figure}
    \centering
    \includegraphics[width=\textwidth]{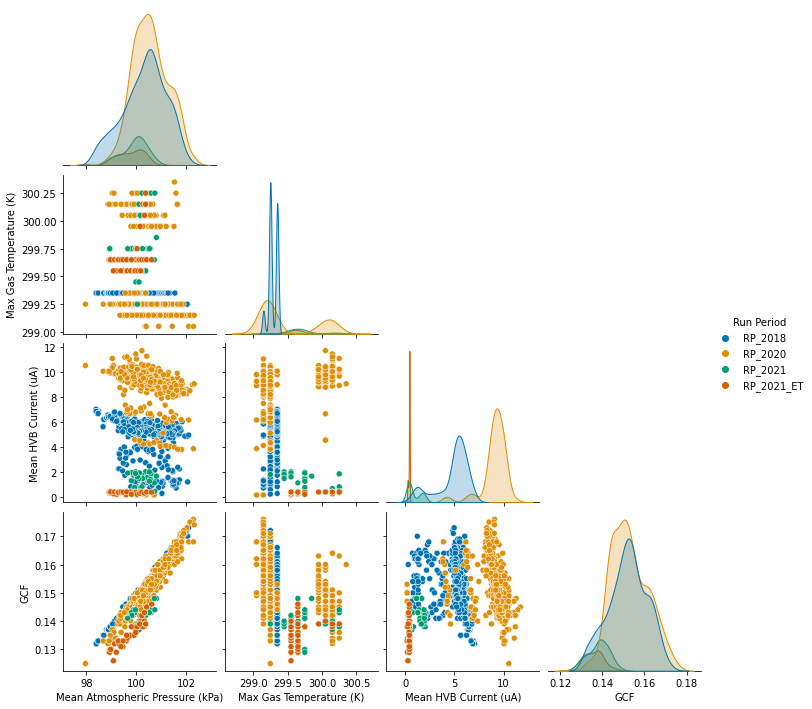}
    \caption{Pair plot displaying the relationship of input features - mean atmospheric pressure, maximum gas temperature, and mean high voltage board current - to the target value, GCF, for multiple different run periods}
    \label{allFeaturesPairPlot}
\end{figure}

A pair plot illustrating the final input features from multiple Run Periods is shown in Fig. \ref{allFeaturesPairPlot}. 

\begin{table}
\centering
\scriptsize
\begin{tabular}{rlccccc}
\hline

Model type & Run period(s) & \# features & MAPE & Max \% err   & Ratio $>$ 1\% err  \\ \hline
LR & 2018 \& 2020  & 11 & 1.3\% & 19.1\% & 97\slash164  \\
LR & 2020 & 11 & 0.7\% & 2.0\% & 30\slash106  \\
LR & 2018 \& 2020  & 5 & 2.3\% & 20.3\% & 96\slash164   \\
LR & 2020 & 5 & 0.7\% & 2.6\% & 26\slash106 \\ \hline

RF & 2018 \& 2020  & 82 & 1.7\% & 18.5\% & 83\slash164 \\ \hline

ANN-7 layers & 2018 \& 2020  & 122 & 1.8\% & 11.4\% & 74\slash164 \\
ANN-4 layers & 2018 \& 2020  & 122 & 1.9\% & 10.8\% & 84\slash164 \\ 
ANN-3 layers & 2018 \& 2020  & 122 & 1.9\% & 11.9\% & 90\slash164 \\ \hline

XGBoost & 2018 \& 2020  & 122 & 1.6\% & 10.2\% & 76\slash164 \\ 
XGBoost & 2018 \& 2020 & 82 & 1.4\% & 11.8\% & 68\slash164 \\
XGBoost & 2018 \& 2020  & 71 & 1.6\% & 11.2\% & 71\slash164 \\
XGBoost & 2018 \& 2020  & 12 & 1.8\% & 11.1\% & 72\slash164 \\ \hline

GPR & 2018 \& 2020  & 25 & 1.7\% & 10.9\% & 80\slash164 \\
GPR & 2018 \& 2020  & 14 & 1.5\% & 9.7\% & 66\slash164 \\
GPR & 2018 \& 2020  & 11 & 1.5\% & 10.1\% & 72\slash164 \\

\textbf{GPR} & \textbf{2020} & \textbf{11} & \textbf{0.5\%} & 4.1\% & \textbf{17\slash106} \\
GPR & 2018 \& 2020  & 5 & 1.5\% & 9.1\% & 70\slash164 \\
GPR & 2020 & 5 & 0.7\% & 3.6\% & 28\slash106 \\
GPR & 2018 \& 2020  & 3 & 1.8\% & 12.0\% & 74\slash164 \\
\textbf{GPR} & \textbf{2020} & \textbf{3} & 0.7\% & \textbf{1.9\%}& 25\slash106 \\

GPR & 2018 \& 2020  & 1 & 3.4\% & 18.6\% & 129\slash164 \\
GPR & 2020  & 1 & 1.1\% & 4\% & 49\slash106 \\
 \hline
\end{tabular}
\caption{Results from algorithm comparison, neural architecture search, and hyperparameter optimization. Mean Absolute Percent Error (MAPE), the maximum percent error, and the ratio of observations with greater than 1\% error are reported. The features included in the comparisons are presented in Table \ref{tab:model_inputs}. }
\label{tab:model_results}
\end{table}

\begin{table}
\centering
\scriptsize
\begin{tabular}{p{.9in}p{.5in}p{4in}}
\hline

Model type & \# features   & Description  \\ \hline \\
LR \& GPR & 11 & Mean and max beam current, Mean and max pair spectrometer rate, Mean and max atmospheric pressure, Max temperature from D1 thermocouple, Max temperature from D5 thermocouple, Mean A HV board current, Max A HV board current, (Max A HV board current - Max I HV board current)  \\ \\
LR \& GPR & 5 & Mean beam current, Mean pair spectrometer rate, Mean atmospheric pressure, Max temperature from D1 thermocouple, Max A HV board current, Max A HV board current \\ \\

RF \& XGBoost & 82 & Features with Pearson correlation coefficient $>$ 0.2 with GCF. Features do not include variables only available post-track reconstruction but do include beam current, engineered features, and features from sub-detectors other than the CDC.  \\ \\

ANN \& XGBoost & 122 & All EPICs-only features. Features do not include variables only available post-track reconstruction but do include beam current, engineered features, and features from sub-detectors other than the CDC.  \\ \\

XGBoost & 122 & All EPICs-only features. Features do not include variables only available post-track reconstruction but do include beam current, engineered features, and features from sub-detectors other than the CDC.  \\ \\

XGBoost & 71 & Features with Pearson correlation coefficient $>$ 0.3 with GCF. Features do not include variables only available post-track reconstruction but do include beam current, engineered features, and features from sub-detectors other than the CDC.  \\ \\

XGBoost & 12 & Features with Pearson correlation coefficient $>$ 0.4 with GCF. Features do not include variables only available post-track reconstruction but do include beam current, engineered features, and features from sub-detectors other than the CDC.  \\ \\

GPR & 25 & Max, mean, min, and standard deviation of the beam current; max, mean, min, and standard deviation of the pair spectrometer rate; max, mean, min, and standard deviation of the atmospheric pressure; max, mean, min, and standard deviation of the temperature of D1 thermocouple; max, mean, min and standard deviation of the temperature of the D5 thermocouple; max, mean, min, and standard deviation of the current of the CDC A HV board, and the engineered feature that is the difference of the max of the current of the A CDC HV board and the max of current of the I CDC HV board  \\ \\

GPR & 14 & Max and mean beam current; max and mean pair spectrometer rate; max and mean atmospheric pressure; max and mean temperature from the D1 thermocouple; max from the D5 thermocouple; engineered feature of the difference between the means of the temperatures of the D1 and D5 thermocouples; max and mean current from the A CDC HV board, an engineered feature the difference between the max current from the A and I CDC HV boards, and an engineered feature the sum of the means of the A, B, C, D, E, F, G, H, and I CDC HV boards. \\ \\

GPR & 3 & Mean atmospheric pressure, Max temperature from the D1 thermocouple, engineered feature of the sum of the max temperature from the D1 thermocouple and the max HV board current from the CDC's A HV board.  \\ \\

GPR & 1 & Mean atmospheric pressure  \\ \\

 \hline
\end{tabular}
\caption{Descriptions of the input features used for models shown in Table \ref{tab:model_results}}. 
\label{tab:model_inputs}
\end{table}

\begin{figure}
    \centering
    \includegraphics[width=0.6\textwidth]{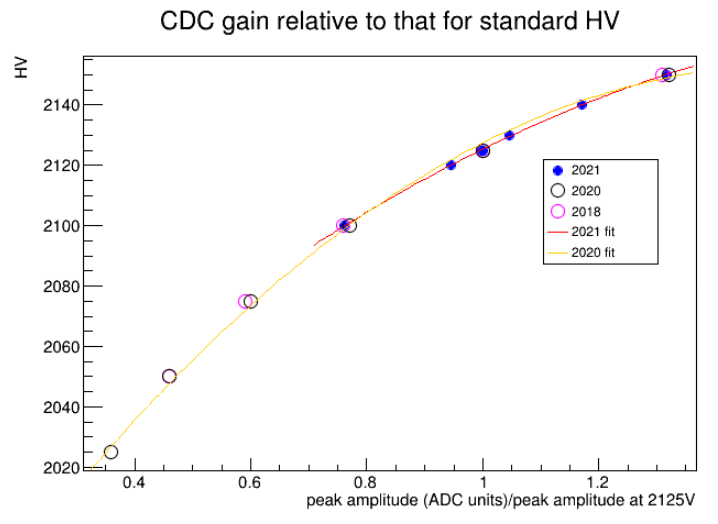}
    \caption{Results of high voltage scans for 2018, 2020, and 2021 run periods.}
    \label{fig:HVScans}
\end{figure}

Using the predicted GCFs, the HV setting is determined from a second-order polynomial fit to the HV as a function of the peak amplitude relative to that obtained at the nominal operating voltage. For this, historical data from high voltage scans are used. High voltage scans are taken at the start of each new Run Period. Example high voltage scans are shown in Fig. \ref{fig:HVScans}.

By exploiting the empirically derived correlation of peak height and HV we use the GPR-predicted GCF and an ``ideal" GCF to obtain the HV setting that stabilizes the CDC gain. \\

\subsubsection{Uncertainty Quantification}

Uncertainty quantification has been identified as a priority research opportunity by the US Department of Energy \cite{energygov} and is an important step in creating trustworthy AI. 
Furthermore, the novel use of a data-driven approach to control a particle detector requires the GlueX collaboration and the CDC expert to trust that the system will not interpolate or extrapolate unexpectedly. While a physics-driven algorithm understands the underlying dynamics and even causality, a data-driven approach only understands training data correlations.
At a minimum, we require a system that at inference can inform users when current conditions are well represented within training data and when the system is merely guessing, i.e., uncertainty.
This uncertainty should be distance-aware and well-calibrated such that when conditions stray further from training data, the system becomes less certain of predictions.

The GPR provides a prediction and the associated uncertainty based on the model's training data \cite{NIPS1995_7cce53cf}. 
A well-calibrated GPR will provide high uncertainty estimates in input parameter regions where extrapolation is required \cite{NIPS1995_7cce53cf, bishop2006pattern}. 
A control system can act based on uncertainty thresholds. 
While unlikely due to operational constraints of the Thomas Jefferson National Accelerator Facility accelerator, if environmental or experimental conditions stray into parameter regions resulting in a high or low predicted GCF with high uncertainty, it is critical that we do \emph{not} adjust the high voltage of the CDC to a correspondingly uncertain, or \emph{unsafe}, value. 
Moreover, the novel use of a data-driven approach to control a particle detector requires physicists to trust the data-driven algorithm; otherwise, it would not be implemented. A detailed description of the control system and the policies established is discussed in Section \ref{sec:ControlSystem}.

We investigated the GPR's predictions and standard deviations, i.e., uncertainties, for possible environmental and experimental conditions using combinations of atmospheric pressure between 96.0 kPa and 102.9 kPa, HV board current between 0.2 $\mu$A and 12.0 $\mu$A, and gas temperatures between 297.0 K and 300.9 K. The input features were incremented in steps of 0.2. We expected a correlation of atmospheric pressure to GCF with the correlation varying independently with both track rate and gas temperature; this behavior is typical of drift chambers \cite{Sauli}. 

\begin{figure}[h]
    \centering
    \hspace*{-18mm}   
    \includegraphics[width=1.2\linewidth]{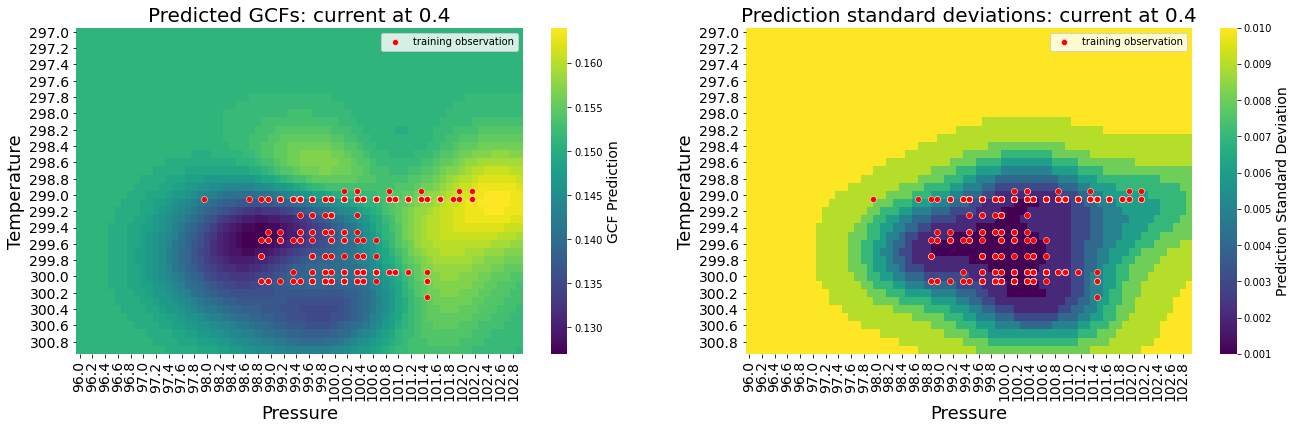}
    \caption{Left: the 2D surface plot of predicted GCF v. atmospheric pressure and gas temperature with a high voltage board current of 0.4 $\mu$A, including training set observations (in red dots). Right: the 2D surface plot of the prediction standard deviation v. atmospheric pressure and gas temperature with a current 0.4 $\mu$A, including training set observations (in red dots). Regions of gas temperature v. atmospheric pressure without training data (the surrounding green areas on the left) correspond to the high uncertainty, yellow areas on the right. }
    \label{fig:heatmap0.4}
\end{figure}

An example of this investigation is shown in Fig. \ref{fig:heatmap0.4}, which illustrates the relationship between atmospheric pressure, gas temperature, and the predicted GCF and associated standard deviations while keeping the HVB current constant (0.4 $\mu$A). These plots demonstrate the behavior for in-training (red points) vs. out-of-training predictions. The predicted GCF reverts to the mean when the standard deviation is over 6\% of the mean GCF. The model standard deviation and illustrations, such as Fig.  \ref{fig:heatmap0.4}, provide useful visuals of the model behavior. Using the uncertainty information, the detector expert developed control protocols for interpolation and extrapolation, such as reverting to the nominal high voltage setting if the model is highly uncertain. This allows the system to collect more training data that could be used for later training to expand the certainty manifold. 

\section{Deployments}
\begin{figure} 
    \centering
    \includegraphics[width=0.7 \textwidth]{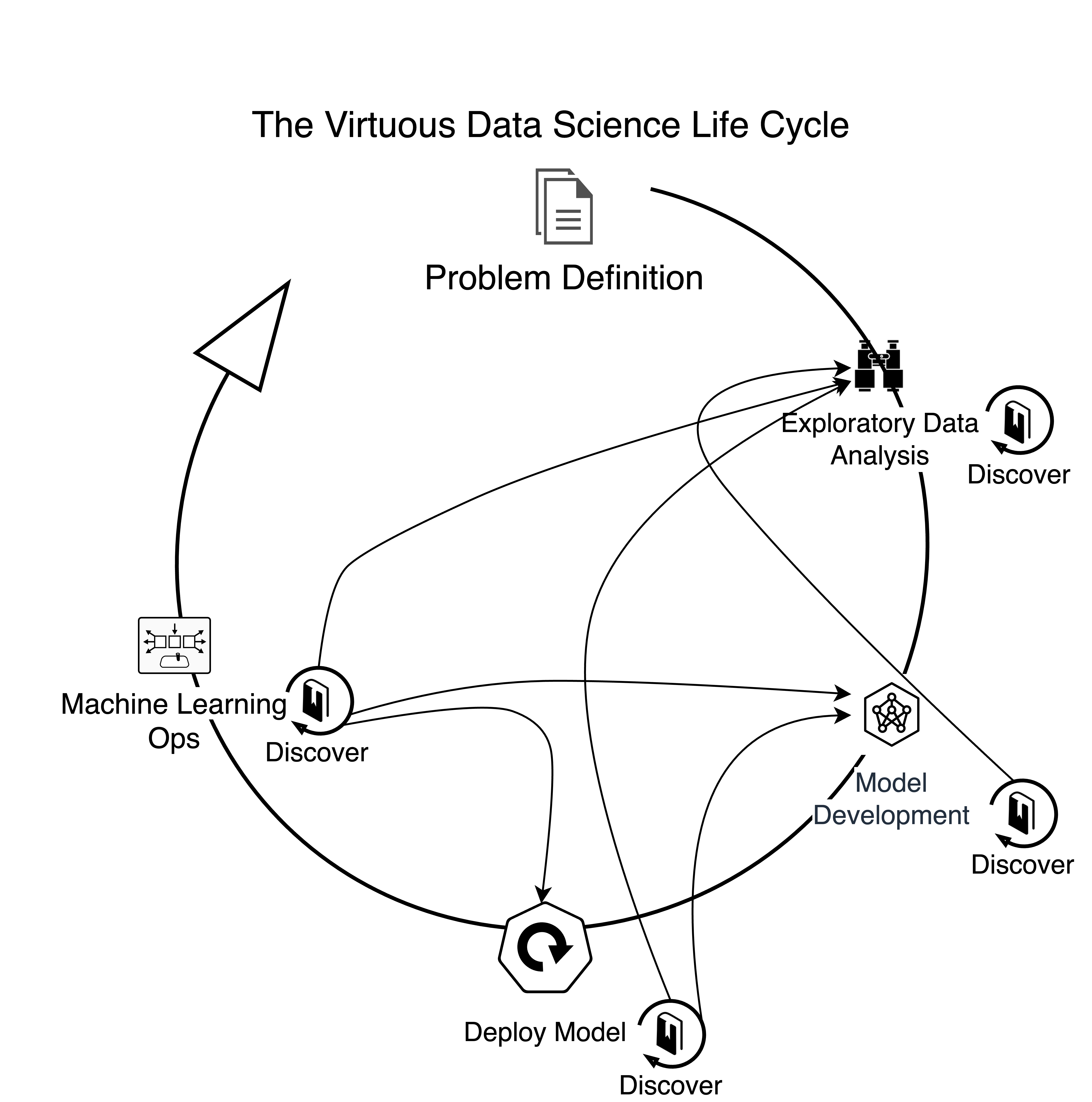}
    \caption{The Virtuous Data Science Life Cycle.} \label{fig:ds_lifecycle}
\end{figure}

A CDC control system was developed using a virtuous cycle of research, model development, model deployment, and evaluation of the in-production model. 
The virtuous development-delivery cycle is used for the collaboration of research and industry in the fields of software and hardware development and data science \cite{Brodie2019}.
Each step in the cycle informs subsequent phases and leads to continuous innovation to address the defined problem. As Fig. \ref{fig:ds_lifecycle} illustrates, discoveries in later phases may deliver insights that result in revisiting previous phases. 

This section summarizes the performance of the GP model and the control policies established based on the uncertainty quantification and insights learned while controlling the GlueX CDC during experiments that took place in Hall-D. 

\subsection{Deployment \#1: PrimEx 2021}
For the first deployment, the shift crew was provided an executable script before the start of each experimental run. The script read the required input features from EPICS, ran inference, and a HV setting was recommended, rounded to the nearest 5V, in order to stabilize the gain. For empty target runs, the shift crew was instructed to return the HV to 2125 V. In some instances, these instructions were ignored. Ideally, we would have required the shift crew to return the CDC to 2125 V prior to running the script to determine the next HV setting. Returning the CDC to 2125 V allows us to obtain the correct reading for the high voltage board current at the traditional operating voltage. The shift crew was not asked to do this to save time and to reduce the complexity of the instructions. 
\subsubsection{Performance}
\begin{figure}[htb]
\centering
    \includegraphics[width=0.9\linewidth]{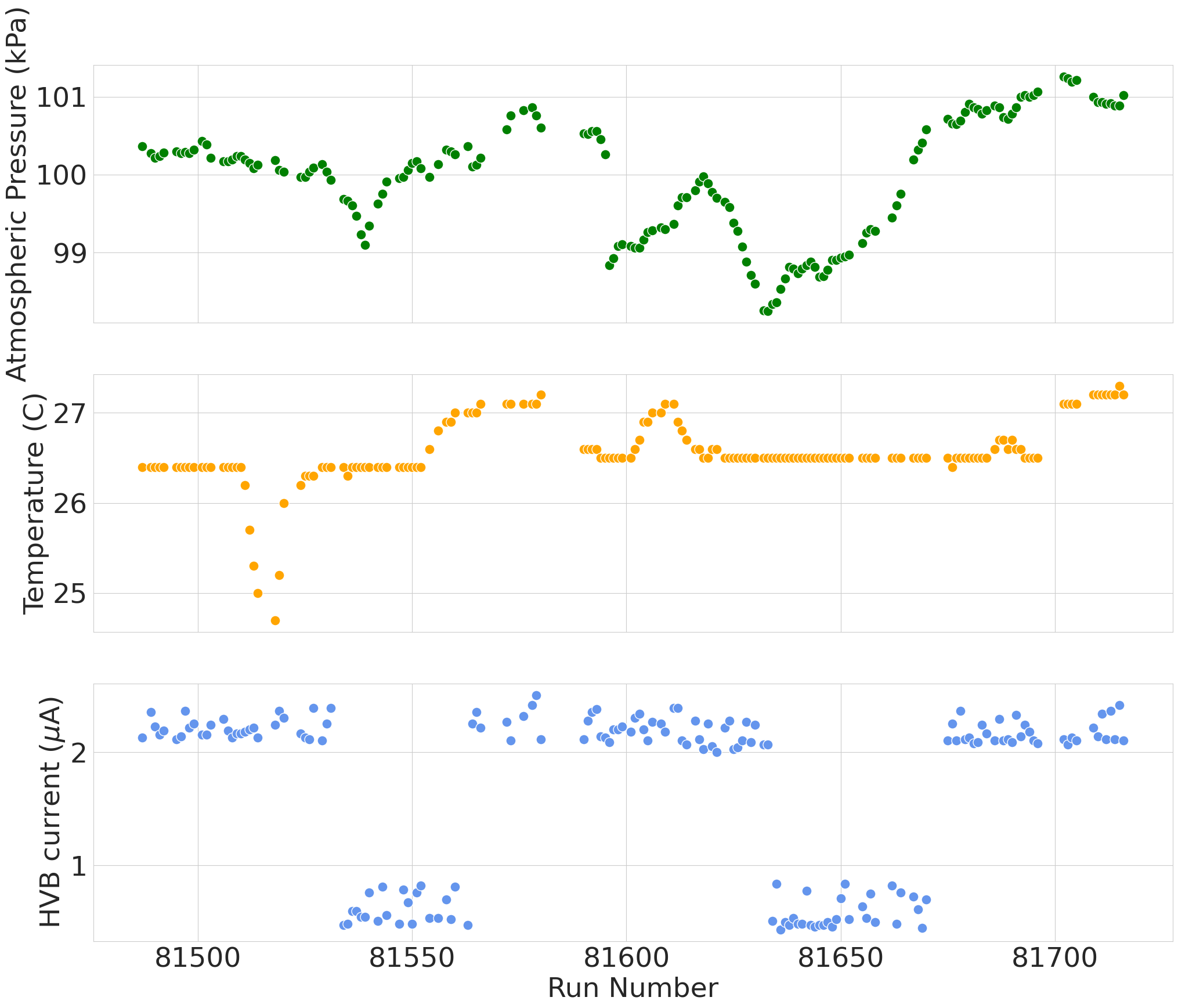}
    \caption{The mean atmospheric pressure (kPa, top row), gas temperature (C, middle row) and current drawn from the CDC HV boards ($\mu$A, bottom row) during the PrimEx 2021 Run Period.}
    \label{fig:epicsVarsPrimex21}
\end{figure}

\begin{figure}[htb]
    \centering
    \includegraphics[width=0.9\linewidth]{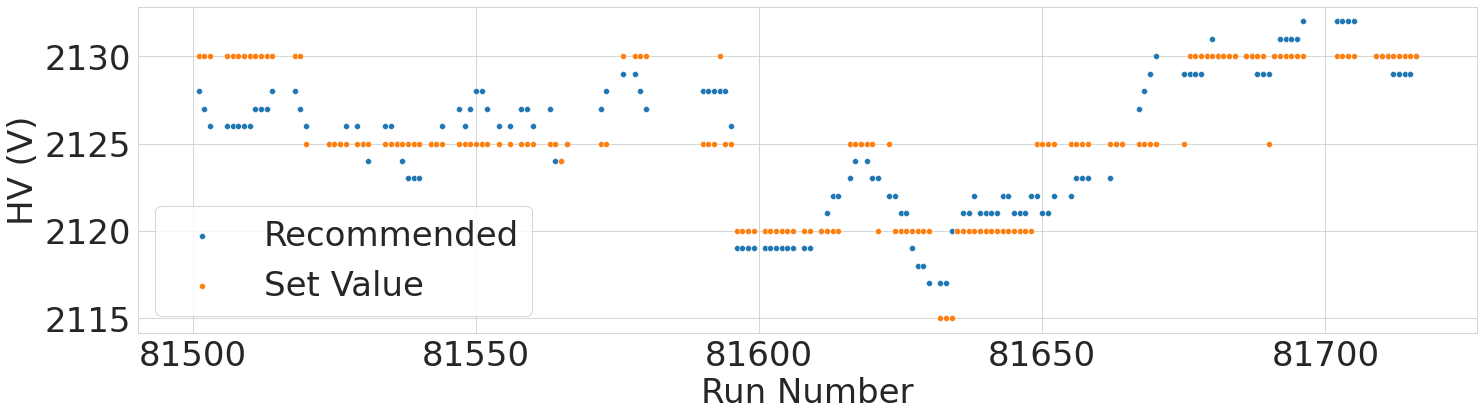}
    \caption{The HV recommendations and set values obtained from the script during the PrimEx 2021 Run Period. }
    \label{fig:hvsetvalues}
\end{figure}

\begin{figure}[htb]
    \centering
    \includegraphics[width=0.9\linewidth]{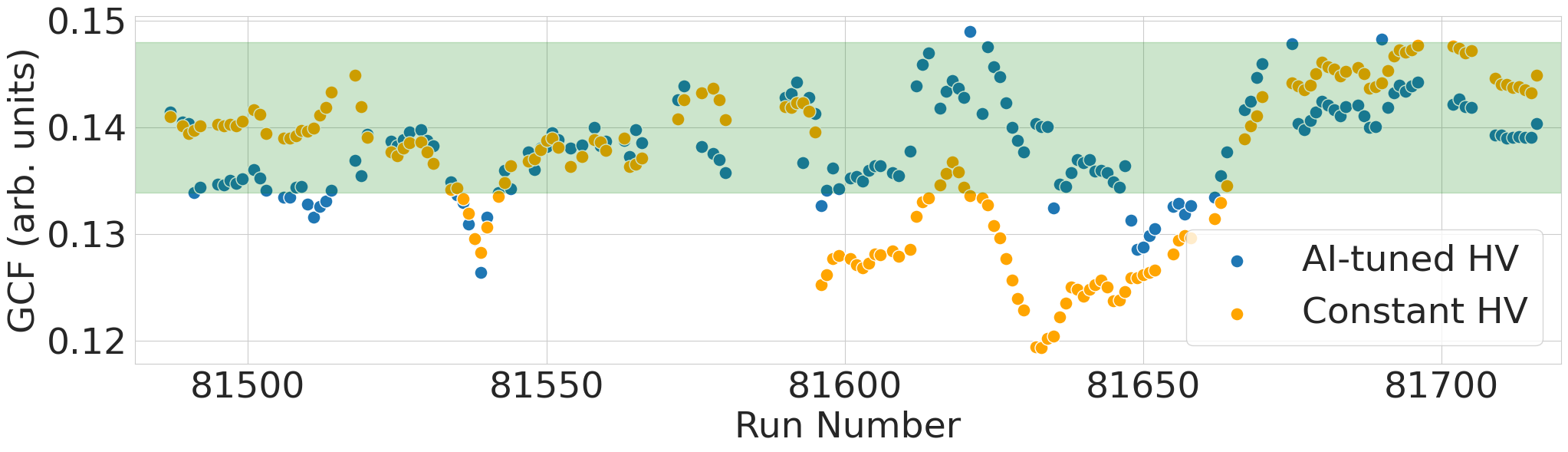}
    \caption{The gain correction factor vs Run Number for runs where the HV was tuned according to the GP versus the gain correction factor obtained from the traditional calibration procedure under similar conditions of atmospheric pressure.}
    \label{fig:AIvsTradGCF}
\end{figure}

The input features used by the GP for inference during the PrimEx 2021 run period are shown in Fig. \ref{fig:epicsVarsPrimex21}. The corresponding HV recommendations from the script and the set values are shown in Fig. \ref{fig:hvsetvalues}. The predicted gain correction factors for the GP-tuned runs and the typical gain correction factor under similar conditions are shown in Fig. \ref{fig:AIvsTradGCF}. The idea behind this figure is to show that the calibration constants predicted from the GP are more stable than those we would have gotten had the traditional calibration been performed. 

\subsubsection{Operational Insights}
It is preferable to not further burden the shift crew with specific operating instructions for specific runs. The shift crews change every 6 hours, and information can be lost from one shift to the next even when clear instructions are present. 

\subsection{Deployment \#2: Cosmics Test, March 2022}
The second deployment implemented a fully autonomous control system for a 2-week collection of cosmic ray data \cite{mcspadden_MLPS}. This made use of cosmic rays passing through the detector instead of charged particles emerging from collisions between the photon beam and target. 
As such, the HV board current given to the model was 9.0 $\mu$A in accordance with the statistical mode of HV board current in our training data. 
In order to compare the behavior of the CDC operating at a constant HV to a GPR-tuned HV, software was used to divide the operation of the CDC into two sections: one was held at 2130 V, and the other was updated to the HV setting determined using the GPR prediction every five minutes. Thomas Britton and Torri Jeske monitored data collection from home primarily to restart the DAQ if necessary and start new runs. Naomi Jarvis participated in The Great Bear Chase, a 50k skiathlon at 0 F in the Upper Peninsula of Michigan. The control of the CDC was truly autonomous. The GP was trained using data from the 2020 and 2021 run periods.

\subsubsection{Performance}
\begin{figure}[htb]
    \centering
    \includegraphics[width=0.95\textwidth]{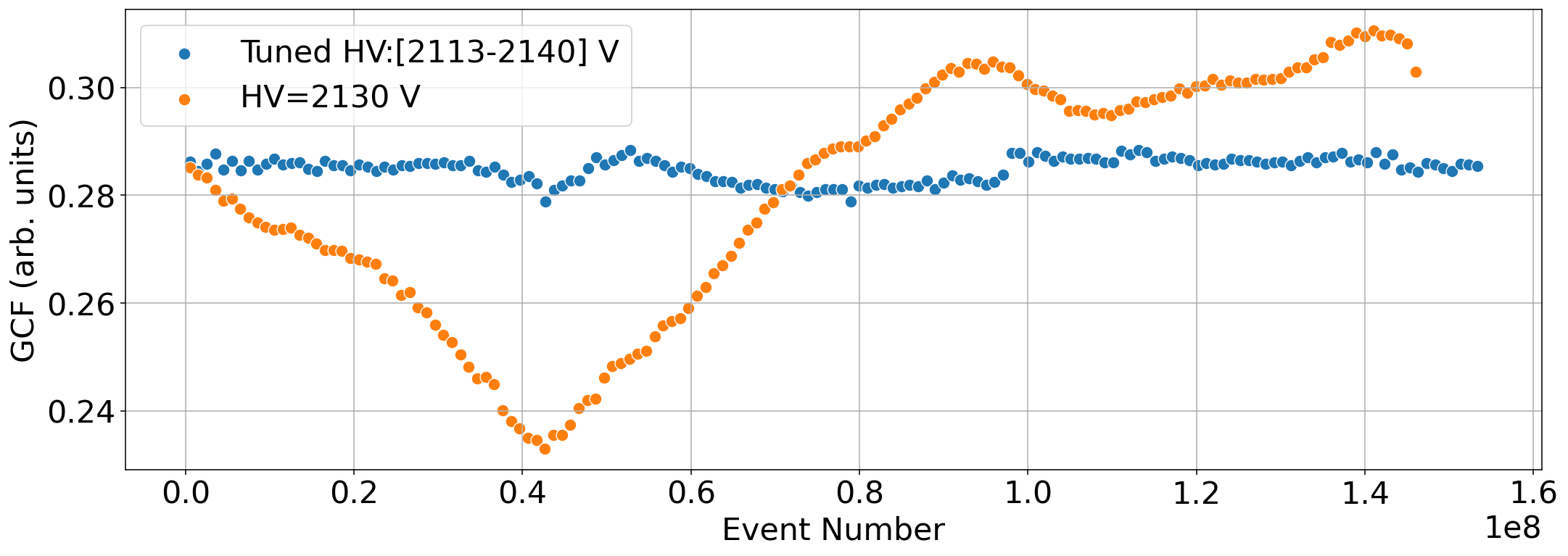}
    \caption{GCF for the model-controlled (blue) and constant 2130 HV (orange) sections of the CDC for a subset of the cosmic ray experiment corresponding to the largest change in atmospheric pressure. The data was taken over approximately four days. The change in atmospheric pressure was due to a large storm that passed through Newport News.}
    \label{fig:cosmics_yay}
\end{figure}

Figure \ref{fig:cosmics_yay} presents the comparison of the two sides of the CDC and the stabilization of gain for the GPR-controlled side of the CDC during a thunderstorm which produced significant changes in atmospheric pressure.

\subsubsection{Operational Insights}
The cosmics test occurred during scheduled downtime, resulting in the temperature of the gas in the CDC straying from the normal range represented in the training data. We opted to artificially fix the temperature input feature to 299 or 300K until the Forward Drift Chamber was turned on. This brought the temperature of the gas back into the trained region. The effect of this can be seen in Fig. \ref{fig:cosmics_yay} around event number 1.0$\times10^8$, where there is a slight jump in the predicted GCF. 

\subsection{Deployment \#3: CPP 2022}
During the Charged Pion Polarizability (CPP) experiment, the script was run automatically at the start of each run using the DAQ ``GO" processes to read EPICS and set the HV. Before predicting the GCF, the high voltage board current was scaled to what it would have been at the default HV value before making its prediction. During the CPP run period, a portion of the electronics from the CDC were removed in order to be used by a different detector system. The target type and position were changed, in addition to running at a much lower beam current. 
Particle identification information from the CDC was not critical for this particular experiment. 
\subsubsection{Performance}
\begin{figure}[H]
    \centering
    \includegraphics[width=0.9\linewidth]{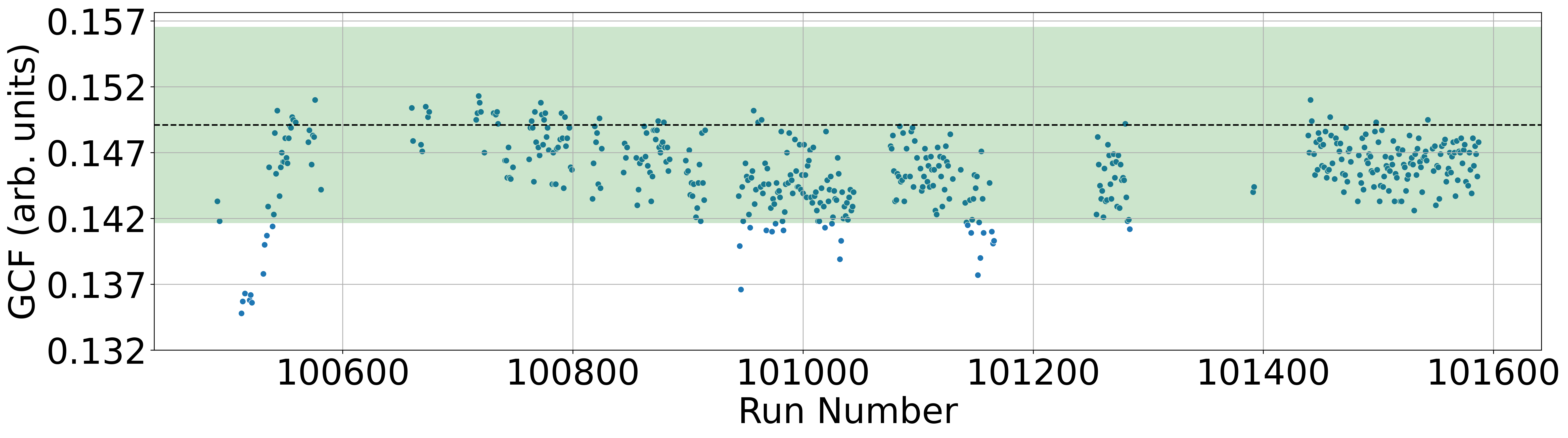}
    \caption{Gain Correction Factors for the CPP run period. The ideal GCF value for the run period is indicated by the horizontal dashed line. The green box indicates $\pm$ 5\%. of the ideal GCF value. }
    \label{fig:cpp_gcfs}
\end{figure}

The GCFs for the CPP run period, excluding empty target runs, are shown in Fig. \ref{fig:cpp_gcfs}. There are only 9 runs that fall outside of the $\pm$ 5\% metric. 


\begin{figure}[H]
    \centering
    \includegraphics[width=0.9\linewidth]{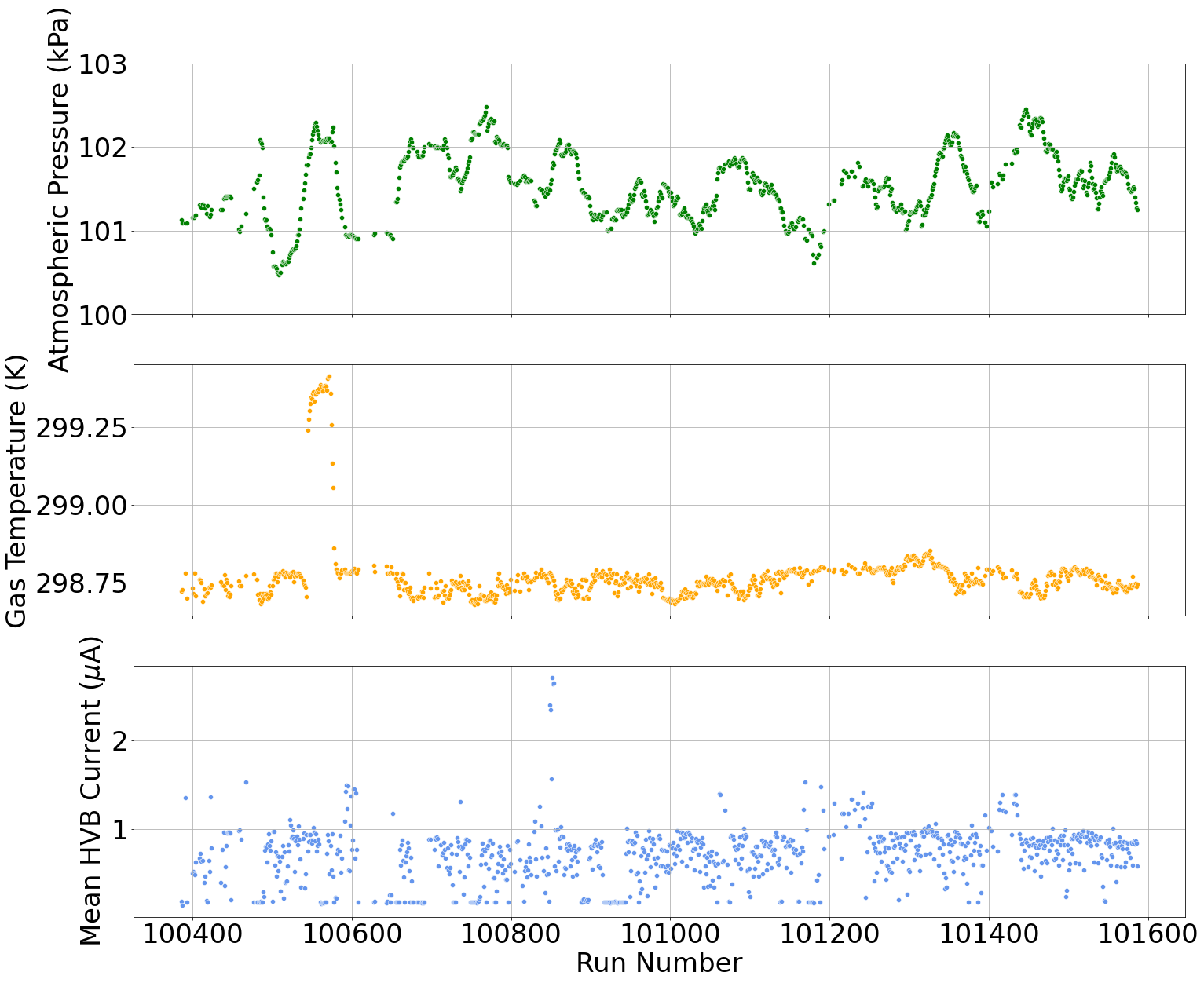}
    \caption{Atmospheric pressure (kPa), mean gas temperature (K), and mean high voltage board current ($\mu$A) during the CPP run period. }
    \label{fig:cpp_epics}
\end{figure}

\subsubsection{Operational Insights}
This was the first run period where an overall control ON/OFF switch was added to the existing CDC HV GUI. This provided shift takers and detector experts the chance to automatically turn off control of the CDC. The shift crew would have to manually reset the HV back to the nominal HV value of 2125 V. Additionally, the configuration file used by the script was updated to include additional parameters 
relating to:
\begin{enumerate}
    \item how EPICS data is gathered via a lookback windows, 
    \item a confidence threshold to determine when to trigger the uncertainty quantification correction,
    \item and whether to control the CDC using the point of closest confidence. 
\end{enumerate}

\subsection{Deployment \#4: PrimEx 2022} 
\begin{figure}[ht]
    \centering
    \includegraphics[width=0.8\textwidth]{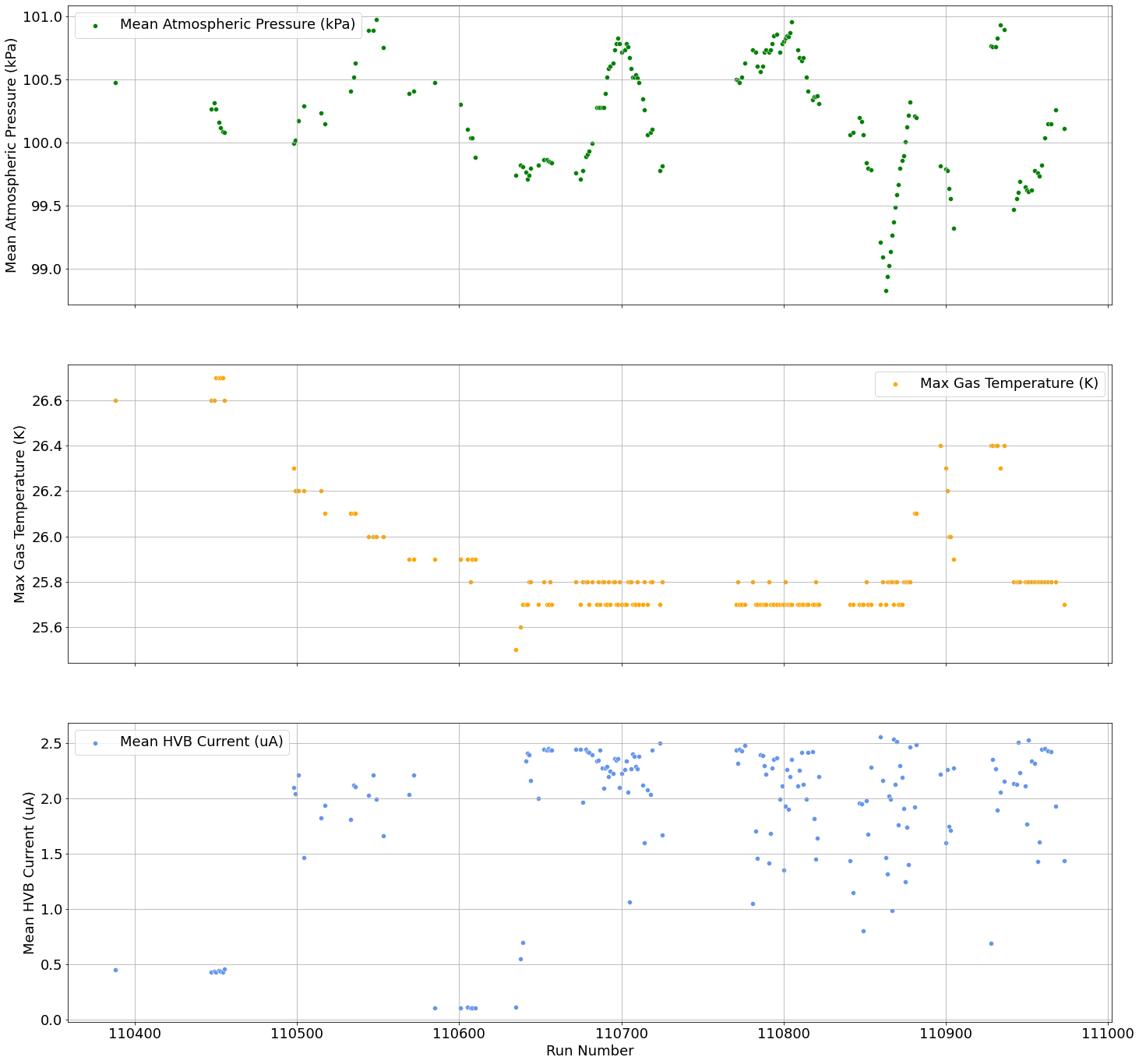}
    \caption{Mean atmospheric pressure (kPa), maximum gas temperature (K), and mean high voltage board current ($\mu$A) during the PrimEx-2022 Run Period.}
    \label{fig:inputs_prime22}
\end{figure}

The PrimEx run period took place from August 27, 2022 to December 19, 2022. The input features used for the model during the run are shown in Fig. \ref{fig:inputs_prime22}.

\label{sec:deployment4}
\subsubsection{Performance}
Figure 15 shows the ratio of the gain correction factor and the ideal gain correction factor for the PrimEx-2022 run period. Nearly all of the runs with an AI-tuned HV fall within our $\pm 5\%$ threshold. 

\begin{figure}[H]
    \centering
    \includegraphics[width=0.9\linewidth]{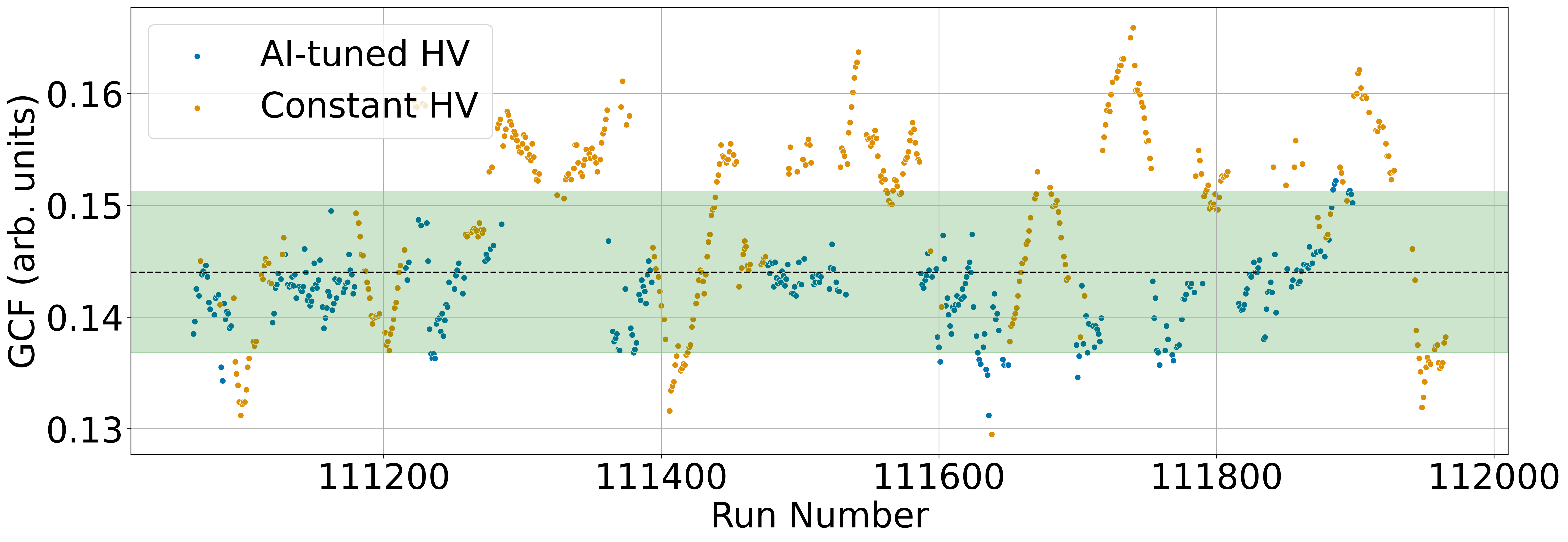}
    \caption{Gain correction factor as a function of Run Number. The orange points were runs taken at a constant 2125 V and the blue points correspond to runs taken with an AI-tuned HV setting. The dashed line indicates the ideal GCF. The green box corresponds to $\pm$ 5\% of the ideal GCF value.}
    \label{fig:primex22performance}
\end{figure}
\subsubsection{Operational Insights}
Of note, the mean high voltage board current during this run period is much smaller compared to the initial training set. During this deployment, an auto-off switch was enabled when the target state is not full and ready. This sets the default operating HV to 2125 V. The shift crew was responsible for changing the switch to on once the target was full again. From September 9th to September 20th, the AI switch was in the off state to investigate the unexpected model behavior when the atmospheric pressure and recommended HV setting were moving in opposite directions, which was contrary to the expectations of our detector expert. On September 20th, the system was re-initialized and control was resumed. On October 12th, the system crashed in a way that it changed the HV during the empty target runs. On October 14th, the system was unable to run inference as the model file was missing. This was rectified and control was resumed. On November 18th, the run coordinator was not aware that the system would automatically revert to 2125 V for empty target runs and turned the switch to off. Control remained off until Thomas Britton noticed the switch was off during a full target run. As the HV was set to 2125 V, this essentially reverted the CDC operation to the traditional method. On December 9th, the system tried to set the HV on the CDC while one of the HV modules was not communicating. This caused a major alarm, as the setpoint (received by EPICS) did not match the setpoint readback, indicating the module did not receive the new setpoint. The instructions for the shift crew were modified to provide instructions to rectify the communication errors in the event this happens again. 

\subsection{Deployment \#5: GlueX 2023}
The most recent deployment took place during the GlueX 2023 run period which took place from January 12th to March 20th, 2023. The control feature was mostly off as a result of complications arising from the atmospheric pressure sensor. On February 16th, RoboCDC control was switched on. According to EPICS, the AI switch for control was enabled, but the HV remained fixed at 2125V until run number 120739 (about 6 hours of data taking). This was due to the control script running on a computer without EPICS channel access. This was fixed promptly by the Hall-D slow controls expert on February 16th. On February 23rd, the high voltage board current was fixed as it was frequently unable to be read from the EPICS archive. On February 24th, during runs 120822-120830, the control was switched off because it was setting the HV after the run had started. 
\subsubsection{Performance}

\begin{figure}[H]
    \centering
    \includegraphics[width=0.9\linewidth]{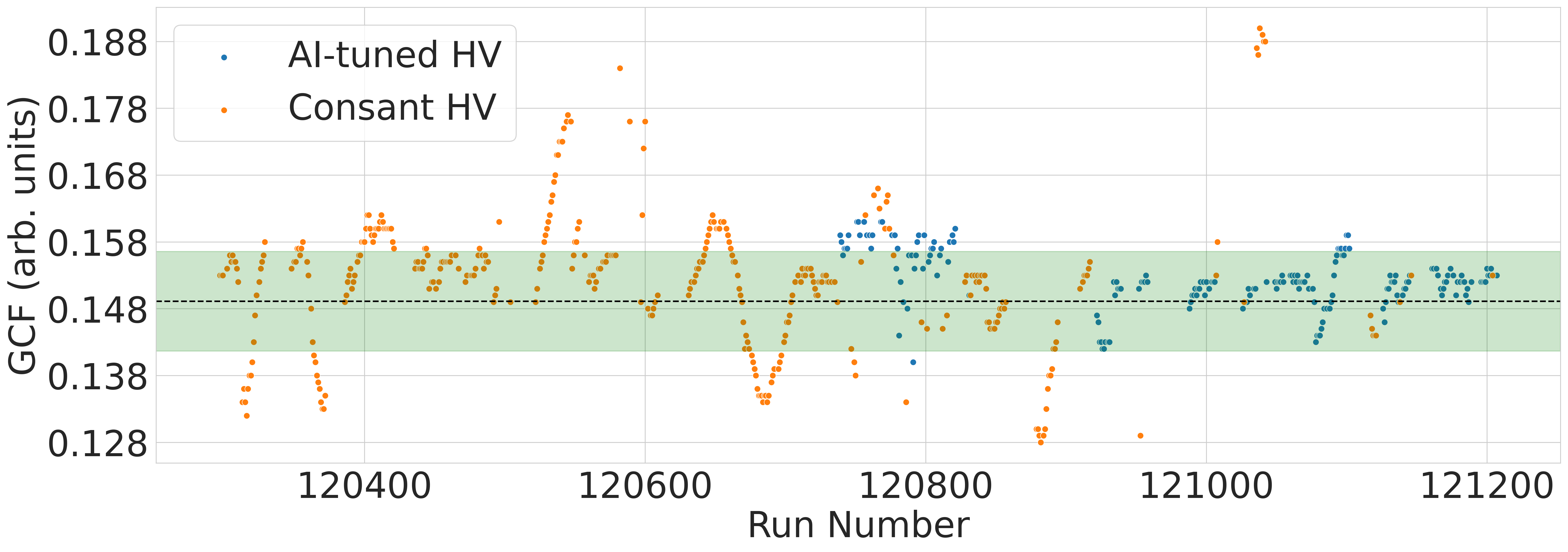}
    \caption{Gain correction factors for the GlueX 2023 run period. A significant number of runs used the default 2125 V setting (indicated in orange), whereas the runs where the HV was adjusted are shown in orange. The dashed line indicates the ideal GCF. The green box corresponds to $\pm$ 5\% of the ideal GCF value.}
    \label{fig:gluex23gcfs}
\end{figure}

\subsubsection{Operational Insights}
The GlueX 2023 run period did not utilize roboCDC during the start of the run period due to a faulty pressure sensor. On March 1st, 2023, for runs 120921-120931, the control script failed and the HV remained at 2123 overnight. Fig. \ref{fig:gluex23gcfs} shows the gain correction factors as a function of run number. For runs 120921-120931, it appears the HV setting was suitable for some, but not all, runs during this time. As such, it is fortunate we were at a HV that is \textit{close} to both the default HV setting and what would have been recommended based on similar environmental conditions. Stable running was achieved during runs 121122 until run 121214. For some runs, the HV had changed after data taking had started. This is to be avoided, as we do not want the raw data files to include data taken with multiple high voltage settings in the same raw data file. The runs where the high voltage changed after the start of the run were identified such that the corresponding raw data files could be excluded from the calibrations.

\section{Control System and User Interface} \label{sec:ControlSystem}
\begin{figure}[htb]
    \centering
    \includegraphics[width=0.9\textwidth]{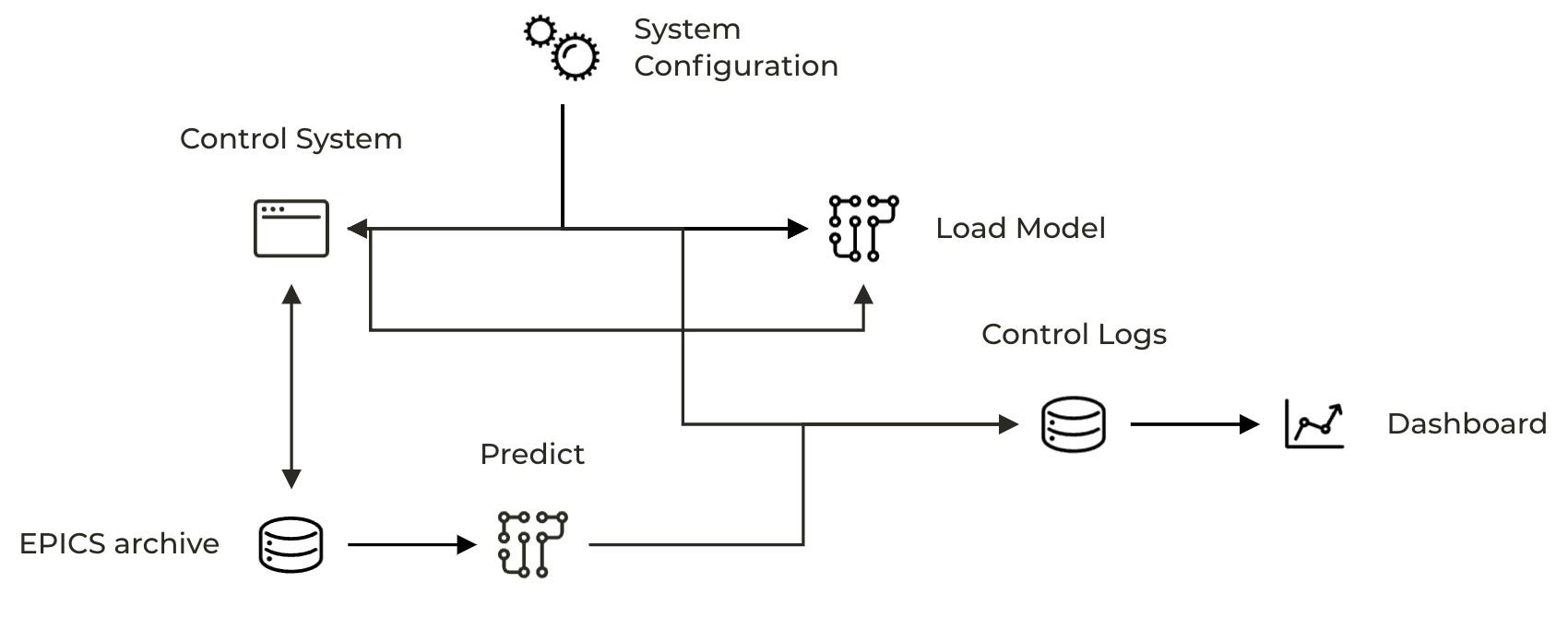}
    \caption{Diagram outlining the main processes for HV control.} \label{fig:robocdc}
\end{figure}

A modular system, named RoboCDC, was developed to utilize the machine learning model and control the CDC. The complete controls harness is comprised of a control system, which issues requests to change CDC HV settings, a module which handles the loading and running of the model, and a decision engine which gathers input data and performs actions depending on the results of model inferences. The entire system utilizes a MySQL database to store model information as well as the input features, inferences, and control decisions throughout data taking and is configurable by modifying parameters in a single configuration file.  A diagram of the system is shown in Fig. \ref{fig:robocdc}. Shift crews can enable/disable the system as deemed necessary by experts by using the standard slow controls GUI \cite{kasemir2007control}. A Grafana dashboard displays visualizations of RoboCDC's inputs, actions, and performance allowing for real-time monitoring of the RoboCDC system by shift crews and experts. 

In certain situations, such as high beam current tests, the system is disabled and begins ``trusting the humans".  
While in the disabled state, the system will not attempt to change HV settings nor take any controlling actions.  
The system will, however, continue to take in data and produce predictions.  
These ``non-controlling" predictions are still recorded in the database for analysis of potential model behavior. 
Furthermore, out-of-distribution input parameter sets can be identified in this manner and potentially used to train future models.

The configuration file contains many parameters which control RoboCDC's behavior while running.  These variables are able to be edited without impacting data taking. The configuration parameters are as follows:

\begin{itemize}
  \item \textbf{model\_name} - The full path to the saved model which will be used.
  \item \textbf{poll\_time} - The time between each running of the model in minutes. 
  \item \textbf{rec\_scale} - The number of poll\_time intervals before the system recommends a HV change. This number can be less than 0.  In the case of negative values of rec\_scale RoboCDC will run its inference-control loop exactly N times where \(N=-1 \times \text{rec\_scale}\) with a poll\_time wait between each pass through the inference-control loop.  For discrete running, like the typical production configuration of RoboCDC, rec\_scale was set to -1; making the system recommend HV settings only once (at run start).  In this specific configuration the poll\_time variable serves little to no purpose.
  \item \textbf{control\_window} - A two element array representing the lower and upper bound on the HV voltage recommendation.  If RoboCDC recommends setting the HV to a value below the lower bound or above the upper bound it instead recommends setting the HV to the value of the nearest bound.  For the CDC the detector expert, Naomi Jarvis, determined [2105.0, 2145.0] to be the safe operational envelope.
  \item \textbf{ideal\_gcf} - The ideal gain correction factor.  This value is used to calculate the recommended HV setting for the CDC based on the fitted curve in Figure \ref{fig:HVScans} and is set during initialization. Alternatively, this value can be set manually through this configuration parameter.
  \item \textbf{default\_hvbi} - The high voltage board current that was found when initialization was performed.  Alternatively, this value can be manually set to a negative value.  When this value is negative it acts as an override, internally setting the scaled high voltage board current to the fixed value given by the magnitude of the default\_hvbi variable.  When this value is greater than 0 then the variable is not used and exists as a record of the conditions present during initialization only.
  \item \textbf{require\_human} - Either 0 for automatic action or 1 to trigger a protocol for including human-in-the-loop action.  Presently, this variable is only connected to a hook in the code and does not impact operation.  This was left included in the case where it was deemed necessary to require such human action.  In the course of this work it was unanimously agreed that the involvement of humans in the action/control loop would place undue burden on shift crews; The system had enough safe guards in place to safely operate autonomously.
  \item \textbf{control\_mask} - This string dictates what parts of the CDC will actually be under the control of RoboCDC.  The variable is passed to the actual control script, utils\/cdc\_tools.py, alongside the recommended HV and was utilized to functionally divide the CDC in half in software for testing.  During normal running it is set to "ALL".
  \item \textbf{default\_temp} - The temperature in Kelvin that was found when initialization was performed. Alternatively, this value can be manually set to a negative value.  When this value is negative it acts as an override, internally setting the temperature used by the model to the magnitude of the default\_temp variable.  When this value is greater than 0 then the variable is not used and exists as a record of the conditions present during initialization only.
  \item \textbf{default\_pressure} - The pressure in kiloPascals that was found when initialization was performed. Alternatively, this value can be manually set to a negative value.  When this value is negative it acts as an override, internally setting the pressure used by the model to the magnitude of the default\_pressure variable.  When this value is greater than 0 then the variable is not used and exists as a record of the conditions present during initialization only.
  \item \textbf{baseline\_V} - The nominal HV setting for all HV boards in the CDC.  The control system is designed to set, or attempt to set, the CDC to the baseline voltage in the event of any part of the control module throws an error.  It is also used in conjunction with "fail safe" protocols and with autoOFF.
  \item \textbf{fail\_safe\_timeout} - The control system relies on EPICS to obtain the model's input parameters.  In the event that EPICS fails to return valid input data for some number of minutes equal to or greater than fail\_safe\_timeout the "FailSafe" protocol is run.  This protocol involves setting the CDC boards described the control\_mask to the setting defined by baseline\_V and logging the incident.
  \item \textbf{confidence\_threshold} - The value of uncertainty in percent of ideal\_gcf that when crossed above triggers an uncertainty correction when the UQcorrection variable is enabled.
  \item \textbf{UQcorrection} - Either 0 to disable uncertainty correction or 1 to enable Uncertainty correction.  This correction involves finding the nearest recommended HV, by Euclidean distance, on the surface created by the input parameters which yield uncertainty values equal to the confidence\_threshold.
  \item \textbf{autoOFF} - This value is either 0 to disable automatically turning off the control system or 1 to enable the autoOFF procedure.  This procedure is run when the system detects the GlueX target is not in the full and ready state.  This procedure was also run when the system desires to set the HV above or below the control window.  This was done in efforts to gather more training data and extend the ``in distribution" set of input parameters.  When the auto-off procedure is run this action is also logged.
  \item \textbf{lookback\_dict} - Because the model's input features are engineered (e.g. the mean) the question of look-back time comes into play.  This variable defines the differing scales, in seconds, each variable should ``look back".  While able to take in an arbitrary dictionary of look-back times only the current is used presently.  This was due to the observations of overshoots in high voltage board currents; correlated with overshoots in beam current, especially after beam trips.
  \item \textbf{max\_lookback\_scale} - This variable defines how many blocks of lookback should be looked back as a multiple of the length defined in lookback\_dict. For example, if the value of a variable ``I" is 60 in lookback\_dict and the max\_lookback\_scale is 2 then when, say, mean is computed it is computed over the last 120 seconds.
\end{itemize}

Once configured appropriately, the control system is initialized when the shift crew starts a run via GlueX's standard data acquisition (DAQ) software.  
Assuming the AI control system is turned on it gathers up the necessary input variables from EPICS and feeds them to the model.  
The model returns a result dictionary, which is appropriately logged in the database, and actions are taken to control the CDC HV in accordance with the model's output and the system's configuration. 
Once input data is fed into the model a prediction is obtained within milliseconds.  
To obtain the target HV setting a ratio of the model's predicted gain correction factor to the ideal gain correction is fed into a polynomial fit of HV to the ratio of peak ADC amplitude to the peak amplitude at the nominal 2125 Volts (essentially reducing to the ratio of gain correction factors).  
This polynomial was derived from data taken during HV scans of the CDC. 
The HV of each of the CDC's 72 boards is adjusted before data-taking begins.
The full control workflow of all functions that make up the entirety of the control system can be found in Appendix \ref{apx:flowcharts}.


\section{Physics results with ML-based calibration constants}

\subsection{Effect of online control on gain calibration}

\begin{figure}[htb]
    \centering
    \includegraphics[width=0.9\linewidth]{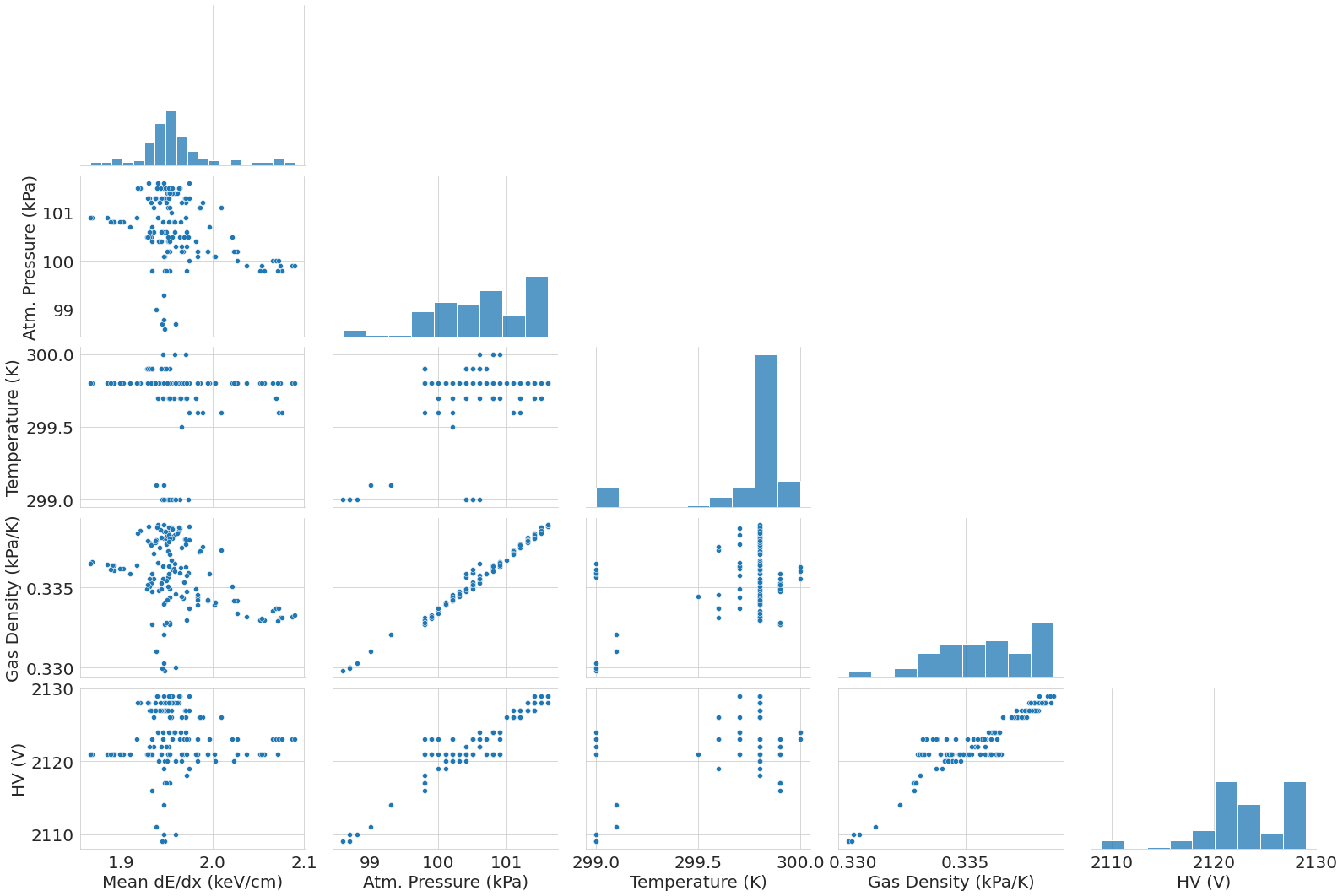}
    \caption{Pair plot of the mean dE/dx, atmospheric pressure, gas temperature, gas density, and HV setting during the GlueX 2023 run period. }
    \label{fig:ttod_pairplot_2023}
\end{figure}

By adjusting the HV of the CDC during the run, the gain is stabilized and the correlations between the atmospheric pressure, gas temperature, and HV board current and the calibration values has been removed. This is shown in Fig. \ref{fig:ttod_pairplot_2023}. In the top plot of the first column, the atmospheric pressure is no longer correlated with the mean dE/dx position (and thus the gain), except for the runs where the AI system was on but unable to adjust the HV to the appropriate value. 

\subsection{Effect of online control on time-to-distance calibration}
\begin{figure}[htb]
    \centering
    \includegraphics[width=0.9\linewidth]{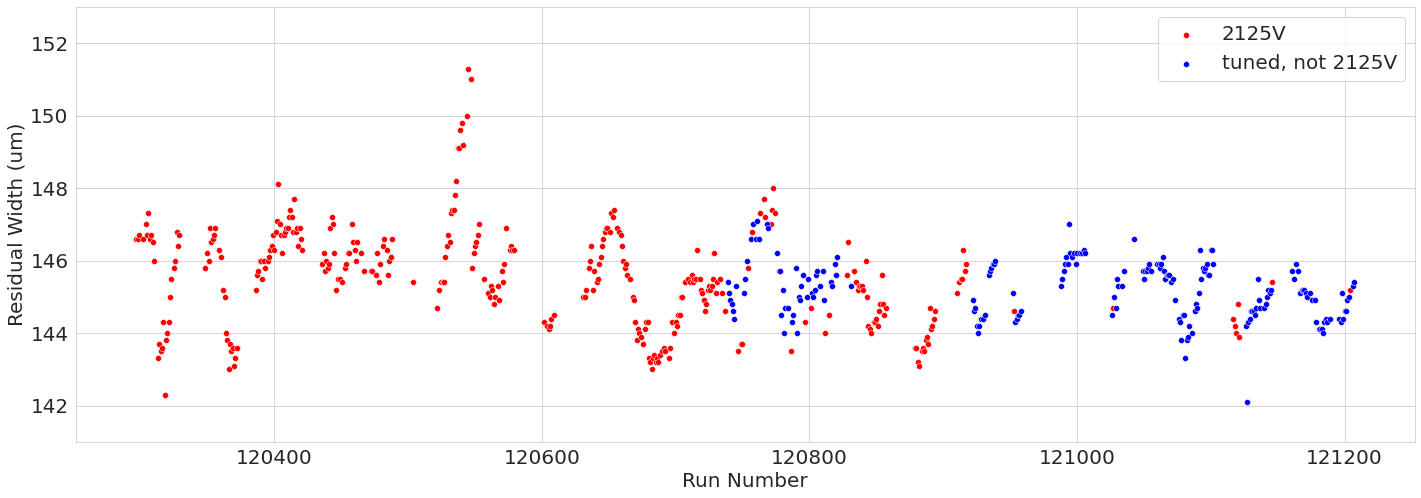}
    \caption{The residual widths obtained before calibration for runs taken at 2125 V (red) and a HV setting determined by the ML system (blue).}
    \label{fig:width_before_calib}
\end{figure}

Up until the GlueX 2023 run period, it was assumed that adjusting the HV of the drift chamber would not affect the time-to-distance calibration in a significant way. For the GlueX 2023 run period, we attempt to analyze the effects of online control on the time to distance calibration. In Fig. \ref{fig:width_before_calib}, the width of the residuals obtained before the traditional time to distance calibration are shown as a function of run number for runs at 2125 V and those with a voltage set according to ML. 

\begin{figure}[htb]
    \centering
    \includegraphics[width=0.9\linewidth]{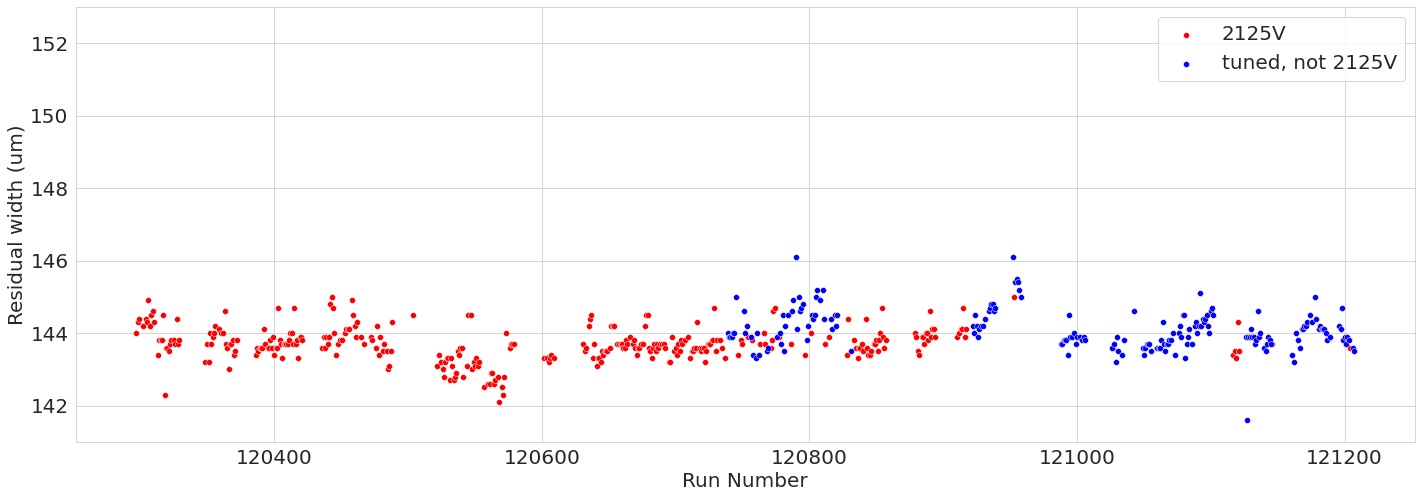}
    \caption{Residual widths obtained using calibration constants obtained from the fits of the calibration constants to the gas density. The runs at 2125 V are shown in red, other voltage settings correspond to the blue data points.}
    \label{fig:widths_PTfits}
\end{figure}

In comparison, the residual widths obtained from using starting values from the fits to the gas density are shown in Fig. \ref{fig:widths_PTfits}. Utilizing the fits of the calibration constants to gas density from the 2020 run period to obtain starting values for the GlueX 2023 run period results in much less variation of the residual width over the entire run period, regardless of whether the data was taken with the CDC HV set to 2125 V. 

\begin{figure}[htb]
    \centering
    \includegraphics[width=0.9\linewidth]{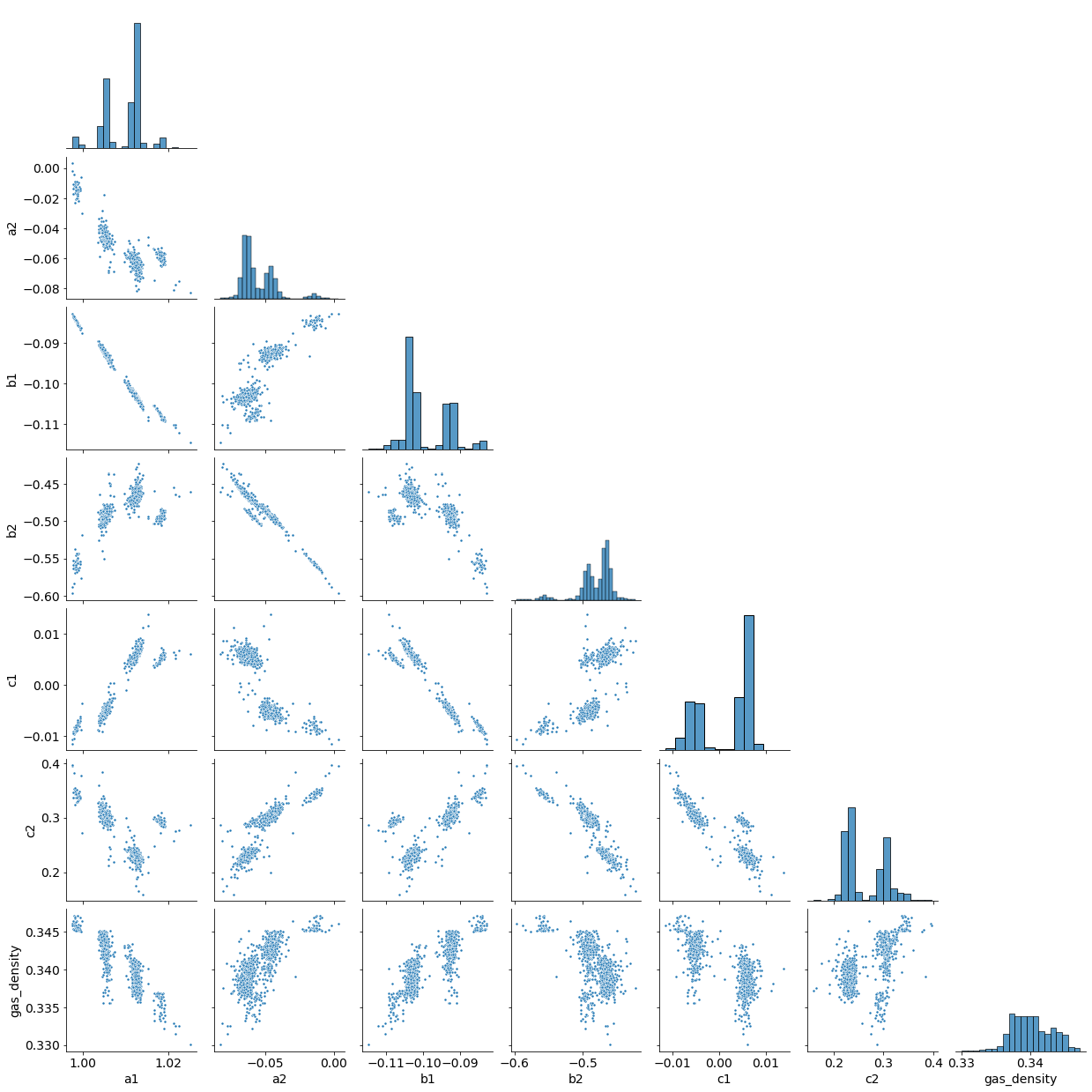}
    \caption{Pair plot of the time to distance fit parameters and the gas density for the 2020 Run Period.}
    \label{fig:ttod_density_2020}
\end{figure}

The time to distance function parameters are correlated with the density of the gas in the CDC. As such, it is possible to obtain starting values for the time to distance fit function at the start of each run using the measured atmospheric pressure and temperature from EPICS. An example of the relationship between the fit parameters and the gas density for the 2020 Run Period is shown in Fig. \ref{fig:ttod_density_2020}. Using this, we can generate functions that produce a set of starting values for each of the fit parameters just by measuring the gas density. This could potentially save a lot of computing resources as there would be less iterations to perform to achieve the desired quality.

If the 5\% metric on the dE/dx values is obtained from calibration using the ML-generated calibration values, then the traditional calibration procedure can be replaced. If a future experiment requires more precise calibrations then more iterations of the software can be used until the designated metric is obtained. 
\section{Other detector Systems} \label{sec:other_systems}
\subsection{CLAS12 Drift Chambers and ECAL}
\begin{figure}[htb]
    \centering
    \includegraphics[width=1.2\linewidth]{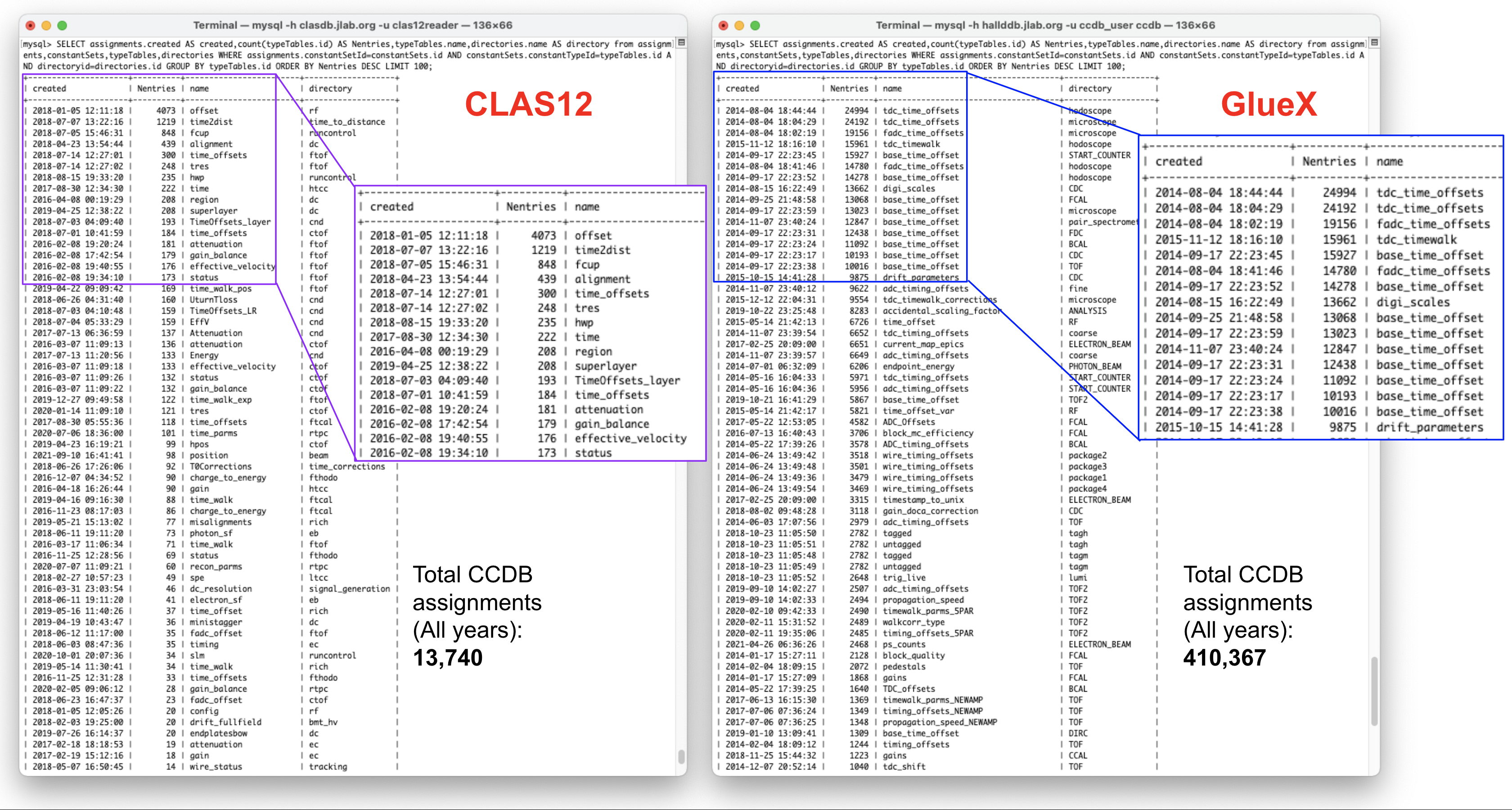}
    \caption{Snapshot of the number of entries in calibration DB for CLAS12 and GlueX. This shows GlueX has roughly 30 times the number of entries as CLAS12 across a number of detectors.}
    \label{fig:CLAS12_CCDB}
\end{figure}

The original proposal highlighted the CLAS12 Drift Chambers (DC) as a natural application for the technology that would first be developed for the GlueX CDC. 
When control of the CDC HV was successfully implemented and attention was turned towards the CLAS12 DC we found that there was more than an order of magnitude fewer calibration entries for CLAS12 DC than for the GlueX CDC. 
This was unexpected by the detector experts in both collaborations. 
Upon looking further, it was seen that this was not unique to just the Drift Chambers, but was a common occurrence across subdetectors (see Fig. \ref{fig:CLAS12_CCDB}). 
This led us to conclude that either cultural differences between the collaborations were notable when it came to determining the frequency of calibration, or the subdetectors themselves were designed differently. Regardless of the reason, the small number of entries for CLAS12 made it impractical to train an AI/ML model. 

\subsection{Forward Drift Chamber}
The Forward Drift Chamber (FDC) was not previously calibrated to control for drifts in the chamber gain. `There is some initial motivation to calibrate the chamber gain in order to improve the PID figure of merit. We did not pursue controlling the FDC, as we do not have the relevant HV scan data in addition to the detector being more complicated in terms of HV settings than the CDC. 

\begin{figure}[htb]
    \centering
    \includegraphics[width=1\linewidth]{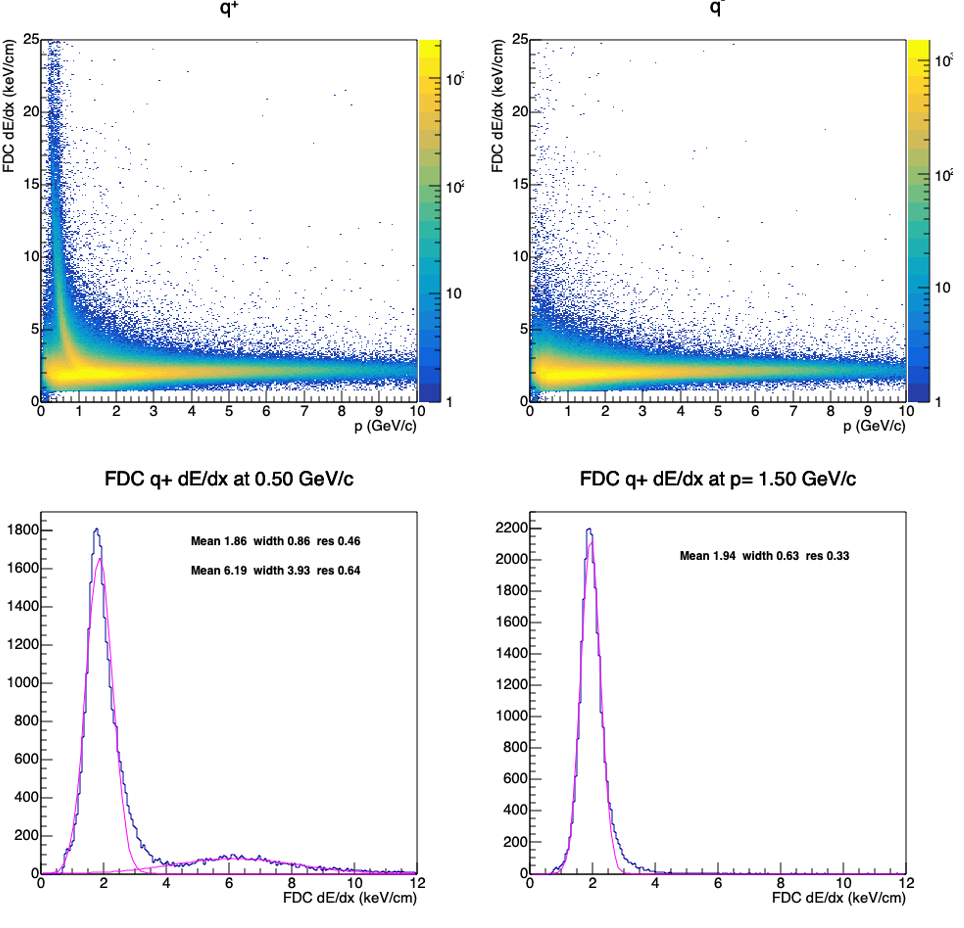}
    \caption{Top row: FDC dE/dx for positive (left) and negative (right) tracks vs momentum. Bottom row: The one-dimensional projections of dE/dx at p=0.5 GeV/c  (left) and p=1.5 GeV/c (right).}
    \label{fig:FDC_idealpeak}
\end{figure}

The first gain calibrations were performed using a similar procedure to that used for the CDC. Namely, we obtain the position of the first ionizing peak at $p=1.5$ GeV/c using data from Run Number 11621. 
In Fig. \ref{fig:FDC_idealpeak}, we show dE/dx vs momentum for positive and negative charged tracks followed by the one-dimensional projections of dE/dx at $p=0.5$ GeV/c and $p=1.5$ GeV/c. The ideal peak position is obtained from the mean of a Gaussian fit to dE/dx at $p=1.5$ GeV/c. 
This ``ideal" peak position is equal to 1.9335 keV/cm. The gain calibration value is then obtained from 

\begin{equation}
    g_n = \frac{mean_{ideal}}{mean_{actual}}\times g_{n-1}
\end{equation}
where n indicates the iteration number and $mean_{actual}$ corresponds to the mean value obtained from the fit for the current iteration. 

Similarly to the CDC gain calibration, the gain calibration values obtained from this procedure are strongly correlated with the atmospheric pressure. Thus, it is possible to use a similar system to the CDC for the FDC gain calibrations. We would not initially include a control aspect, but would have calibration constants available at the start of each run. 

\begin{figure}[htb]
    \centering
    \includegraphics[width=0.9\linewidth]{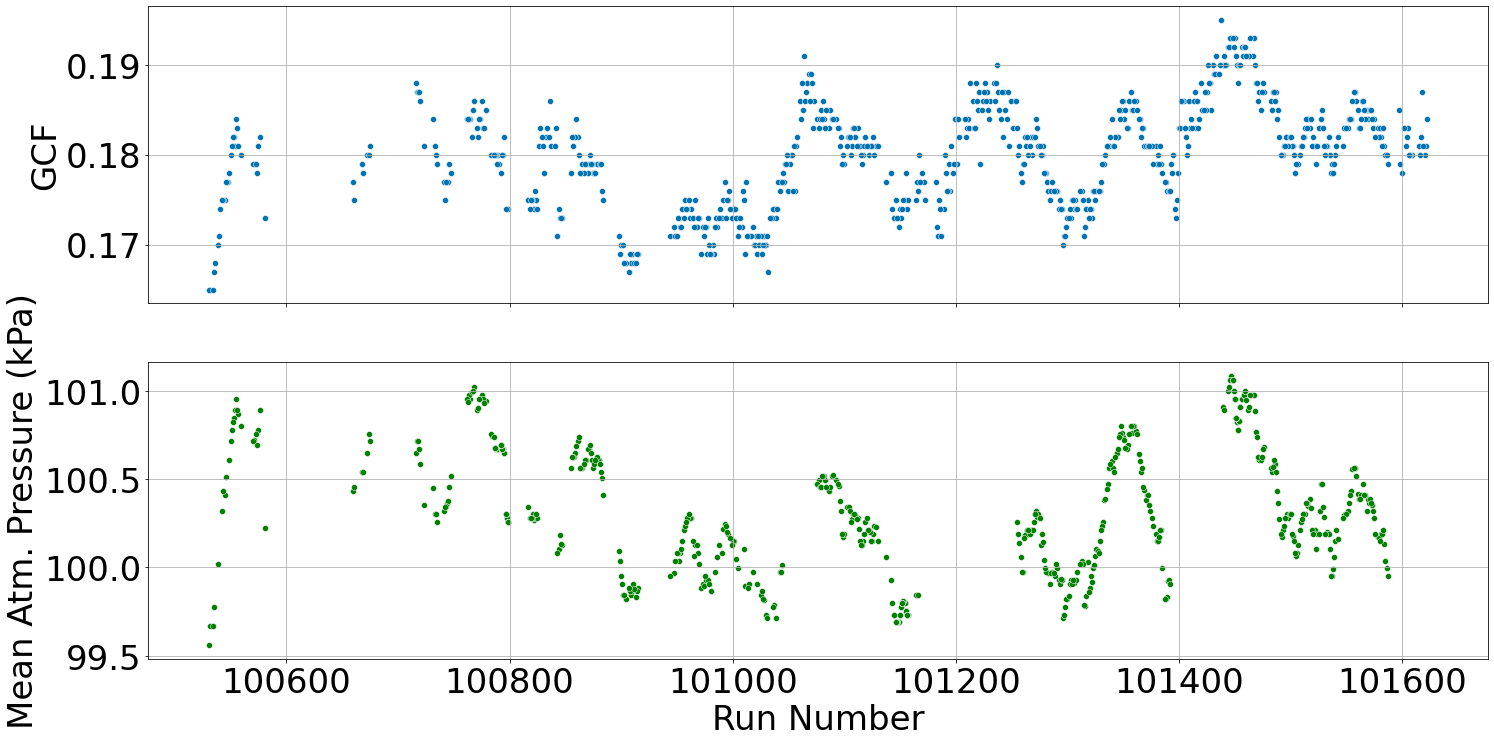}
    \caption{Top: Gain Correction Factor as a function of Run Number for the CPP Run Period. Bottom: Mean atmospheric pressure as a function of run number.}
    \label{fig:fdc_gcf_pressure}
\end{figure}

\subsection{Barrel Calorimeter}
The Barrel Calorimeter (BCAL) is used to detect photon showers with energies of at least 0.05 GeV, within 11-216 degrees in $\theta$, and 0-360 in $\phi$. The pedestal values for the BCAL fluctuate with temperature. The readout of the temperature sensors is not fine-grained enough to be useful in identifying correlations between the pedestal and temperature values. This effort was not pursued further as it would have been costly (both in terms of time and equipment) to install a new temperature sensor with a more fine-grained readout.

\begin{figure}[H]
    \centering
    \includegraphics[width=1\linewidth]{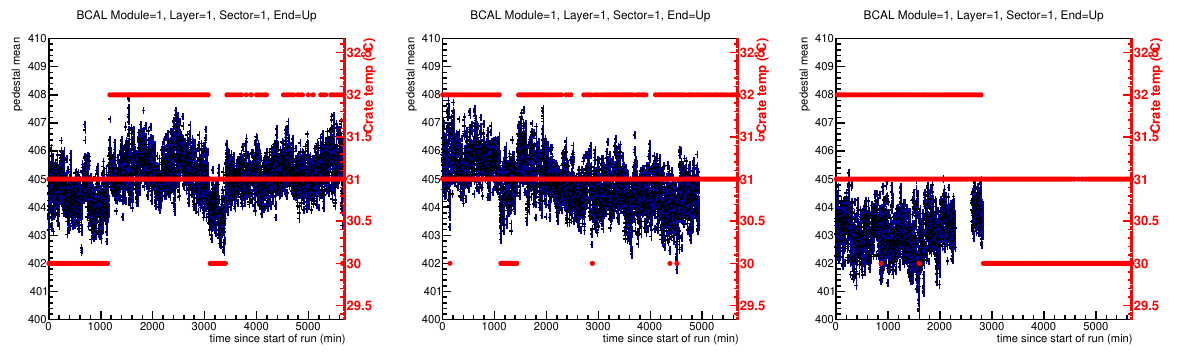}
    \caption{BCAL pedestal vs. time for a single channel. The measured pedestal value(blue) is plotted vs. the left axis and the measured temperature(red) is plotted vs. the right axis.
    }
    \label{fig:bcal_ped_vs_t_all}
\end{figure}

\subsection{Forward Calorimeter} \label{sec:fcal}

Here, we describe work pertaining to Hall-D's FCAL. 
First, we discuss a capstone project conducted with the University of Virginia, which explored a Convolutional Neural Network (CNN) to predict gain calibration constants for the FCAL's 2800 lead glass detectors. 
The finding that the gain calibration constants were only weakly correlated with the PMT measurements of the LED monitoring system's pulses led to a simulation study. 
The study found that a Long Short-Term Memory neural network is capable of learning from the evolving time series of a simulated LED pulse and that the gain calibrations can be predicted with a mean absolute error of $<$ 1\%, even with an injection of up to 7\% noise.
Finally, we examine the correlation of the change of the gain calibration constants and luminosity in experimental data to better understand the effect of radiation damage caused by proximity to the beamline.

\subsubsection{Forward Calorimeter Gain Calibration Using A Convolutional Neural Network}
\begin{figure}[h]
    \centering
    \includegraphics[width=0.4\linewidth]{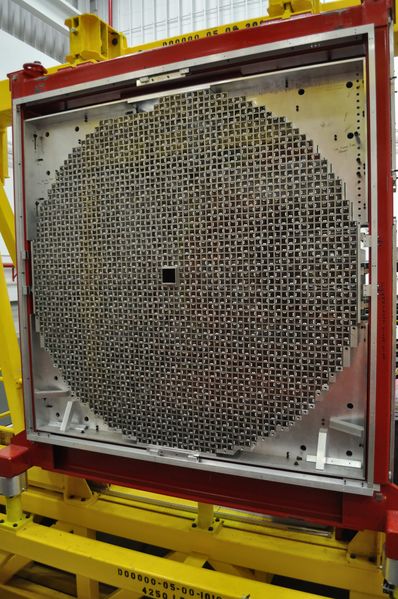}
    \caption{The FCAL, under construction, as viewed looking downstream. Not pictured are the plexiglass plates, nor the components of the LED monitoring system.}
    \label{fig:FCAL}
\end{figure}

In partnership with the University of Virginia School of Data Science, a master's capstone project explored whether the existing LED monitoring system data could be used to train a CNN to produce the gain calibrations for each FCAL block for an experimental run. The self-selected team of four students included Cullan Bedwell, Abhijeet Chawhan, Julie Crowe, and Diana McSpadden (JLab). The project was conducted over one term; weekly meetings were held with the AIEC investigators.

The JLab GlueX FCAL is a 2.4 \si{\meter} circular array of 2800 4 \si{\centi\meter~\times} 4 \si{\centi\meter~\times} 45 \si{\centi\meter} lead glass blocks; each block is attached to a photomultiplier tube (PMT) \cite{ADHIKARI2021164807}.
Due to variations in the lead glass blocks and the PMTs, their responses to photons are inconsistent. 
This behavior is expected \cite {leo2012techniques, JONES2006366, ADHIKARI2021164807, Hanagaki2022}. 
An offline gain calibration is necessary to correct for variations in response to photons.  

Traditionally, a gain calibration constant is applied offline to raw experimental data to correct for the varying PMT measurements over time and the variations between PMTs. 
Thus, for each experimental run, a gain calibration for each of the FCAL's 2800 PMTs must be calculated. 
A reference PMT and corresponding reference light source, such as a predictable radioactive light source, are unavailable for the GlueX experiment. 
When available, reference PMTs and radioactive light sources require non-negligible maintenance. 
Thus, the FCAL calibration procedure must rely on other known quantities or measures. 
For GlueX FCAL calibration, the theoretical $\pi^0$ mass is used.

The FCAL gain calibration constant, a unitless scaler, minimizes the difference in the theoretical $\pi^0$ mass and the measured $\pi^0$ mass summed over run events \cite{jaegle2009pi}. 
To calculate the gain calibrations, an iterative method, the ``Gain Balancing Procedure" \cite{moriya2013measurement}, has been employed by fitting raw experimental FCAL data, applying gain calibrations until the measured mass of $\pi^0$ is within a tolerated threshold of the theoretical mass, as shown in Eq. \ref{eq:fcal_calib}. 
This method is computationally expensive and time-intensive for scientists. FCAL experts must also contend with the lack of $\pi^0$ particles in the outer blocks of the FCAL due to a lack of $\pi^0$'s at larger angles from the beam line.

\begin{equation} \label{eq:fcal_calib}
GCF_{new} = GCF_{old} \cdot \left(\frac{m_{meas}}{m_{theory}}\right)^2,
\end{equation}

where $m_{meas}$ is the mass of the neutral pion ($\pi^0$) obtained from the experimental data, and where $m_{theory}=134.9786 \pm  0.0005$ MeV \cite{particle2020review}.

\begin{figure}[htb]
    \centering
    \includegraphics[width=0.75\linewidth]{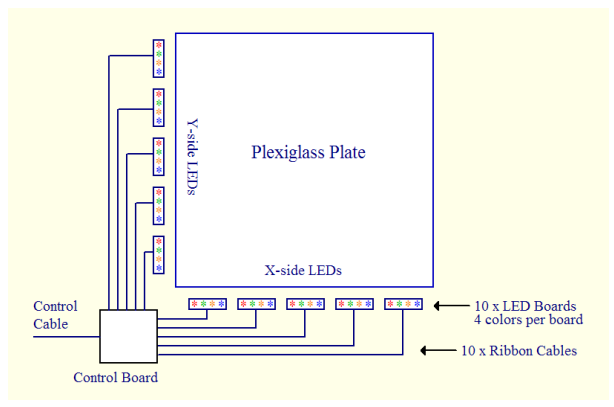}
    \caption{The FCAL monitoring system for one of the quadrants. The four LED colors chosen were violet, blue, and green. (Image taken from original publication \cite{anassontzis2014relative}).}
    \label{fig:led_system}
\end{figure}

The FCAL's blocks are divided into four quadrants. 
Plexiglass sheets are coupled to each quadrant.
As shown in Fig. \ref{fig:led_system}, ten LED boards are arranged in two sets of five along the two external edges of each square plexiglass sheet. 
The LED monitoring system utilizes five distinct color-strength flashes to monitor the 2800 PMTs:
\begin{enumerate}
    \item the small violet pulse: 390 nm at 12 V,
    \item the large violet pulse: 390 nm at 22 V,
    \item the green pulse: 574 nm at 29 V,
    \item the small blue pulse: 470 nm at 10 V,
    \item and the large blue pulse: 470 nm at 15 V.
\end{enumerate}

Each LED color and strength pulses for a 10-minute duration. 
These pulses are monitored to ensure that PMT readings are nominal.
Additional details of FCAL monitoring system, including the verification, spatial variability, and measures taken to minimize the spatial variability, are available from \cite{anassontzis2014relative}.
The system was not designed to be used for calibration; however, the measurements from the LED monitoring system are stored and readily available for offline or online use, and it is reasonable to investigate if the LED readings could be used for calibration.

\begin{figure}[h]
  \centering
  \includegraphics[width=0.32\textwidth]{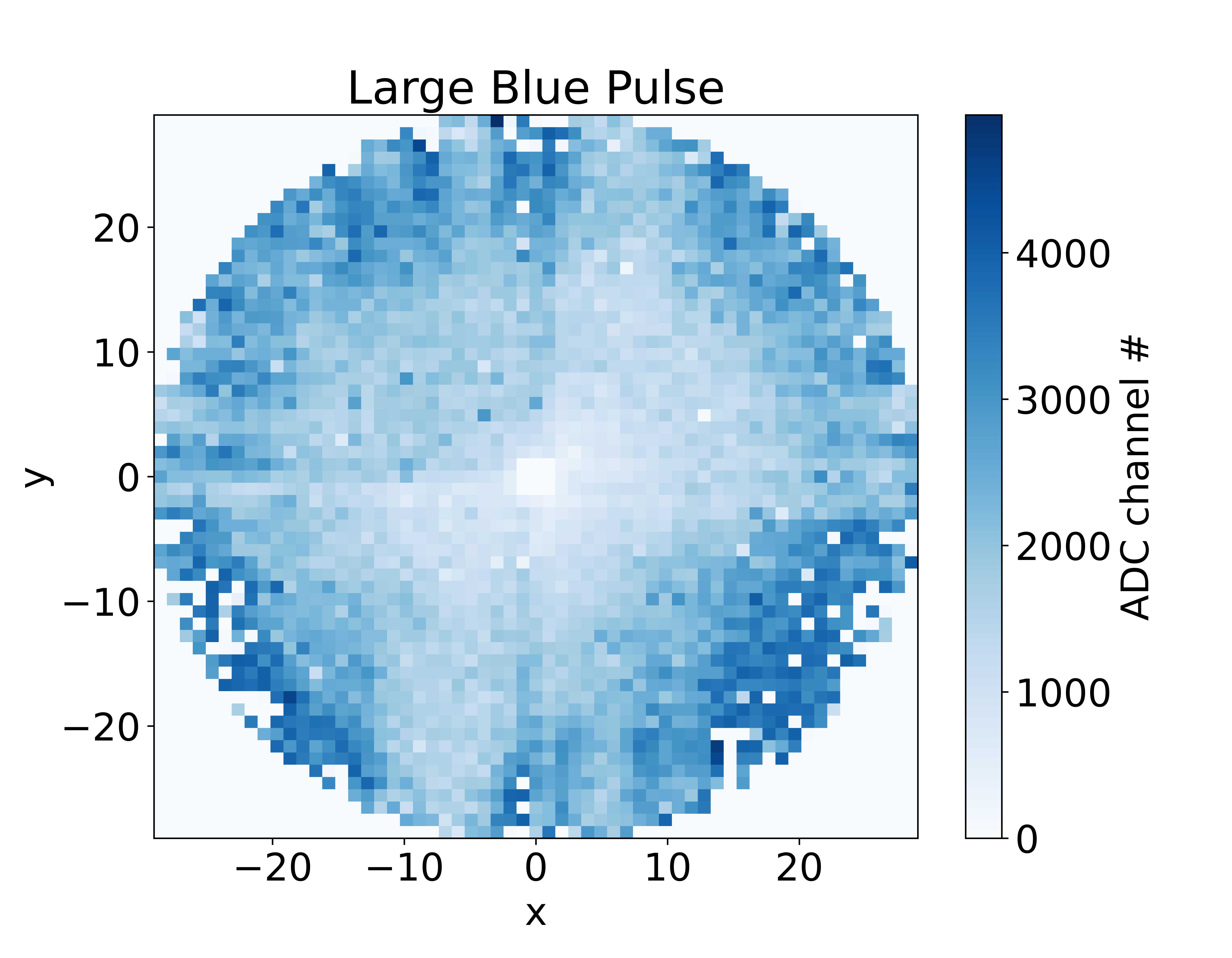}
  \includegraphics[width=0.32\textwidth]{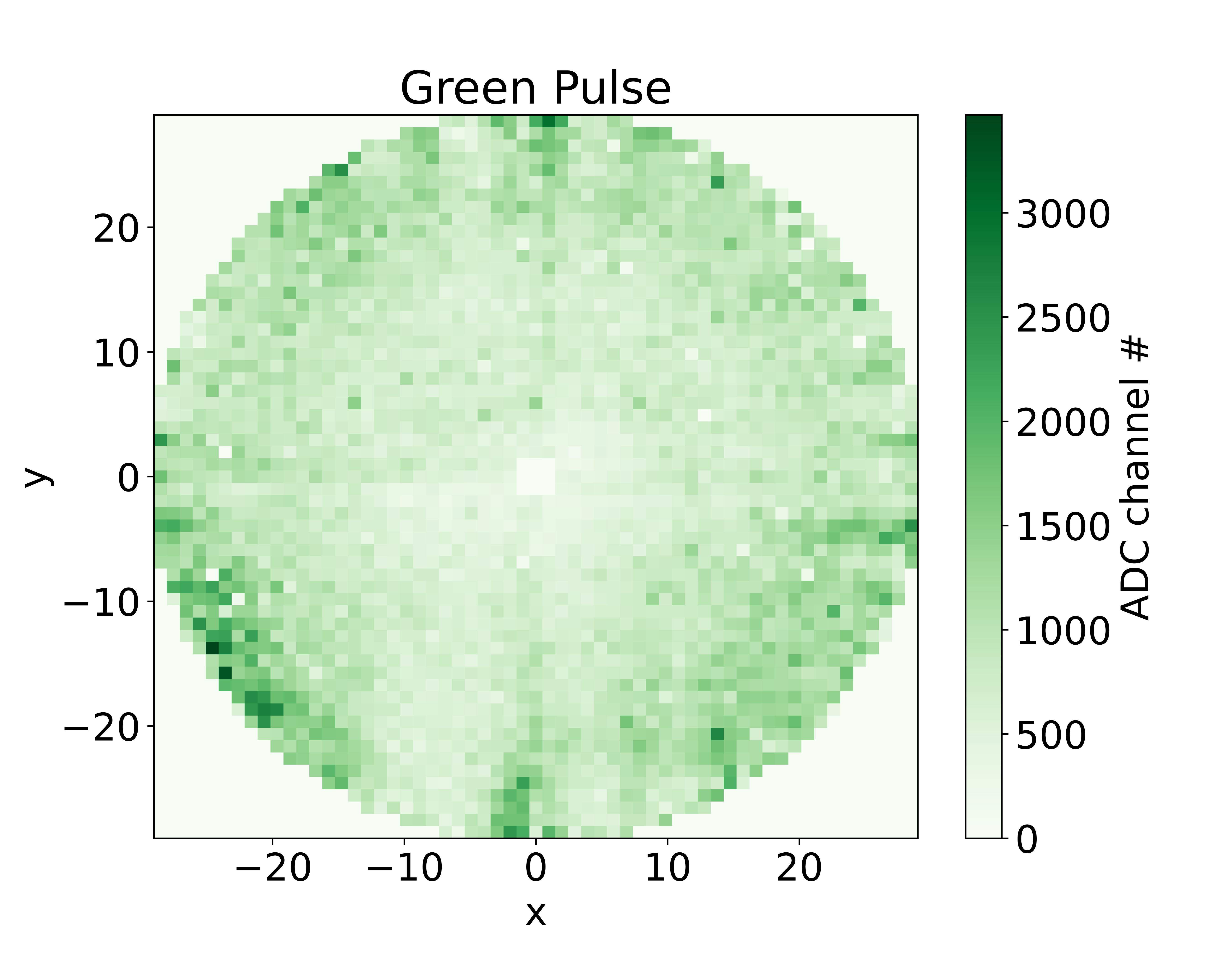}
  \includegraphics[width=0.32\textwidth]{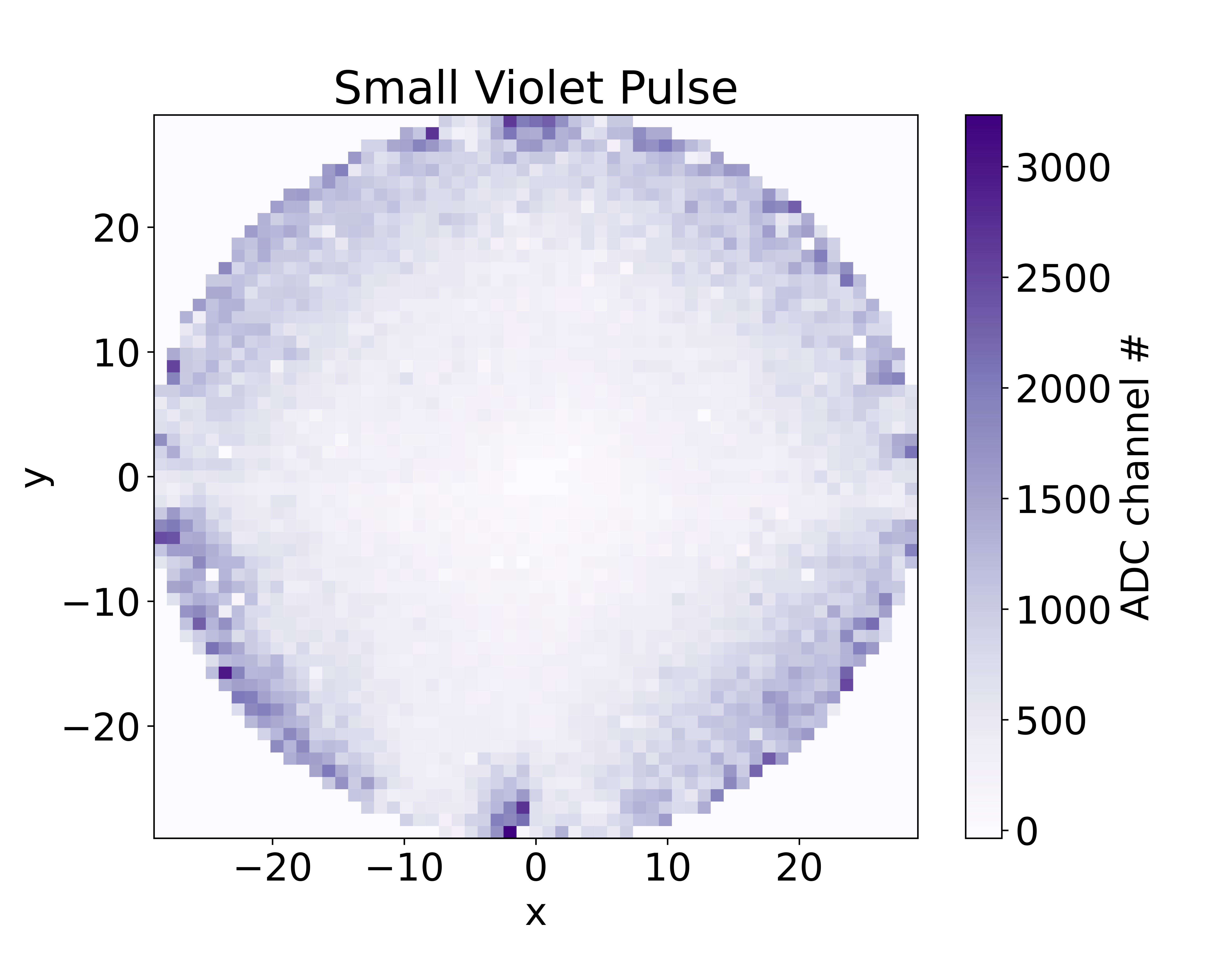}
  \caption{The peak amplitude recorded during a PrimEx run for the large blue, green, and small violet pulses. The x, and y axis are the locations of the FCAL blocks from beamline. Within the boundaries of the FCAL, blocks with zero values were either 1) too near large pulses and were oversaturated, 2) too distant from small pulses to record, or 3) not responding, possibly due to degrading PMT power supply bases. These plots represent input data for three of the five pulse types for a single run.}
\label{fig:pulses}
\end{figure}

Measurements from these pulses include the signal peak fitted with a Gaussian, the peak’s width, and the statistical mean. A visualization of the ADC peak fits for the large blue and green pulses is shown in Fig. \ref{fig:pulses}. Additionally, the $\chi^2$ values of each of these fits are recorded for each run and each block/PMT. 
In principle, these LED signals can be correlated to the individual detector gains when used as inputs to an ML algorithm.

FCAL blocks are each placed on a grid within a quadrant; additionally, the FCAL blocks also have a radial distance from the center beamline and each can be assigned an FCAL ``ring number" starting from the center of the FCAL.
As ``ring 1" of the FCAL contains no blocks, ``ring 2" is the center-most ring.
A block's ring number is important in two ways related to gain calibration.
First, it affects the utility of the LED pulses, i.e., the input data.
With respect to the monitoring system, the large blue and violet pulses often oversaturate the blocks in the outer ring of the FCAL. 
This can be seen in an example of a run's ADC channel number for the large blue pulses  shown in Fig. \ref{fig:pulses}. 
Several of the outer ring blocks show no value, indicating that the peak amplitude could not be fit due to oversaturation. 
Conversely, the small blue and violet pulses are used to monitor the middle and outer rings, and the pulses often do not reach the inner rings, again leading to no available input data. 

\begin{figure}[h]
  \centering
  \includegraphics[width=0.95\textwidth]{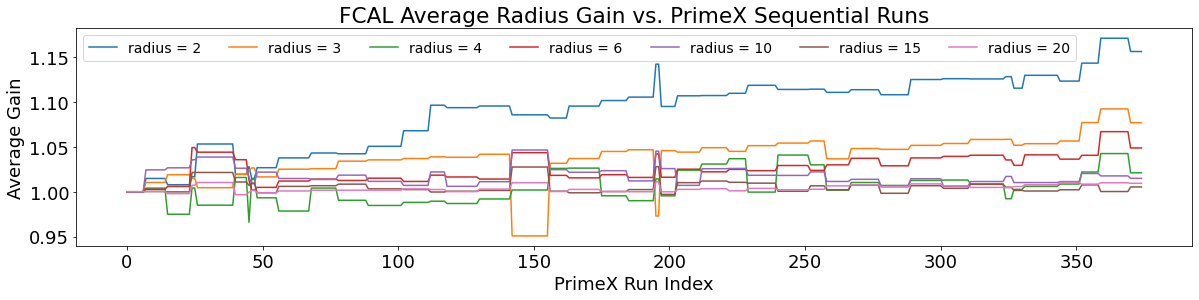}
  \caption{For the PrimEx runs included the dataset, the average gain for the blocks in each FCAL ring is shown sequentially throughout the run period. The gain increases throughout the run period for the inner rings and, on average, remains stable for the middle and outer rings of the FCAL.}
\label{fig:average_gain_ring_line}
\end{figure}

\begin{figure}[h]
  \centering
  \begin{subfigure}{0.45\linewidth}
    \includegraphics[width=0.98\textwidth]{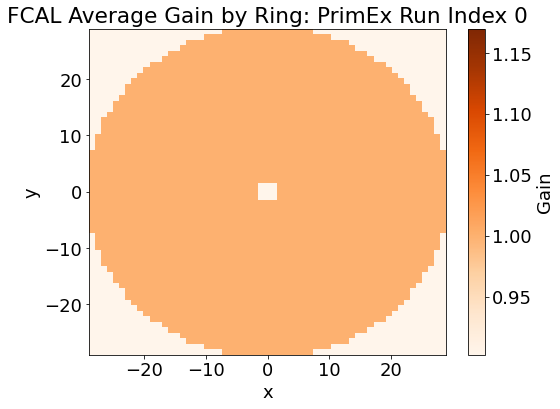}
    \caption{First PrimEx run}
  \end{subfigure}
  \begin{subfigure}{0.45\linewidth}
      \includegraphics[width=1.\textwidth]{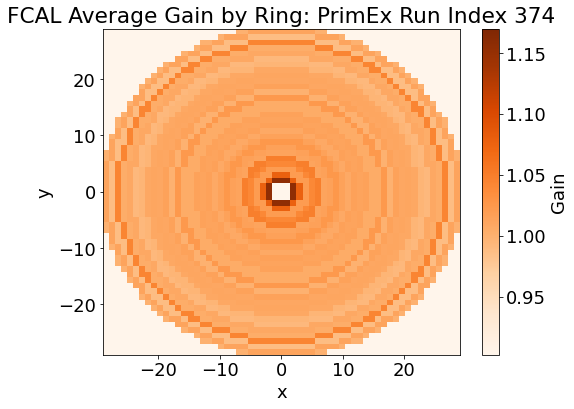}
      \caption{Last PrimEx run}
  \end{subfigure}
    
  \caption{The figures illustrate the average gain calibration factor (unitless) for the FCAL rings for the first PrimEx run, and for the last PrimEx run in the capstone project data set. 
  The average gain calibration factor is 1.0 for the first run, which is treated as a reference run. 
  In the inner rings of the FCAL, the gain increases throughout a run period due to the browning of the lead glass blocks caused by exposure to the beam.}
\label{fig:average_gain_first_last_runs}
\end{figure}

Second, a block's radial distance from the beam line leads to changes in the gain calibration values throughout a run period.
Continued exposure to the beam causes the FCAL's lead glass to brown,
resulting in an increase in the gain calibration factor.
The correlation between integrated beam current and an FCAL block's gain calibration value holds true only for blocks exposed to enough beam current to brown the lead glass. The effect can be seen in Figs. \ref{fig:average_gain_ring_line} and \ref{fig:average_gain_first_last_runs}.

%

Analytically combining the color and strength pulses allows a more complete picture of the FCAL, but an ML model needs a mechanism to avoid learning erroneous information from over and undersaturated blocks. 
Additionally, during the 2019 PrimEx run period, the PMT’s power supply bases were not frequently monitored, resulting in a progressive degradation in performance. 
Experts observed that the bases failed intermittently, with run failures becoming more frequent as the bases degraded. 

\begin{figure}[h]
  \centering
    \includegraphics[width=0.4\textwidth]{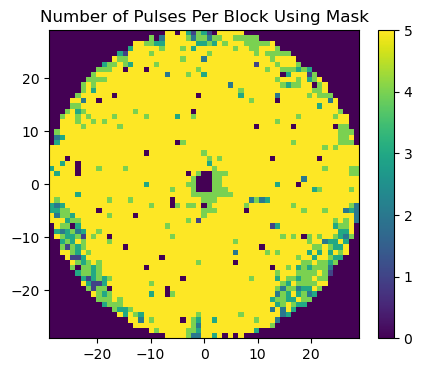}
  \caption{With the mask, 2287 blocks have information from all five pulse types, 299 blocks from four pulse types, 108 from three pulse types, 38 from two pulse types, 16 from one pulse type, and 52 blocks with no input data due to mask.}
\label{fig:fcal_mask}
\end{figure}

A method was explored to mask blocks that were under or oversaturated or had poorly performing power supplies.
To generate the masks for each pulse type, the range of the fitted peak amplitude position was examined from the total range of peak values.
The selected ranges for peak amplitudes for the pulses were as follows: 100 - 4000 ADC for the large violet and large blue pulses and 100 - 2500 ADC for the small violet, small blue, and green pulses. 
After the fitted peak amplitude cuts, blocks with a negative $\chi^2$ fit value were masked.
The results of the mask are shown in Fig. \ref{fig:fcal_mask}.
Notably, 2287 of the 2800 FCAL blocks have input from all five types; however, using the mask results in 52 blocks with no input available for an ML-based gain calibration.

Due to the spatial nature of the FCAL a CNN approach was explored. 
3-D convolutions were implemented. Intuitively, this allows for the mixing of the information from the colored pulses to predict the gain calibration constants. 
The input for the CNN contained a matrix of the masked LED data, which was represented as values per block, per run, and per channel. 

$\textbf{L} (N \times T)$ denotes the matrix containing LED pulse data collected from the 2800 LED blocks, where the rows represent measurements from $N$ runs, and the columns are measured for the $T$ LED blocks. 
The input of a 3-D convolution layer is a 3-D tensor, denoted as $X(N,T,c)$, where $N$, $T$, and $c$ represent its numbers rows, columns, and channels, respectively. 
Therefore, we convert the measurement matrix $\textbf{L} (N \times T)$ to a tensor $L(N,T,1)$ for each channel, so that it can be fed to a convolution layer.
A convolutional layer consists of several convolutional kernels. 
Each kernel is a filter with the size of $(ks, kt, c)$, which maps a patch of its input tensor to a scalar, given by
           
\begin{equation}
x_{out} = \sigma(W(ks,kt,kd) \circ X(ks,kt,kd) + b)
\end{equation}

where $W(ks,kt,kd)$ and $b$ are the parameters of the convolution kernel and the bias term, respectively; $\sigma$ represents a nonlinear activation function.   
A convolutional kernel slides over the run (row) and block (column) dimension of the input tensor with a $kd$ depth and maps to a 1-D output tensor. 

A neural network consisting of 3-D Convolutional layers in conjunction with 2-D Convolutional layers and finally passing the flattened input through Dense layers was implemented for this experiment. 
A NAS was conducted to compare model architecture and hyperparameters. 
Additionally, the best-performing model architecture was compared with and without masked input. 
With only 375 PrimEx runs available, five-fold cross-fold validation was utilized to evaluate the model performance.
Mean squared error (MSE) was used as the training loss function.
Results are presented in Table \ref{tab:fcal_results}.

\begin{table}[ht]
\centering
\small
\caption{5-fold performance comparison. Reported metrics are averaged over the FCAL blocks and over a 10\% holdout/test set of PrimEx runs.}
\label{tab:dataset_performance_metrics}
\begin{tabular}{lcS[table-format=3.3]S[table-format=3.3]S[table-format=3.3]}
\hline
\textbf{Dataset} & \textbf{fold idx} & \textbf{residual} & \textbf{MAPE}  & \textbf{MSE} \\ \hline
masked              & 1 & 0.025 & 2.191 & 0.004 \\
unmasked            & 1 & 0.866 & 81.194 & 25.103 \\
\hline
masked              & 2 & 0.025 & 2.221 & 0.004  \\
unmasked            & 2 & 0.057 & 4.991 & 0.015 \\
\hline
masked              & 3 & 0.031 & 2.663 & 0.004  \\
unmasked            & 3 & 0.081 & 7.021 & 0.033 \\
\hline
masked              & 4 & 0.033 & 2.912 & 0.007  \\
unmasked            & 4 & 0.228 & 20.764 & 0.747 \\
\hline
masked              & 5 & 0.021 & 1.861 & 0.002  \\
unmasked            & 5 & 0.060 & 5.272 & 0.016 \\
\hline
masked   & avg & 0.027 & 2.370 & 0.004  \\
unmasked & avg   & 0.258 & 23.848 & 5.183  \\
\hline
\end{tabular}
\label{tab:fcal_results}
\end{table}

Calibrations were evaluated based on their utility to calibrate the FCAL's measurement of $\pi^0$ mass.
By utilizing block masking, we obtained average results within 2.37\% of the target gain constants, and our best fold's test set was within 1.86\%.
However, scientists are interested in whether predicted calibrations result in physics data worthy of detailed analysis. 
Thus, the ML calibrations are compared to traditional calibrations using the same criteria.
Masked blocks were given a predicted gain calibration of 1.0, as is the practice of the FCAL expert for blocks that are unable to be calibrated in the traditional method due to too few detected $\pi^0$ particles.  

For the best fold's predictions, we compare the measured $\pi^0$ mass for our calibrations to measurements obtained with traditional calibrations. These comparisons are shown in Fig. \ref{fig:FCAL_correlations}.
The traditional calibrations are more accurate.

\begin{figure}[h]
\centering
\includegraphics[scale=0.25]{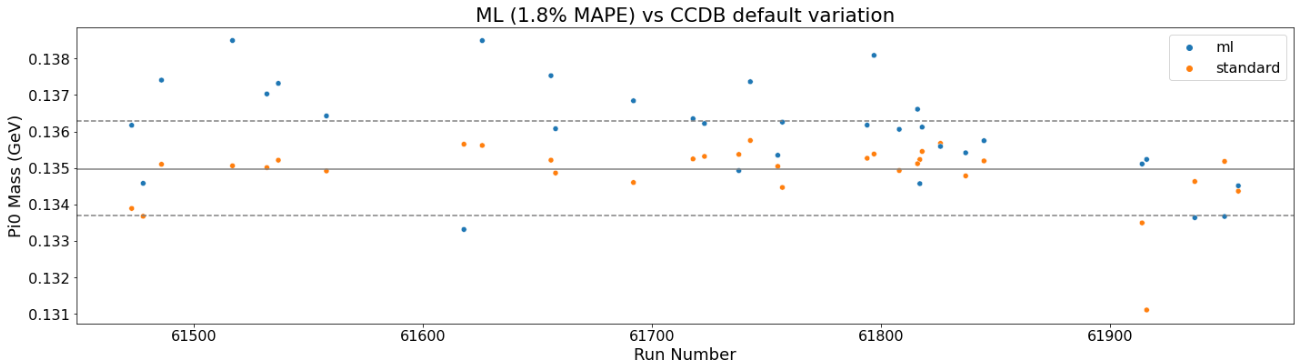}
\caption{Displays the comparison of measured $\pi^0$ mass using the gain calibrations predicted by the best performing CNN (blue) and using standard/traditional calibrations (orange). The solid line indicates the theoretical $\pi^0$ mass. The dashed lines indicate the $1\%$ thresholds. The traditional calibrations are generally more accurate.}
\label{fig:analysis_manyruns}
\end{figure}


Continued analysis of the LED pulse amplitudes and the traditional gain calibrations makes it clear that other than the radiation-damaged blocks at the center of the FCAL, there is only weak-to-moderate correlation between the measured LED pulse amplitudes and an FCAL block's gain calibration value. 
This lack of strong correlation is illustrated in Fig. \ref{fig:FCAL_correlations} illustrating the Pearson correlation coefficients for the PrimEx 2019 run period. 
Moreover, a contemporaneous and independent analysis of correlations between the LED monitoring system and FCAL gains by a Hall-D scientist did not find a clear relationship. This motivated the following simulation study described in Section \ref{sec:sim}.

\begin{figure}[h]
\centering
\begin{subfigure}{0.9\linewidth}
\includegraphics[width=1.\textwidth]{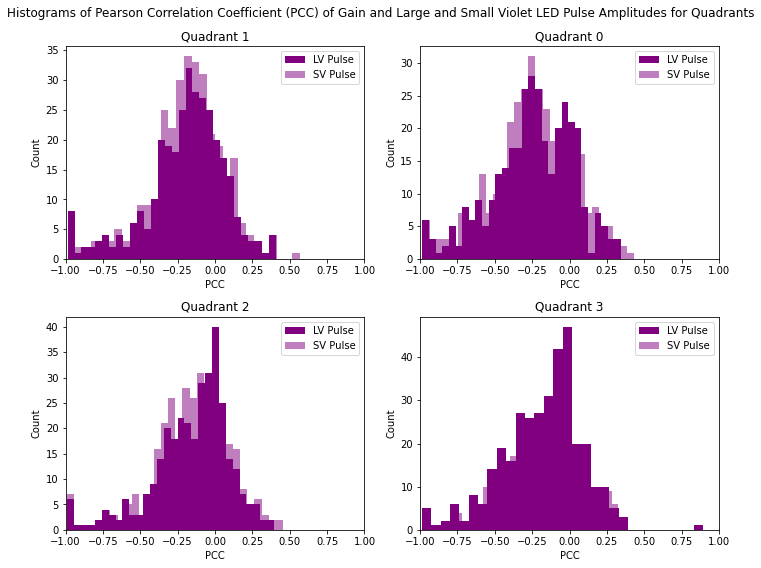}
\caption{Large violet and small violet pulses}
\label{fig:LV_SV}
\end{subfigure} \\

\begin{subfigure}{0.9\linewidth}
\includegraphics[width=1.\textwidth]{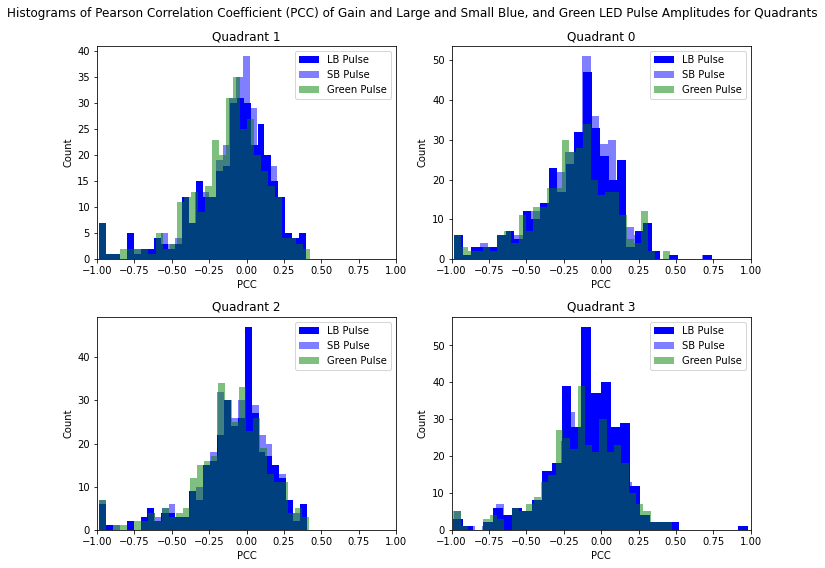}
\caption{Large blue, small blue, and green pulses}
\label{fig:LB_SB_G}
\end{subfigure}
\caption{The Pearson correlation coefficient (PCC) of the FCAL block gain calibration values and the LED pulse amplitudes. An anti-correlation (values $<$ 0) is expected; however, the scale of the anti-correlation is smaller than experts expect.}
\label{fig:FCAL_correlations}
\end{figure}

\subsubsection{FCAL Simulated Machine Learning} \label{sec:sim}

The lack of a strong relationship between LED pulse amplitudes and the gain calibrations constants motivated a simulation study. 
With simulated data, the ability to extract the gains from the amplitudes can be verified. 
This study demonstrates that in the presence of a correlation between the measured LED pulse amplitude and gain correction factor, a machine-learning approach could ultimately lead to ML-predicted gain calibration constants.  This in turn could be translated into an autonomous control system for the GlueX FCAL and other calorimeter systems. 

For our simulation, the LED pulse amplitude, $A$, is modeled by the following Eq.:
\begin{equation}\label{eqn:bigA}
        A_{b,t} = g_{b,t}^{-1} \cdot \omega_b \cdot \alpha_t \cdot R_{b,t} + \epsilon_{b,t},
\end{equation}
where:\\
\indent b is the block index\\
\indent t is the time index\\
\indent $A_{b,t}$ is the amplitude read for block $b$ at time $t$\\ 
\indent $g_{b,t}$ is the PMT gain for block $b$ at time $t$\\
\indent $\omega_b$ is the optical coupling of block $b$\\
\indent $\alpha_t$ is the LED pulser amplitude at time $t$\\
\indent $R_{b,t}$ is the radiation damage of block $b$ at time $t$\\
\indent $\epsilon_{b,t}$ is a stochastic measurement error for block $b$ at time $t$\\
\newline
For each block, a superposition of two sine functions of arbitrary frequency and phase was selected for the simulation of $g_{b,t}$,
\begin{equation} \label{eqn:gain}
    g_{b,t} = 1 + C_b(\sin(\eta_{b,1} t + \phi_{b,_1}) + \sin(\eta_{b,2} t + \phi_{b,_2}))
\end{equation}
where the block constant $C_b$ scales the sine wave superposition independently for each block and the frequencies($\eta_{b,1},\eta_{b,2}$) and phases($\phi_{b,_1},\phi_{b,_2}$) values are independently drawn from a normalized random distribution for each block. The gain correction factor values obtained from Eq. \ref{eqn:gain} fluctuate around one to reflect the experimental data. The optical coupling constant, $\omega_b$, is sampled for each block and trial from a common normal distribution,
\begin{equation}\label{eqn:omega_b}
    \omega_{b}=\mathcal{N}(\mu, \sigma^2)
\end{equation}
 with $\mu=1$ and $\sigma=0.1$. For each block and trial, $\omega$ was held constant over $t$. The amplitude of the LED pulser, $\alpha_t$, is algorithmically modeled as piecewise line segments that reverse direction randomly with a probability of 1\% and also upon reaching the boundaries of a $\pm 5\%$ envelope centered on a value of 1.0. 

$R_{b,t}$ is the radiation damage for each block. For the purposes of this study, the radiation factor, $R_{b,t}$, is set equal to one. An additional noise parameter, $\epsilon$, is modeled as Zero Mean Gaussian White Noise, and is possibly block dependent such that when

\begin{equation}\label{eqn:epsilonx}
    \epsilon \leftarrow N(0,x)
\end{equation}

where x is the noise level expressed as a percentage.

\begin{figure}[H]
    \centering
    \includegraphics[width=0.8\linewidth]{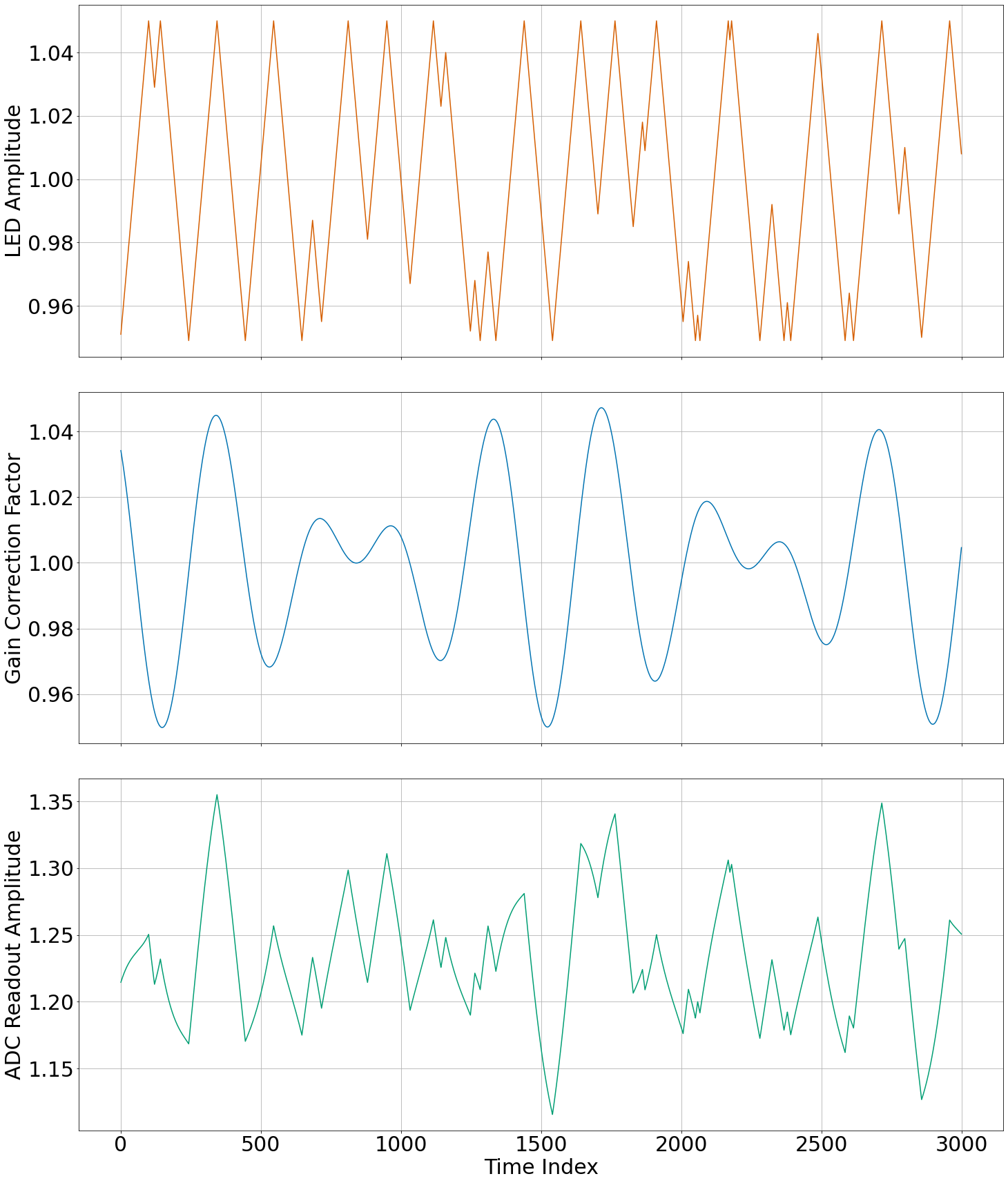}
    \caption{For a single simulated FCAL block, in orange(top), $\alpha$, the amplitude of the LED pulser; in blue(middle), 
    $g$, the gain correction factor for a single block; and, in green(bottom), $A$, the ADC readout amplitude for the block. }
    \label{fig:gf}
\end{figure}

Fig. \ref{fig:gf} shows an example of $\alpha_t$, $g_{t},$ and $A_t$ that are used to calculate the LED pulse amplitude at each time step without stochastic noise.

The ADC readout amplitude, $A$, (example shown at bottom of Fig. \ref{fig:gf}) is the only input data available for training ML models to predict the PMT gain $g$, in order to be consistent with the approach described in Section \ref{sec:fcal}.

The first method was to extract the gain calibration values directly from Eq. \ref{eqn:bigA}. The gain calibration values extracted are then given as input features to a Generalized Linear Model (GLM), Three Layer Perceptron (TLP), Random Forest (RF), and Long Short Term Memory (LSTM). The approach can be summarized in the steps below: 

\begin{itemize}
    \item Select some block $b$ and use $g_{b,t}$ as the target to model.
    \item For each repeated trial at a selected noise level ($\epsilon$) use Eq.\ref{eqn:bigA} to calculate $A$ for all blocks.
    \begin{itemize}
        \item Select the two most positively and negatively correlated $A$ blocks with the target $g_{b}$, for a total of four predictors, using the training portion of the data.
        \item Train a selected model type on these $A$ blocks and report, for each trial, both the \% Mean Absolute Error (MAE) and Max Absolute Error (AE) across the time index using the validation data.
    \end{itemize}
    \item  Report Mean \% MAE and  Mean \% Max AE over all trials and the corresponding standard error of the mean values.
\end{itemize}

A trial refers to a distinct trained solution obtained and evaluated. All model results are based on 100 trials for each noise level except for the LSTM models, which used 25 trials for each noise level, due to differences in the time required for training. Training for each model used an 80\%/20\% split for training/validation without shuffling to preserve the time order of the data. LSTM models used a history window depth of 50 time unites. 

Model performance for an individual trial was evaluated using the MAE and maximum AE across the time index for the validation portion of each trial. Total model performance was evaluated using mean MAE and mean Maximum AE and associated Standard Errors of the Means over all trial evaluation scores.
Tables (\ref{tab:fcal_raw_model_results}) and (\ref{tab:fcal_rawmax_model_results}) summarize these results for each noise level and model type. As expected, as the noise level increases, the performance of each model degrades. 

\begin{table}[H]
\small
\centering
\caption{Mean \% MAE Results from Modeling for PMT Gain (using $A$).}
\begin{tabular}{lllllllll}
\hline
\textbf{Model} & \textbf{GLM} & & \textbf{TLP} & & \textbf{RF} & & \textbf{L(50)} & \\
\hline
Noise & Mean & StdErr & Mean & StdErr & Mean & StdErr & Mean & StdErr \\
\hline
0\% & 0.66 & 0.00 & 0.65 & 0.00 & 1.2 & 0.0027 & 1.2 & 0.026 \\
1\% & 1.1 & 0.0025 & 1.1 & 0.0032 & 1.3 & 0.0036 & 1.2 & 0.029 \\
2\% & 1.5 & 0.0054 & 1.5 & 0.0068 & 1.6  & 0.010 & 1.2 & 0.027 \\
3\% & 1.8 & 0.0089 & 1.8 & 0.010 & 1.8 & 0.014 & 1.3 & 0.036 \\
4\% & 2.0 & 0.010 & 2.0 & 0.010 & 2.0 & 0.014 & 1.3 & 0.056 \\
5\% & 2.1 & 0.010 & 2.1 & 0.010 & 2.2 & 0.011 & 1.3 & 0.059 \\
6\% & 2.2 & 0.010 & 2.3 & 0.010 & 2.2 &  0.010 & 1.4 & 0.055 \\
7\% & 2.3 & 0.010 & 2.3 & 0.0082 & 2.3 & 0.010 & 1.5 & 0.063 \\
8\% & 2.3 & 0.010 & 2.3 & 0.0094 & 2.3 & 0.0079 & 1.5 & 0.040 \\
9\% & 2.2 & 0.0069 & 2.3 & 0.0067 & 2.3 & 0.0067 & 1.6 & 0.059 \\
10\% & 2.3 & 0.0061 & 2.3 & 0.0056 & 2.3 & 0.0059 & 1.5 & 0.074 \\
\hline
\end{tabular}

\label{tab:fcal_raw_model_results}
\end{table}

\begin{table}[H]
\small
\centering
\caption{Mean \% Max AE Results from Modeling for PMT Gain (using $A$).}
\begin{tabular}{lllllllll}
\hline
\textbf{Model} & \textbf{GLM} & & \textbf{TLP} & & \textbf{RF} & & \textbf{L(50)} & \\
\hline
Noise & Mean & StdErr & Mean & StdErr & Mean & StdErr & Mean & StdErr \\
\hline
0\% & 1.3 & 0.00 & 1.3 & 0.00 & 2.8 & 0.0094 & 2.4 & 0.089 \\
1\% & 3.6 & 0.028 & 3.8 & 0.043 & 3.8 & 0.028 & 2.8 & 0.072 \\
2\% & 5.1 & 0.043 & 5.1 & 0.040 & 5.2 & 0.048 & 3.0 & 0.056 \\
3\% & 6.1 & 0.053 & 5.7 & 0.064 & 5.9 & 0.050 & 3.3 & 0.063 \\
4\% & 6.6 & 0.052 & 6.1 & 0.070 & 6.2 & 0.040 & 3.5 & 0.12 \\
5\% & 6.9 & 0.052 & 6.3 & 0.076 & 6.3 & 0.036 & 3.4 & 0.12 \\
6\% & 6.8 & 0.050 & 6.2 & 0.079 & 6.4 & 0.038 & 3.4 & 0.12 \\
7\% & 6.8 & 0.054 & 6.1 & 0.077 & 6.3 & 0.034 & 4.1 & 0.11 \\
8\% & 6.6 & 0.045 & 6.0 & 0.088 & 6.3 & 0.033 & 4.3 & 0.13 \\
9\% & 6.6 & 0.045 & 5.9 & 0.074 & 6.3 & 0.034 & 4.6 & 0.15 \\
10\% & 6.6 & 0.040 & 5.9 & 0.070 & 6.2 & 0.030 & 4.4 & 0.16 \\
\hline
\end{tabular}

\label{tab:fcal_rawmax_model_results}
\end{table}


The LSTM(50) outperforms all of the other models for nearly all noise levels for both mean and maximum absolute errors. The GLM, TLP and RF models have similar performance across noise levels for both mean and max AE.

If we establish a notional acceptance criteria for model performance of 1\% Mean MAE, that is to say {\it average performance} over all trials, we see that GLM and TLP are acceptable at 0\% and marginally at 1\% noise (Table \ref{tab:fcal_raw_model_results}). If instead the metric is  1\% Mean Max AE, or {\it worst-case performance}, we see (Table \ref{tab:fcal_rawmax_model_results}) that no models would be acceptable. 

Poor performance of direct modeling efforts on the available block amplitude data $A$ as indicated by Tables (\ref{tab:fcal_raw_model_results}) and (\ref{tab:fcal_rawmax_model_results}), suggest pursuing an alternate approach using some feature engineering. 

Since mean values across blocks for the block optical coupling constants $\omega$ and block PMT gains $g$  are very close to unity and likewise since the mean of the overall system noise $\epsilon$ is close to zero, we expect to benefit from averaging over blocks (and in some cases time) in bootstrapping a crude first model into one that we may be able to improve with machine learning techniques.  Thus, the following feature engineering method was implemented as described below.

Inverting Eq. \ref{eqn:bigA} with $\omega$ and $R$ constant for each block and any choice of $\epsilon$, we can write:

\begin{equation}\label{eqn:alpha_hat}
    \hat{\alpha}_t = \frac{1}{B}\sum_{b=1}^{B}A_{b,t}
\end{equation}

\begin{figure}[H]
    \centering
    \includegraphics[width=0.9\linewidth]{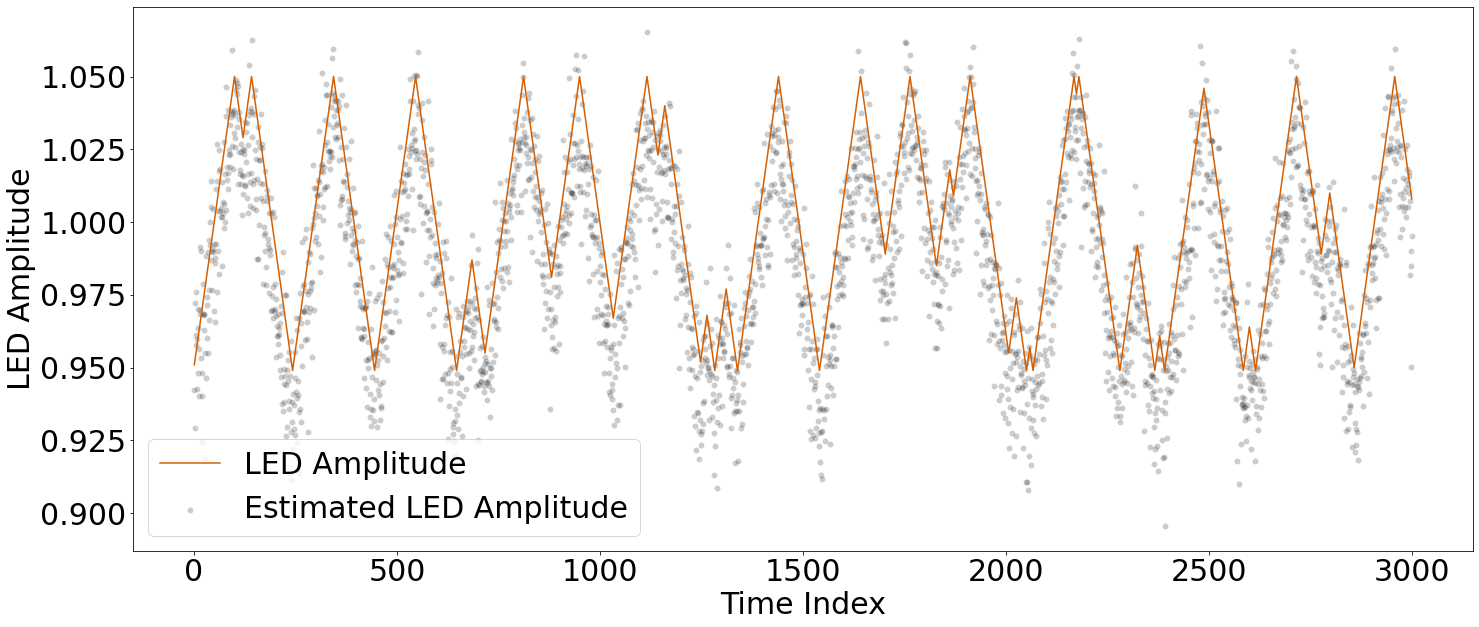}
    \caption{An Example of Estimated $\alpha$ with $\epsilon$ at 10\% noise.}
    \label{fig:ae10}
\end{figure}

Where $\hat{\alpha}_t$ is taken as our estimate of the time varying LED pulser amplitude $\alpha$ and where the mean is taken across blocks for each discrete time step in a trial. 
Fig. \ref{fig:ae10} shows an example of this model with $\epsilon$ at 10\% noise. 

Continuing to bootstrap Eqn. \ref{eqn:bigA}, we have for our estimate of mean $A_b$ for each block:

\begin{equation}\label{eqn:A_b}
    \overline{A}_b =  \frac{1}{T}\sum^T_{t=1}A_{b,t}
\end{equation}

and the mean value of $\alpha$ (for scaling):

\begin{equation}\label{eqn:alpha_bar}
    \overline{\alpha} =  \frac{1}{T}\sum^T_{t=1}\alpha_{t}
\end{equation}
 
where these means are calculated for the chosen block over the training portion of the data. It is possible to estimate the optical coupling constant $\omega_b$ for each block as


\begin{equation}\label{eqn:sclfctr}
    \hat{\omega}_b =  \frac{\overline{A}_b}{\overline{\alpha}},
\end{equation}

where the mean is taken across the time index. 

\begin{figure}[H]
    \centering
    \includegraphics[width=0.9\linewidth]{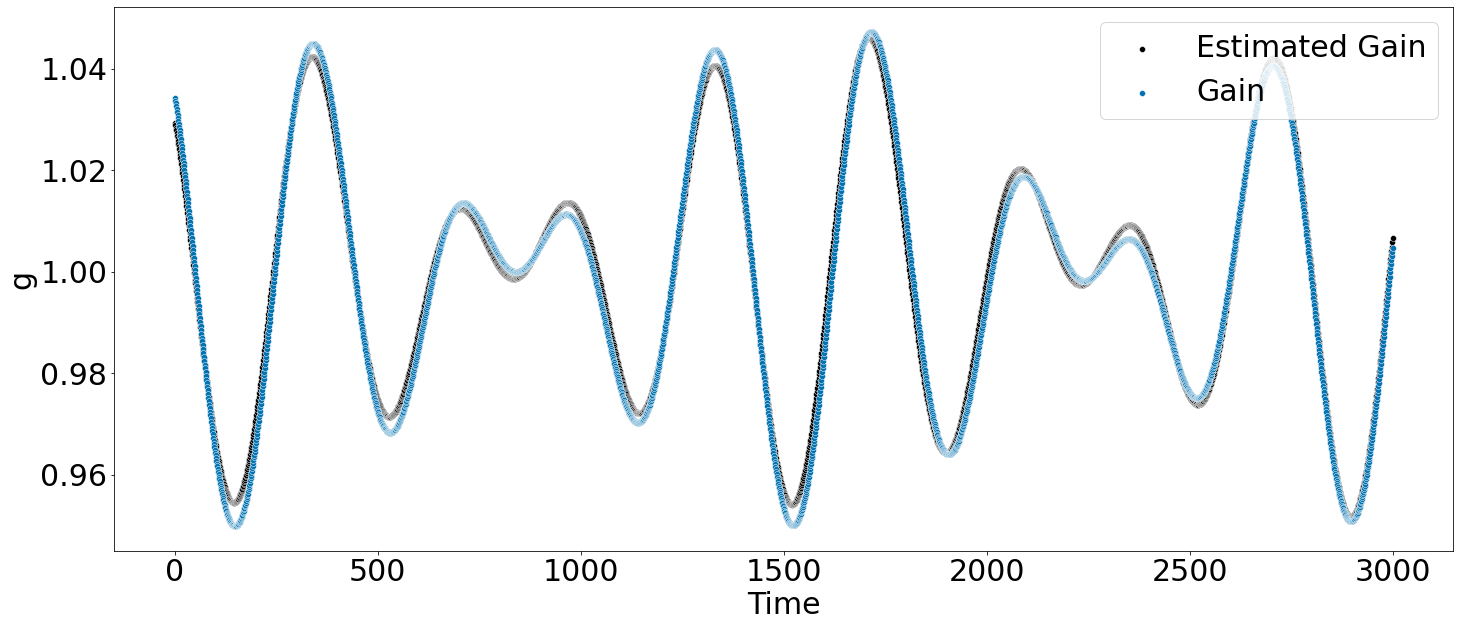}
    \caption{An example of estimated PMT gain with $\epsilon = 0$ and Correlation($\hat{g_{b,t}}, g_b) = 99.6\%$}
    \label{fig:aa0}
\end{figure}

\begin{figure}[H]
    \centering
    \includegraphics[width=0.9\linewidth]{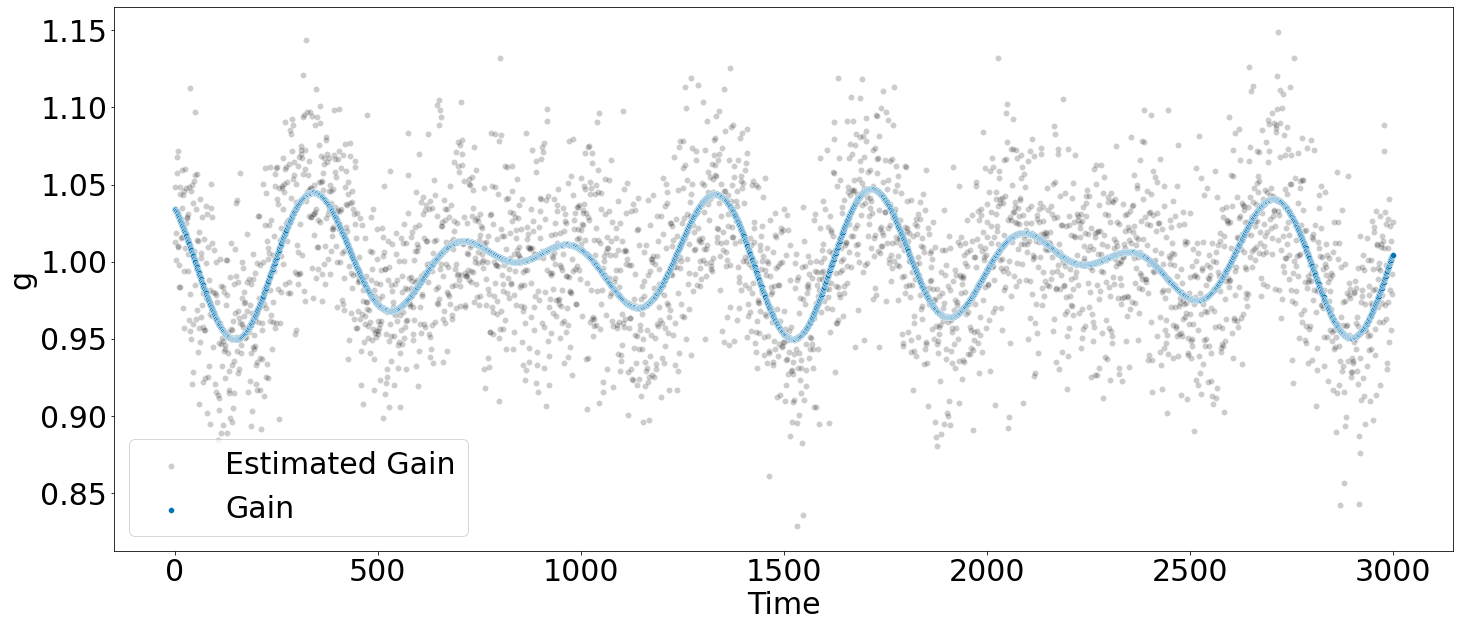}
    \caption{An example of estimated PMT gain with $\epsilon = N(0,5\%)$ and Correlation($\hat{g_{b,t}}, g_b) = 50.7\%$}
    \label{fig:aa3}
\end{figure}

\begin{figure}[H]
    \centering
    \includegraphics[width=0.9\linewidth]{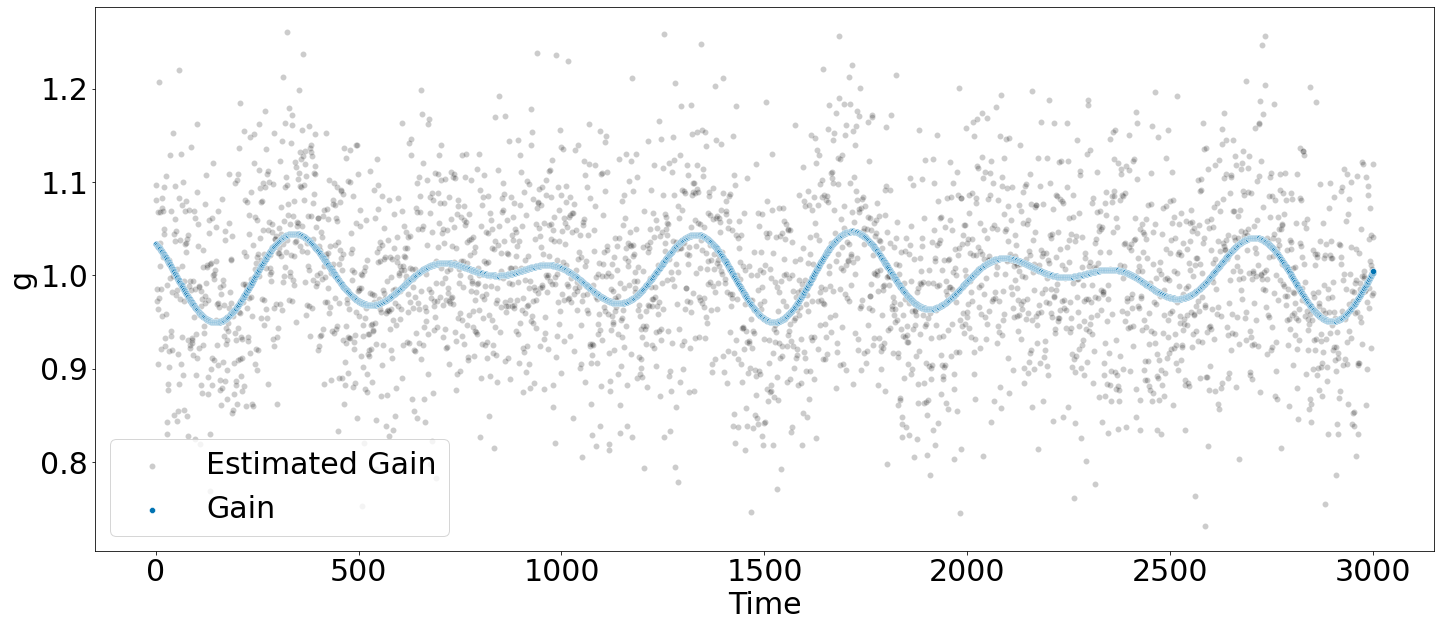}
    \caption{An example of estimated PMT Gain with $\epsilon$ = N(0,10\%) and Correlation($\hat{g_{b,t}}$, $g_b$) = 28.6\%}
    \label{fig:aa10}
\end{figure}

Finally, once again inverting Eq. (\ref{eqn:A_b}) with the previous calculated approximations, the estimated gain value for a given block at time t using the averaged amplitude is written as


\begin{equation}\label{eqn:gb_est}
    \hat{g_{b,t}} = \hat{\omega}_b \cdot \frac{\hat{\alpha}_t}{A_{b,t}}.
\end{equation}
Eq.  \ref{eqn:gb_est} is referred to as an Averaged Amplitude model to extract the gain values. 
Figs \ref{fig:aa0} - \ref{fig:aa10} show that the accuracy of the estimated gain $\hat{g}$ looks good by eye in the absence of noise, but degrades significantly as the noise level increases. 

\begin{figure}[H]
    \centering
    \includegraphics[width=0.9\linewidth]{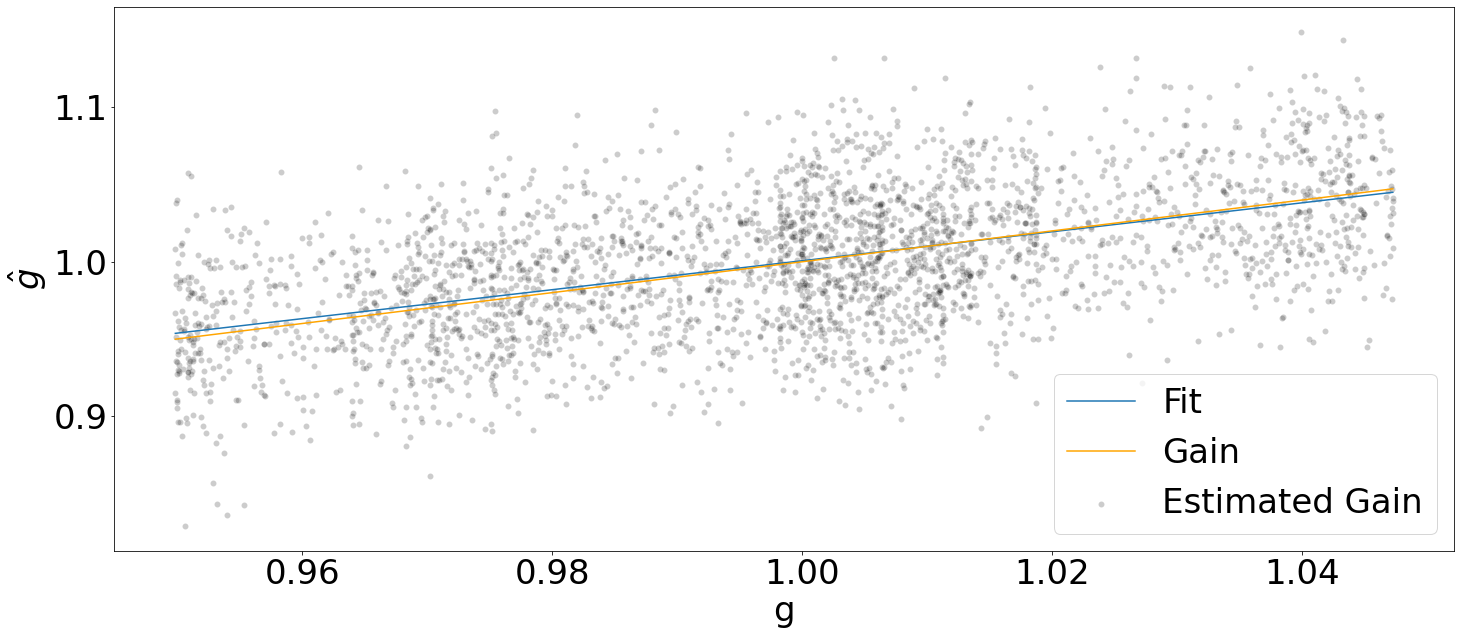}
    \caption{An example of estimated PMT Gain versus Actual $g$ with $\epsilon$ = N(0,5\%).}
    \label{fig:AAvsg5}
\end{figure}

Fig. \ref{fig:AAvsg5} shows that at 5\% noise level the AA calculation, while a noisy fit to the PMT gain (blue line), nonetheless demonstrates no systematic bias in its figure, the centroid (orange line) of which faithfully represents the deterministic PMT signal.  Other noise levels demonstrate this as well.  It is then expected that  ML modeling will de-noise this basic result into a final model.

\begin{figure}[H]
    \centering
    \includegraphics[width=0.9\linewidth]{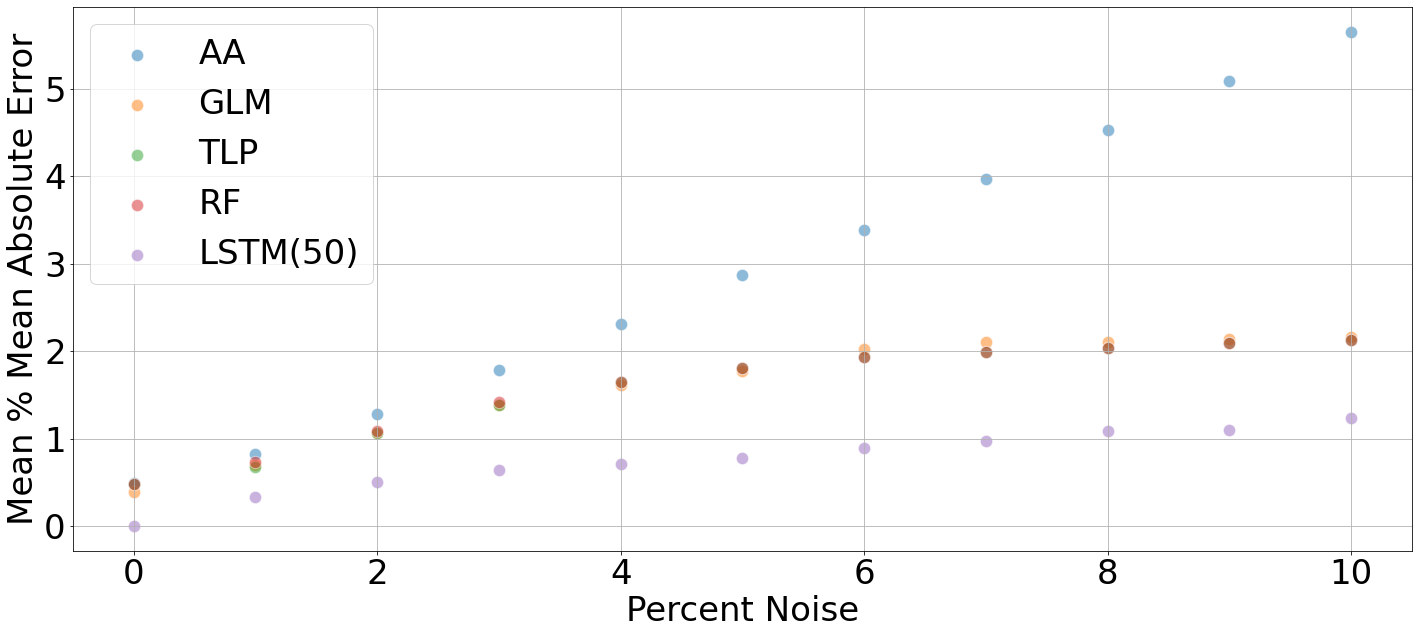}
    \caption{Summary of Trials: Mean \% MAE for Models versus Noise Level. }
    \label{fig:all-mae}
\end{figure}

We use the left-hand-side of Eq. \ref{eqn:gb_est} (the estimated gain for a block at each time index) as the sole feature input to the GLM, TLP, RF, and LSTM models. The GLM, TLP, and RF models had nearly identical performance regardless of noise level, as shown in Fig. \ref{fig:all-mae}.

Using the same notional acceptance criteria as in the previous method, we see that the AA meets the criteria up to 1\% noise, while GLM, and RF models meet the criteria up to 2\% noise whereas the LSTM(50) is acceptable up to about the 8\% noise level, where LSTM(50) is the LSTM model for history depth of 50 time units.  Table \ref{tab:fcal_model_results} (Fig. \ref{fig:all-mae}) summarizes the mean and standard error of the \% MAE calculated from the collection of trials as a function of noise level for each model.

\begin{table}[H]
\small
\centering
\caption{Mean \% MAE Results from Modeling for PMT Gain (using $\hat{g}$).}
\begin{tabular}{lllllllllll}
\hline
\textbf{Model} & \textbf{AA} & & \textbf{GLM} & & \textbf{TLP} & & \textbf{RF} & & \textbf{L(50)} & \\
\hline
Noise & Mean & StdErr & Mean & StdErr & Mean & StdErr & Mean & StdErr & Mean & StdErr \\
\hline
0\% & 0.50 & 0.00 & 0.45 & 0.00 & 1.0 & 0.089 & 0.48 & 0.00051 & 0.28 & 0.0054\\
1\% & 0.83 & 0.0042 & 0.69 & 0.0031 & 1.3 & 0.078 & 0.72 & 0.0034 & 0.35 & 0.010 \\
2\% & 1.3 & 0.0059 & 1.1 & 0.0054 & 1.4 & 0.056 & 1.1   & 0.00047 & 0.50 & 0.022 \\
3\% & 1.8 & 0.0092 & 1.4 & 0.0056 & 1.7 & 0.041 & 1.4 & 0.0067 & 0.64 & 0.017 \\
4\% & 2.3 & 0.011 & 1.6 & 0.0062 & 1.9 & 0.031 & 1.6 & 0.0062 & 0.74 & 0.021 \\
5\% & 2.9 & 0.014 & 1.8 & 0.0069 & 2.0 & 0.023 & 1.8 & 0.0075 & 0.82 & 0.023 \\
6\% & 3.4 & 0.017 & 1.9 &  0.0061 &  2.0 &  0.019 & 1.9 & 0.0064 & 0.91 & 0.026 \\
7\% & 4.0 & 0.018 & 2.0 &  0.0065 &  2.1 &  0.014 & 2.0 & 0.0065 & 0.98 & 0.029 \\
8\% & 4.6 & 0.022 & 2.0 &  0.0055 &  2.1 &  0.012 & 2.0 & 0.0060 & 1.1 & 0.036 \\
9\% & 5.1 & 0.024 & 2.1 &  0.0053 &  2.1 &  0.010 & 2.1 & 0.0059 & 1.2 & 0.031 \\
10\% & 5.7 & 0.031 & 2.1 & 0.0048 &  2.2 &  0.0082 & 2.1 & 0.0055 & 1.2 & 0.048 \\
\hline
\end{tabular}
\label{tab:fcal_model_results}
\end{table}

\begin{figure}[H]
    \centering
    \includegraphics[width=0.9\linewidth]{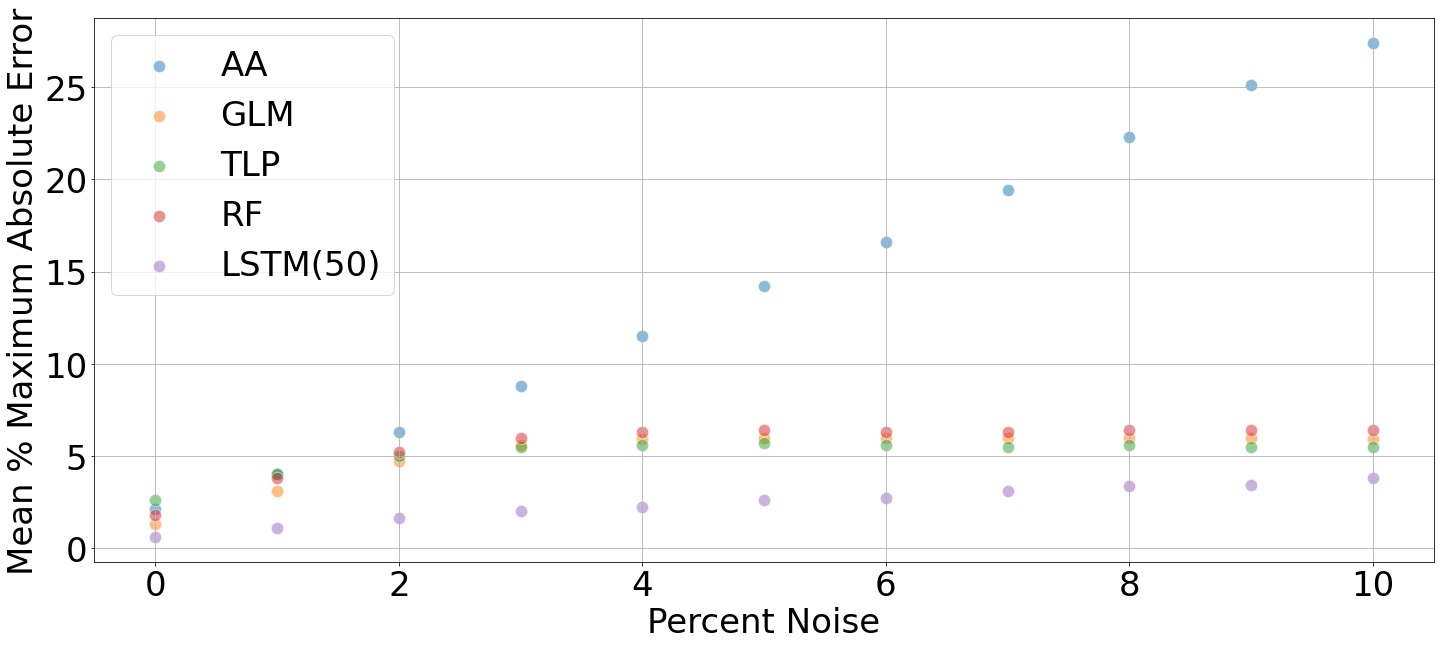}
    \caption{Summary of Trials Mean \% Max AE for Models versus Noise Level. }
    \label{fig:all_mxmae}
\end{figure}

If instead the metric is  1\% mean Maximum AE, or {\it worst-case performance}, we see in Table \ref{tab:fcal_model_max_results} (Fig. \ref{fig:all_mxmae}) that only the LSTM(50) model is acceptable and only for the 0\%, 1\% noise levels.

\begin{table}[H]
\small
\centering
\caption{Mean \% Max AE Results from Modeling for PMT Gain (using $\hat{g}$).}
\begin{tabular}{lllllllllll}
\hline
\textbf{Model} & \textbf{AA} & & \textbf{GLM} & & \textbf{TLP} & & \textbf{RF} & & \textbf{L(50)} & \\
\hline
Noise & Mean & StdErr & Mean & StdErr & Mean & StdErr & Mean & StdErr & Mean & StdErr \\
\hline
0\% & 2.1 & 0.00 & 1.6 & 0.00 & 2.6 & 0.16 & 1.8 & 0.0041 & 0.59 & 0.017 \\
1\% & 4.0 & 0.045 & 3.6 & 0.035 & 4.1 & 0.069 & 3.8 & 0.049 & 1.1 & 0.038 \\
2\% & 6.3 & 0.072 & 5.0 & 0.050 & 5.0 & 0.045 & 5.2 & 0.057 & 1.6 & 0.051 \\
3\% & 8.8 & 0.090 & 5.7 & 0.049 & 5.5 & 0.058 & 6.0 & 0.054 & 2.1 & 0.074 \\
4\% & 12 & 0.12 & 6.0 & 0.043 & 5.6 & 0.067 & 6.3 & 0.049 & 2.3 & 0.083 \\
5\% & 14 & 0.15 & 6.1 & 0.048 & 5.7 & 0.071 & 6.4 & 0.047 & 2.6 & 0.080 \\
6\% & 17 & 0.20 & 6.1 & 0.046 & 5.6 & 0.064 & 6.3 & 0.045 & 2.7 & 0.084 \\
7\% & 20 & 0.20 & 6.0 & 0.038 & 5.5 & 0.062 & 6.3 & 0.046 & 2.9 & 0.096 \\
8\% & 22 & 0.22 & 6.1 & 0.039 & 5.6 & 0.055 & 6.4 & 0.051 & 3.3 & 0.12 \\
9\% & 25 & 0.25 & 6.0 & 0.038 & 5.5 & 0.052 & 6.4 & 0.051 & 3.7 & 0.16 \\
10\% & 28 & 0.33 & 5.9 & 0.031 & 5.5 & 0.050 & 6.4 & 0.049 & 3.7 & 0.13 \\
\hline
\end{tabular}
\label{tab:fcal_model_max_results}
\end{table}

\begin{figure}[htb]
    \centering
    \includegraphics[width=0.9\linewidth]{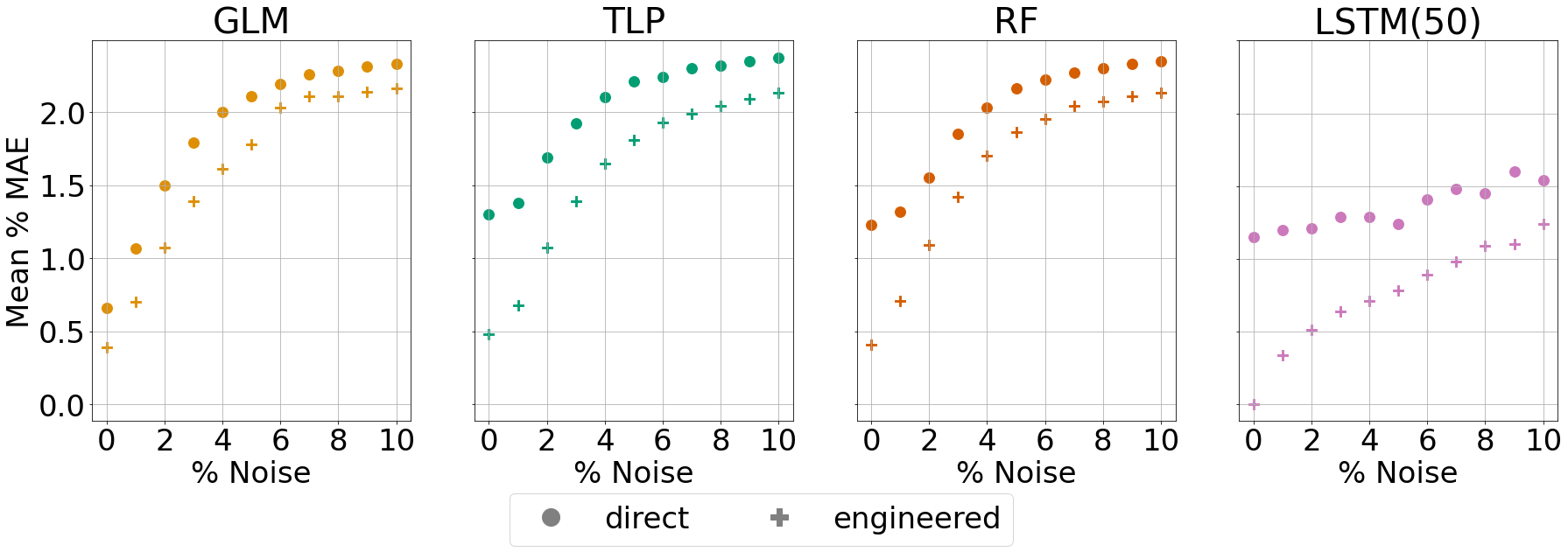}
    \caption{Model performance as a function of percent noise for each model. The circles correspond to using Eqn. \ref{eqn:bigA} directly as inputs whereas the crosses utilize the feature engineered inputs
.}
    \label{fig:model-comparison}
\end{figure}

Lift charts for model performance using the engineered inputs shows the comparison to the first method in Fig. \ref{fig:model-comparison}. For all models, using the feature engineered inputs results in improved performance across all noise levels. 

\begin{figure}[H]
    \centering
    \includegraphics[width=0.9\linewidth]{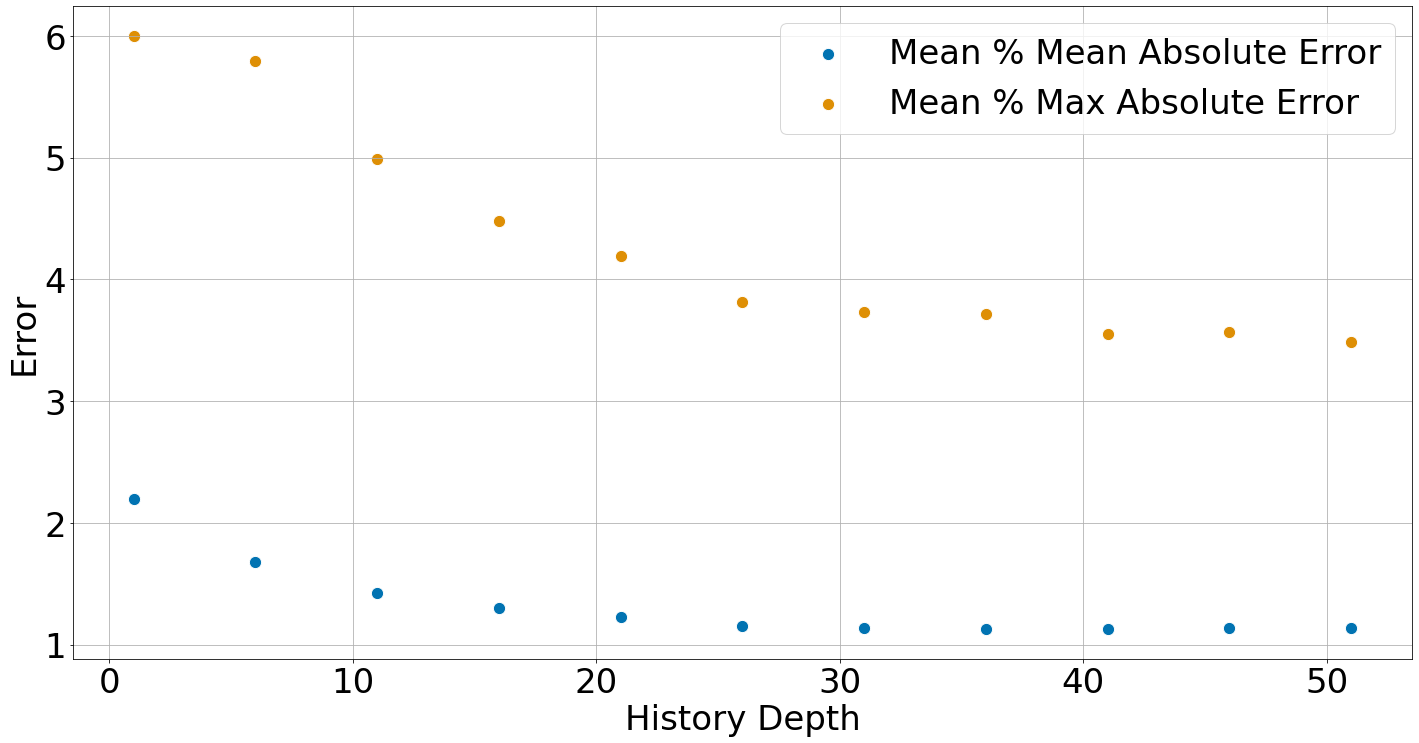}
    \caption{The LSTM model performance using $\hat{g}$ by history depth at 9\% noise level.}
    \label{fig:lstm9}
\end{figure}

The history depth refers to the amount of past data the model uses to make predictions. To further evaluate the performance of an LSTM model, results were generated for various history depths up to 50. Fig. \ref{fig:lstm9} shows the performance of the LSTM as a function of the history depth size at a 9\% noise level. 


Here at an example 9\% noise level, best average performance can be achieved at a history window depth of about 25 whereas best worst-case performance requires a history window depth of about 40. 

Thus, given a rather tight threshold on MAE of 1\%, an LSTM(50) model should be acceptable to predict the gain of individual FCAL blocks given the pulse amplitudes each block sees over time. It is important to note this work ignores the effects of radiation damage. Including the radiation damage would add on a spatial dependence and would affect blocks located near the beam line. The simulation work contained within validates a machine learning approach to extract the gains assuming there exists a correlation between individual block gains and the individual pulse amplitudes detected from LED pulse events.  This correlation is expected but disappointingly has not been observed in the FCAL LED pulse data.

\subsubsection{FCAL Inner Ring Radiation Damage}
\begin{figure}[H]
    \centering
    \includegraphics[width=0.8\textwidth]{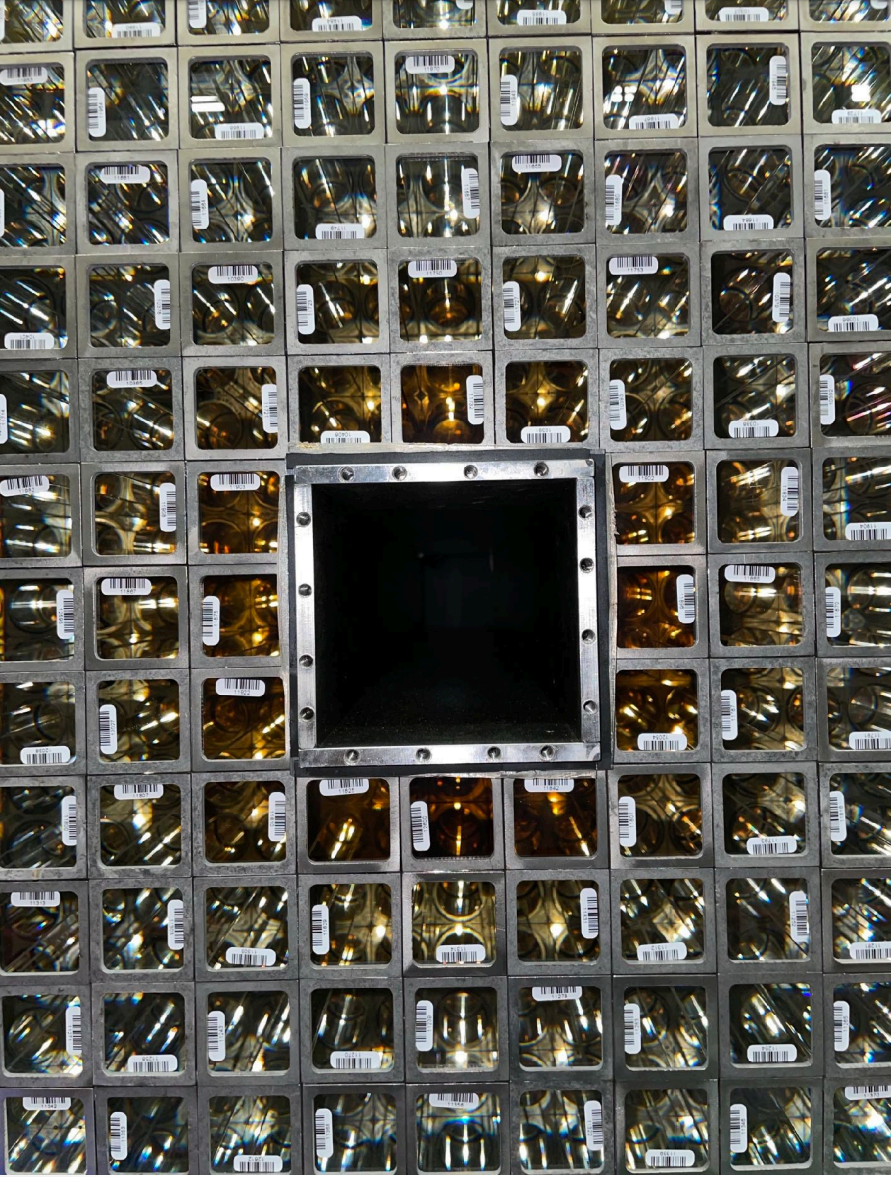}
    \caption{Image of the inner blocks during the disassembly of the detector. Here the radiation damage is clearly visible as a brown tint on the inner blocks.}
    \label{fig:radiationDamage}
\end{figure}

The inner blocks of the FCAL suffer radiation damage due to the proximity to the beam line. This effect is seen visually as "browning" of the lead glass blocks, shown in Fig. \ref{fig:radiationDamage}. As the radiation damage is largely localized to blocks located in inner rings, we investigate the correlation between the change in the existing gain calibration constants from the start of a run period to the luminosity. Hall-D does not have real time measurement of the luminosity. As a proxy, we utilize the scalers from the Time-of-Flight (TOF) system in Hall-D.

\begin{figure}[htb]
    \centering
    \includegraphics[width=0.8\textwidth]{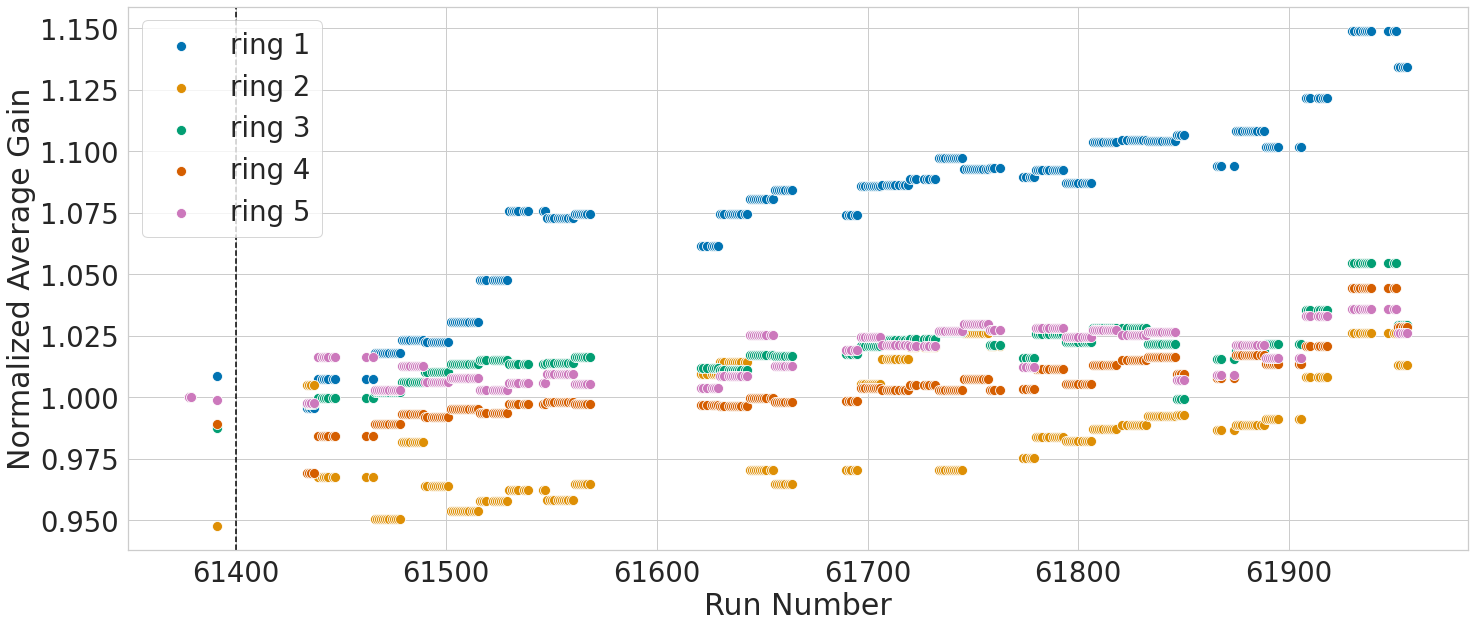}
    \caption{ Normalized Average Gain by Inner Ring for the PrimEx run period. }
    \label{fig:normalizedAvgGain}
\end{figure}

\begin{figure}
    \centering
    \includegraphics[width=0.8\linewidth]{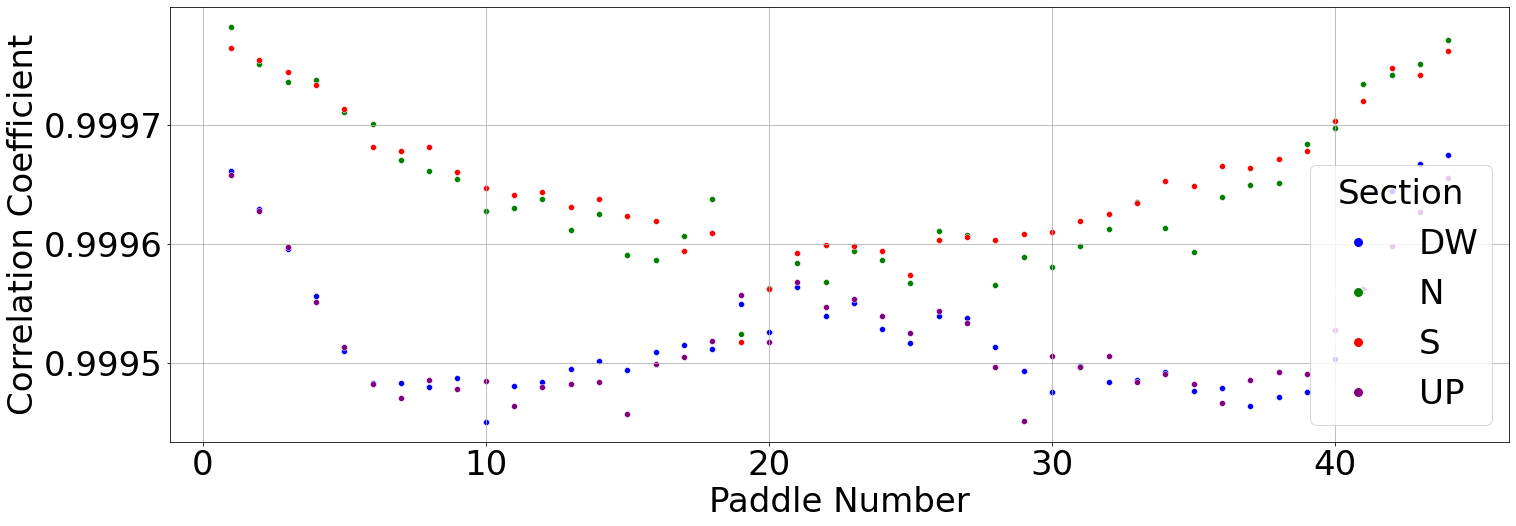}
    \caption{Pearson R correlation coefficients for the integrated TOF scalers and integrated beam current vs the TOF paddle number.}
    \label{fig:CC_beam+TOF}
\end{figure}

The normalized average gain by ring as a function of run number is shown in Fig. \ref{fig:normalizedAvgGain}. The effect is most prominent in ring 1 and gradually declines as the ring number increases. Ring 2 is a clear outlier, and this behavior is not understood. As a sanity check, we observe the correlations between the integrated beam current and the integrated TOF scalers, which should be strongly correlated. This relationship is shown in Fig. \ref{fig:CC_beam+TOF}. 

\begin{figure}[htb]
    \centering
    \includegraphics[width=0.8\linewidth]{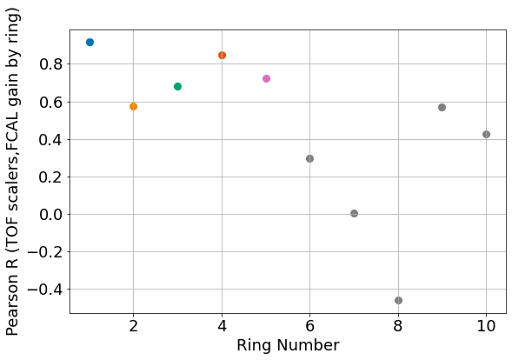}
    \caption{Pearson R correlation coefficients for the TOF scalers and FCAL gain values. The TOF scalers were averaged by paddle and the FCAL gain values were averaged by ring. The color of the dots in rings $<$ 6 correspond to the same rings shown in Fig. \ref{fig:normalizedAvgGain}. }
    \label{fig:pearsonR_ring_TOF}
\end{figure}

In Fig. \ref{fig:pearsonR_ring_TOF}, the correlation coefficients between the normalized average gain values from the inner FCAL rings and the average scaler values from the TOF (averaged over all paddles) is shown. The strongest correlations exist for the inner most ring closest to the beam line, as expected. The remaining fluctuations in the correlation values for rings 2-10 is unexpected. This investigation should be repeated for other Run Periods for comparison when the calibration method utilizing the LEDs is performed. 

\section{Conclusions and Outlook}
The AIEC project resulted in a maintainable, production control system for JLab's Hall-D CDC that autonomously controls the CDC's HV at the start of each run in an uncertainty-aware manner. 
The system allows for the model to be easily trained with additional data taken from previously unseen environmental and experimental conditions such that the model increases its certainty manifold. 
The ML-controlled HV has been shown to stabilize gain without degrading timing resolution. 
Additionally, the project resulted in a calibration process for the Hall-D FDC detector.

It may be obvious, but is important to emphasize that a control system relying on sensors will not function if those sensors are offline, and a contingency plan is required. 
For RoboCDC this occurred in Deployment \#5 when the control feature was disabled due to the atmospheric pressure sensor being offline.
The CDC remained at 2125 V, i.e. uncontrolled, which allowed experiments to continue as planned, and more training data was collected at the nominal HV.

Several additional detectors in JLab's Hall-B and Hall-D were investigated for ML-based calibration and control. 
A barrier to a data-driven control system is the lack of historical calibration data from diverse conditions.
While the original proposal expected the ML techniques applied to the Hall-D CDC to translate to CLAS12's Drift Chambers more than an order of magnitude fewer historical calibration entries made the effort infeasible.
Detectors also require sensors with measurement frequency and precision required for required calibration and control tasks. 
The BCAL detector pedestal is known to fluctuate with crate temperature; however, the only available temperature sensor lacked the precision required.

A ML method to learn the calibration values for the FCAL's 2800 blocks was investigated by a UVa capstone project team. 
There is a lack of correlation between the FCAL monitoring system's LED pulses and the gain calibration values, except for blocks near the center of the FCAL.
A concurrent and independent analysis agrees with the puzzling lack of correlation. 
Results from the capstone project led to a simulation study and an investigation of LSTM models to learn from the time series evolution of a block's gain.

Investigating the varied detectors in Hall-B and Hall-D allowed the project staff to make note of multiple key insights throughout the project. 
We offer these insights that future projects may benefit.

\paragraph{Key Insights}

\begin{enumerate}    
    \item The importance of the active participation of the detector expert cannot be overstated. 
    The fields of nuclear physics and data science have their own jargon, data visualization preferences, and professional cultures.
    The detector expert's inclusion throughout the project planning, data exploration, model development, system implementation, and system evaluation enabled the detector expert and data scientist to share a common project ``language."
    The common language facilitated the communication of complex requirements, for example, requirements for interpretable model output.
    Without the integration of a detector expert within the project team from the inception stage through execution, we expect any implementation of a production ML-based control system to be very challenging.
    \item We recommend a ``self-contained" control system, with minimal reliance on additional systems. 
    The system may still experience interruptions in sensor availability, as we did during the GlueX 2023 Run Period due to the malfunctioning atmospheric pressure sensor; 
    however, the early decision to isolate the CDC control system from other detector subsystems allowed us to meet our project objectives without unneeded dependencies.
    \item The inclusion of uncertainty quantification for safe ML-based control in autonomous high-risk experimental systems is important, as are the decision control plans when uncertainty exceeds selected thresholds.
    Gaussian process regression is the UQ method used for the CDC control system; additional techniques include ensemble ML methods \cite{ZOUNEMATKERMANI2021126266}, MC Dropout \cite{gal2016dropout}, Deep Quantile Regression \cite{koenker2005quantile}, and techniques that incorporate GP's within a deep learning model \cite{damianou2013deep, liu2020simple, rasmussen2006gaussian}.
 \item The historical record and documentation of traditional calibrations, including a record of the iterations of calibration values, proved useful throughout the CDC process and would have been helpful during the evaluation of the FCAL data.
 \item The FCAL simulation exercise yielded several insights:
    \begin{enumerate}    
        \item First, the importance of feature engineering that takes advantage of domain knowledge and specific knowledge of the underlying dynamics of the target signal.
        \item Second, the value of recursive deep learning techniques, specifically, LSTM modeling \cite{LSTM1997}, particularly as the target signal is considered as a sequence.
    \item Third, LSTM requires deeper history processing to curtail larger worst-case deviations of predicted signals from the target signal than average-case deviations..
    \end{enumerate}
\end{enumerate}

This body of work represents the first steps towards a ``self-driving laboratory", one in which intelligent machines are able to gather data, perform analyses, and self-correct. The AIEC program successfully developed and deployed an autonomous system for the GlueX CDC, allowing it to control its own HV, which in turn stabilized the gain correction factor with no significant impact on time-to-distance calibrations. RoboCDC is now in production, embedded as part of a modular control system as part of the standard operating procedure. The work, additionally, took steps beyond the single detector and probed applications to other detectors and halls. And even while those steps did not produce another autonomous detector system it did illuminate challenges with further applications, which helps inform the field on what needs to be done in the future to enable such applications. With the success of this project, another dedicated project, AI for Optimized Polarization (AIOP) is due to begin in 2024. It is the hope that this body of work provides both inspiration and technical details to others as we work towards self-driving laboratories and autonomous discoveries.

\section{AIEC Publications \& Presentations}

\subsection{Publications}

Jeske, T., McSpadden, D., Kalra, N., Britton, T., Jarvis, N., \& Lawrence, D. (2023, February). Using AI to predict calibration constants for the central drift chamber in GlueX at Jefferson Lab. In \textit{Journal of Physics: Conference Series} (Vol. 2438, No. 1, p. 012132). IOP Publishing. 
\\
\\
Jeske, T., McSpadden, D., Kalra, N., Britton, T., Jarvis, N., \& Lawrence, D. (2022). AI for Experimental Controls at Jefferson Lab. \textit{Journal of Instrumentation}, 17(03), C03043.

\subsection{Key Presentations}

\textbf{AI Driven Experiment Calibration and Control} \href{https://indico.jlab.org/event/459/contributions/11374/attachments/9497/13764/AIEC_CHEP_2023.pdf}{Slides} \\ \\
Britton, T., Lawrence, D., Jeske, T., McSpadden, D., \& Jarvis, N. (2023, May 8–12). AI Driven Experiment Calibration and Control [Conference presentation]. Conference on Computing in High Energy \& Nuclear Physics, Norfolk, VA, United States. \url{https://indico.jlab.org/event/459/contributions/11374/} \\ \\

\textbf{Using Machine Learning to control the GlueX Central Drift Chamber} \href{https://indico.cern.ch/event/1158681/contributions/5192980/attachments/2571444/4433709/NSJ_AIEC_HEP2023.pdf}{Slides} \\ \\
Jarvis, N., Britton, T., Jeske, T., Lawrence, D., \& McSpadden, D. (2023, January 9–13). Using Machine Learning to control the GlueX Central Drift Chamber [Conference presentation]. Conference on High Energy Physics in LHC Era, Valparaíso, Chile. \url{https://indico.cern.ch/event/1158681/contributions/5192980/} \\ \\
 
\textbf{Control and Calibration of GlueX Central Drift Chamber Using Gaussian Process Regression} 
\href{https://neurips.cc/media/PosterPDFs/NeurIPS%202022/56949.png}{Poster}  \href{https://ml4physicalsciences.github.io/2022/files/NeurIPS_ML4PS_2022_35.pdf}{Paper} \\ \\
McSpadden, D., Jeske, T., Jarvis, N., Britton, T., Lawrence, D., \& Kalra, N.  (2022, December 3). Control and Calibration of GlueX Central Drift Chamber Using Gaussian Process Regression [Poster Presentation]. Machine Learning and the Physical Sciences workshop at NeurIPS, New Orleans, LA, United States.  \url{https://ml4physicalsciences.github.io/2022/} \\ \\

\textbf{Gaussian process for calibration and control of GlueX Central Drift Chamber} \href{https://indico.cern.ch/event/1106990/contributions/4998092/attachments/2533066/4358817/ACAT_2022_PDF_Slides.pdf}{Slides} \\ \\
Jeske, T., Britton, T., Kalra, N., Jarvis, N., McSpadden, D. \&, Lawrence, D. (2022, October 23-28). Gaussian process for calibration and control of GlueX Central Drift Chamber [Conference Presentation]. Advanced Computing and Analysis Techniques in Physics Research, Bari, Italy.  \url{https://indico.cern.ch/event/1106990/contributions/4998092/} \\ \\

\section*{Acknowledgements}
\addcontentsline{toc}{section}{Acknowledgements}

This work is supported by a grant from the U.S. Department of Energy, Office of Science, Office of Nuclear Physics under the LAB-20-2261 FOA.\\
\newline
\noindent
The Carnegie Mellon Group is supported by the U.S. Department of Energy, Office of Science, Office of Nuclear Physics, DOE Grant No. DE-FG02-87ER40315\\
\newline
\noindent
This research used resources of the Thomas Jefferson National Accelerator Facility, which is a DOE Office of Science User Facility supported by the U.S. Department of Energy, Office of Science, Office of Nuclear Physics under contract DE-AC05-06OR23177.

\bibliographystyle{plainnat}
{
\small
\bibliography{main}
}

\begin{appendix}
\section{Control System Workflows}
\label{apx:flowcharts}
\subsection{Initialize}
\includegraphics[height=1.\textheight]{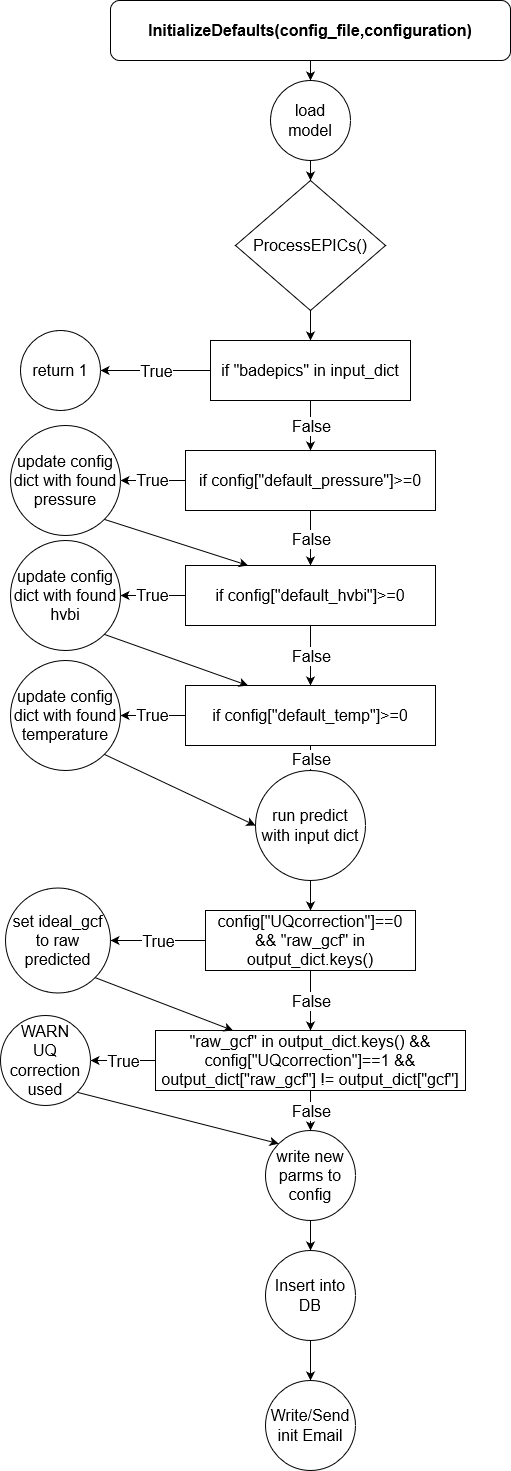}
\subsection{AutoOFF}
\includegraphics[width=1.\textwidth]{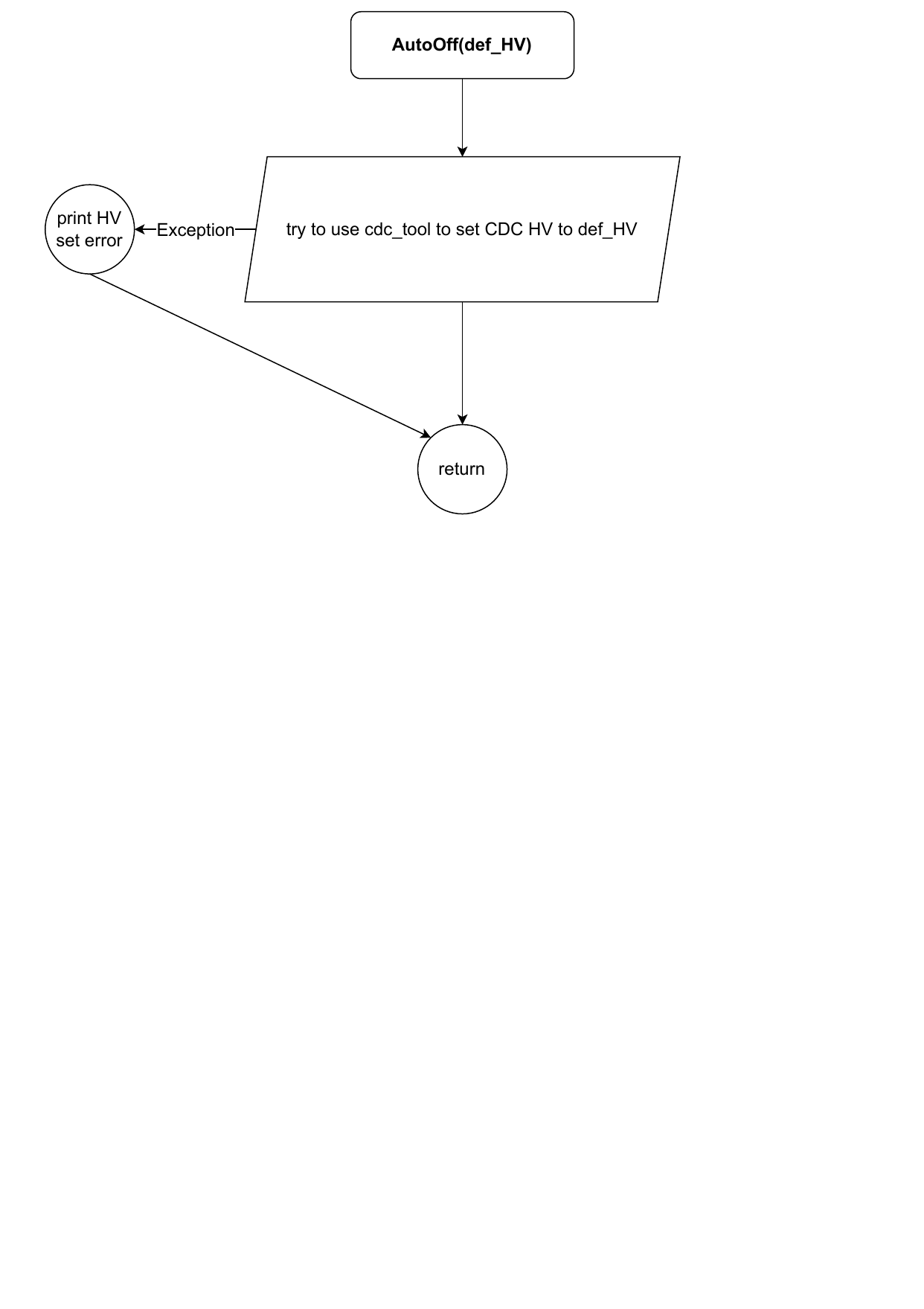}
\subsection{calcHV}
\includegraphics[width=1.\textwidth]{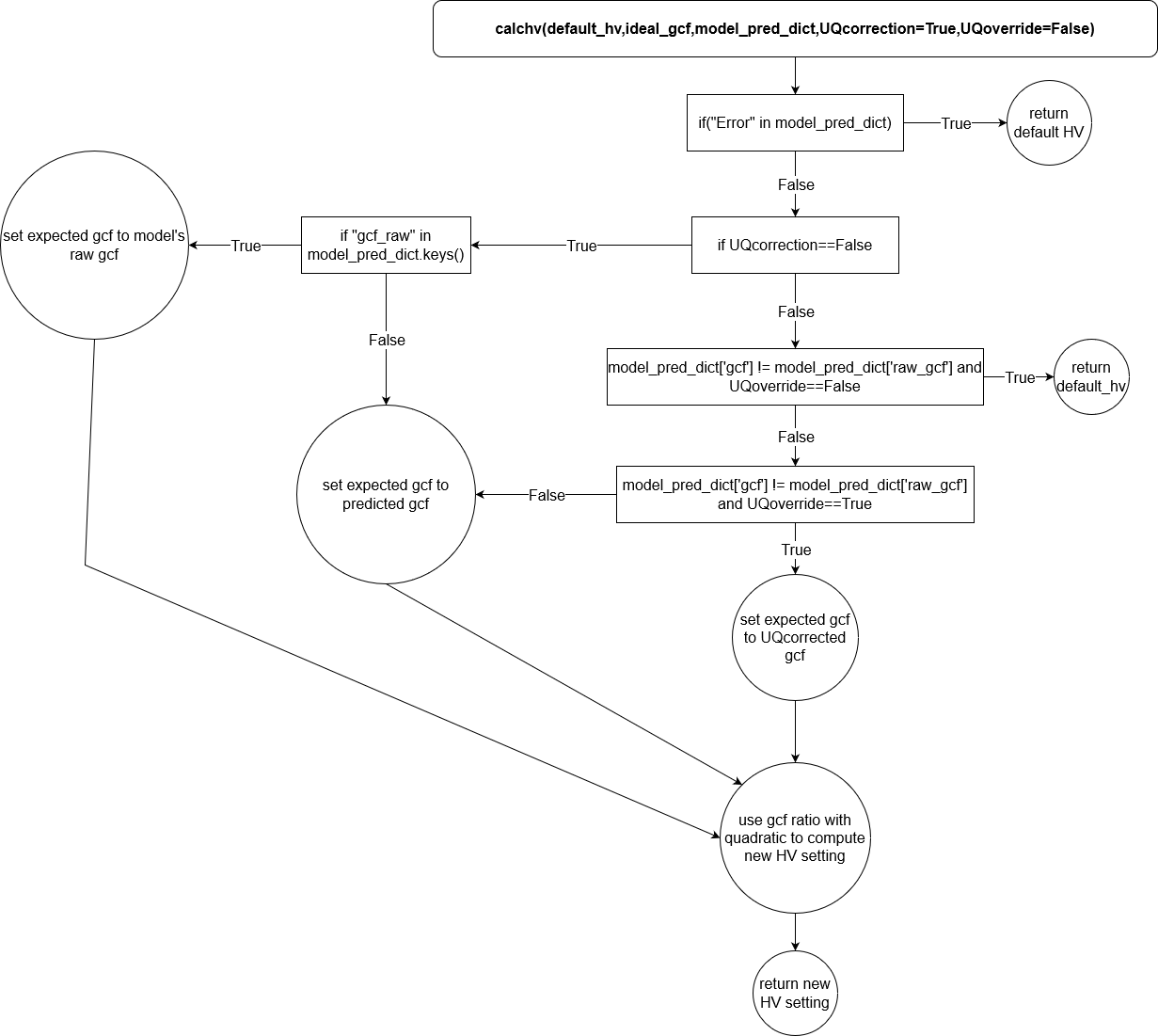}
\subsection{FailSafe}
\includegraphics[height=1.\textheight]{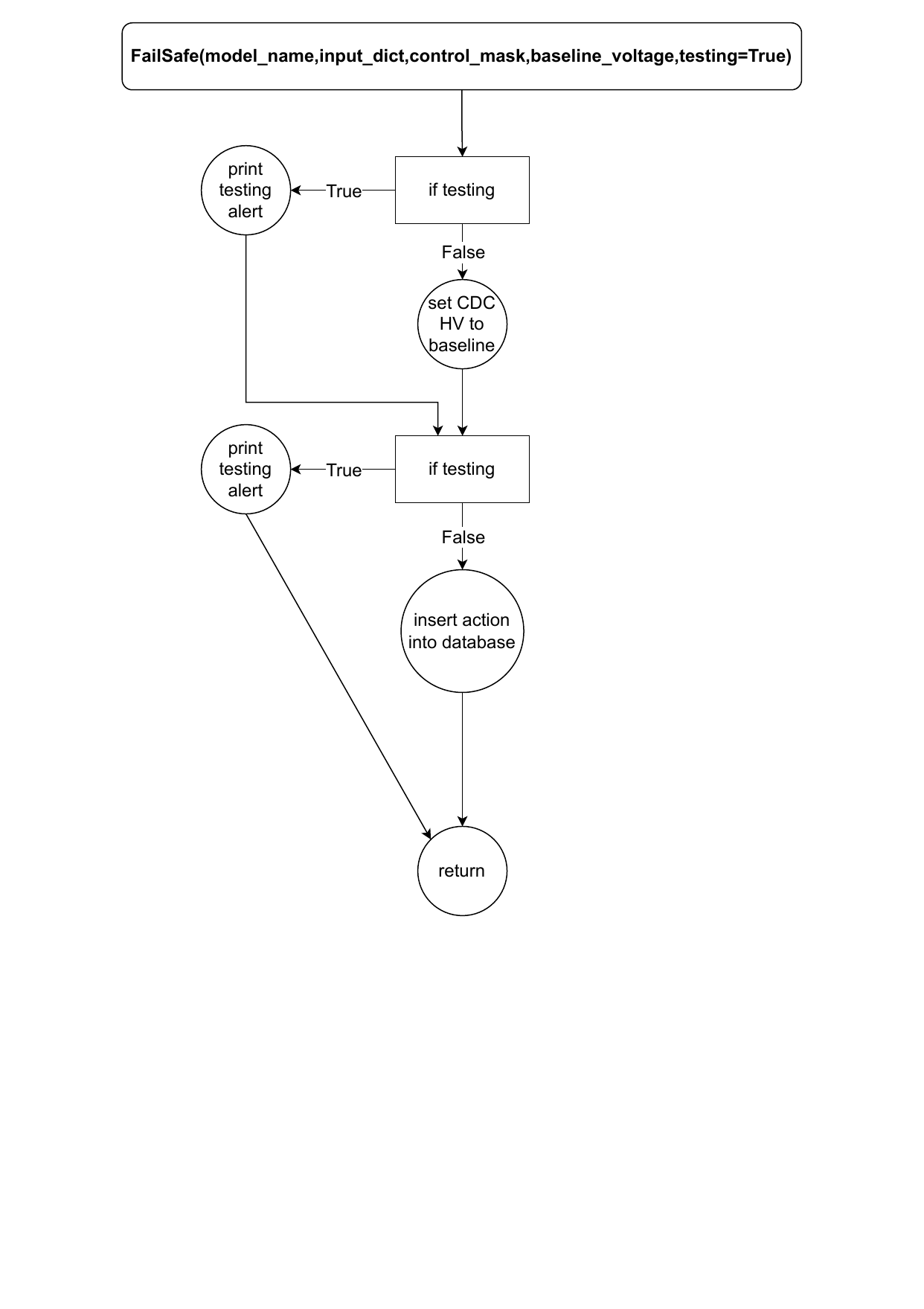}
\subsection{Lookback}
\includegraphics[height=0.9\textheight]{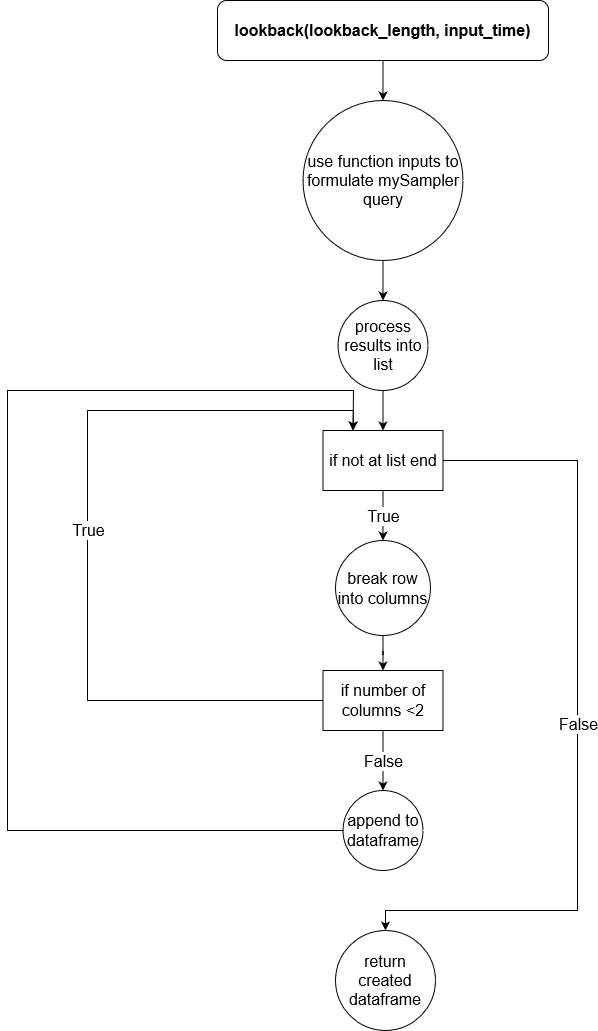}
\subsection{ProcessEPICS}
\includegraphics[height=1.\textheight]{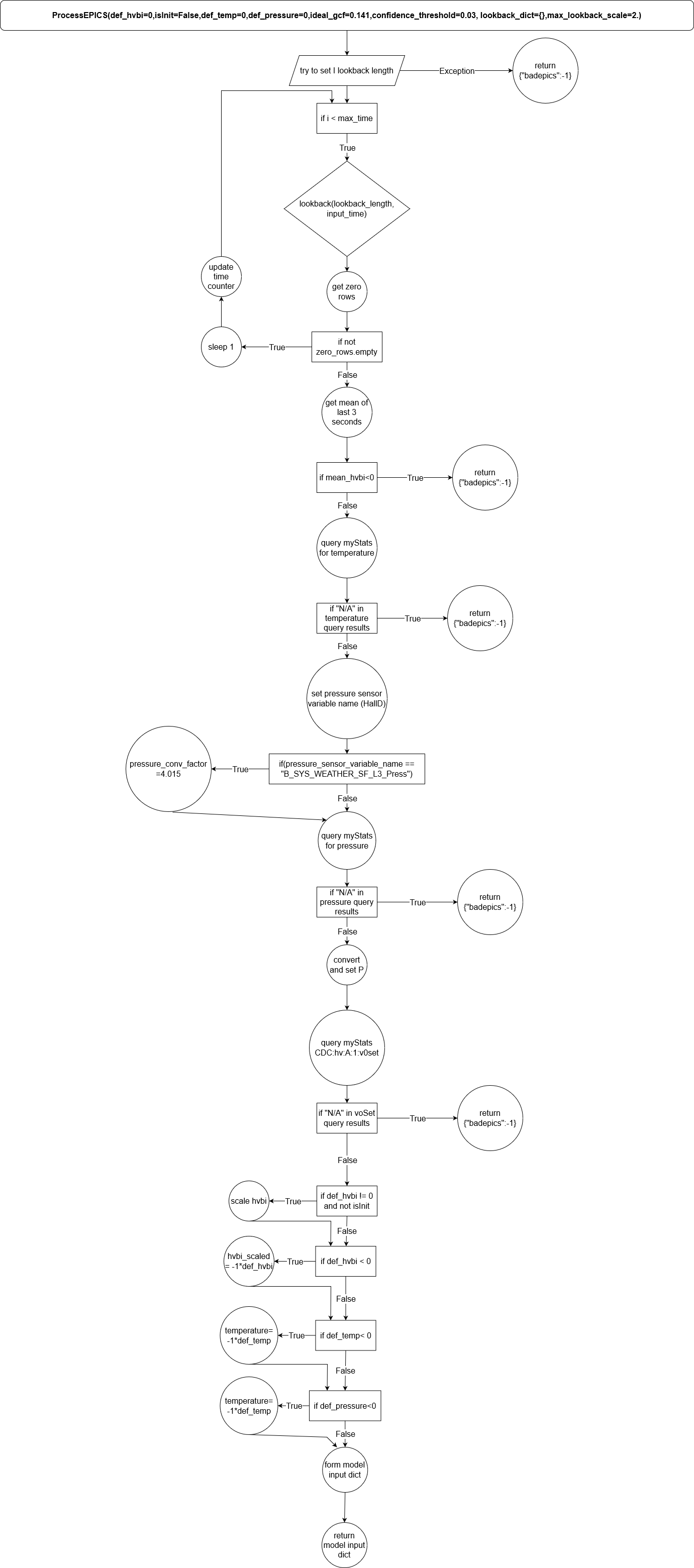}
\subsection{WatchProcessRecommend}
\includegraphics[width=1.\textwidth]{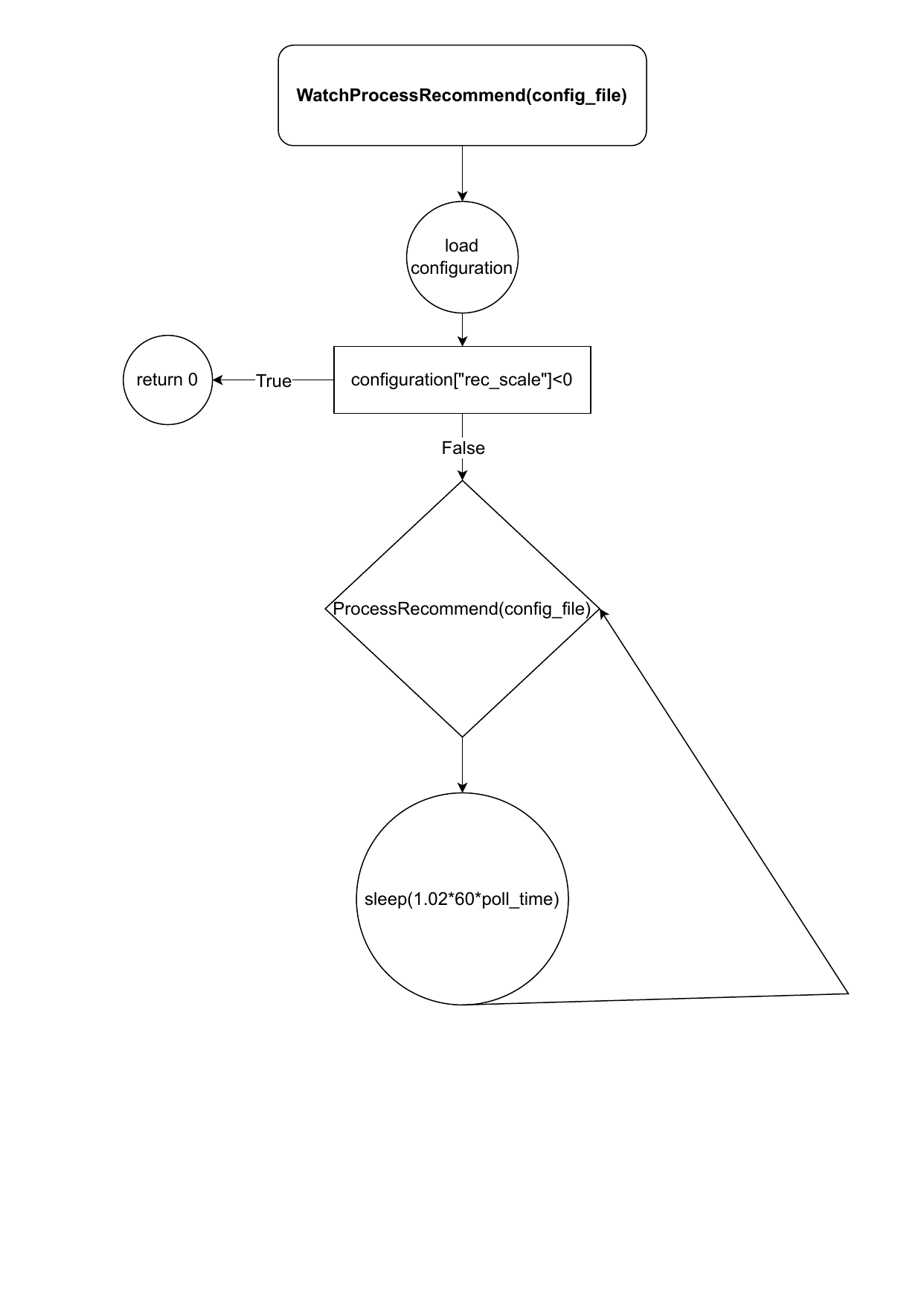}
\end{appendix}

\end{document}